\documentclass[twocolumn]{aastex631}

\usepackage{comment}
\usepackage{graphicx}
\usepackage{txfonts}
\usepackage{hyperref}

\newcommand{\rev}[1]{{\textcolor{black}{#1}}}
\newcommand{\rerev}[1]{{\textcolor{black}{#1}}}
\newcommand{\eqref}[1]{(\ref{#1})}

\shorttitle{On the Origin of Reflective Hazes}
\graphicspath{{./}{figures/}}
\begin{document} 

\title{Photochemical Hazes in Exoplanetary Skies with Diamonds: \\Microphysical Modeling of Haze Composition Evolution via Chemical Vapor Deposition}

\shortauthors{Ohno}

\author[0000-0003-3290-6758]{Kazumasa Ohno}
\affiliation{Division of Science, National Astronomical Observatory of Japan, 2-21-1 Osawa, Mitaka-shi, Tokyo, Japan}

\begin{abstract}
Observational efforts in the last decade suggest the prevalence of photochemical hazes in exoplanetary atmospheres.
Recent \emph{JWST} observations raise growing evidence that exoplanetary hazes tend to have reflective compositions, unlike the conventionally assumed haze analogs, such as tholin and soot.
In this study, I propose a novel hypothesis: diamond formation through chemical vapor deposition (CVD) may be happening in exoplanetary atmospheres.
Using an aerosol microphysical model combined with the theory of CVD diamond and soot formation established in the industry community, I study how the haze composition evolves in exoplanetary atmospheres for various planetary equilibrium temperature, atmospheric metallicity, and C/O ratio.
I find that CVD diamond growth dominates over soot growth in a wide range of planetary parameters.
Diamond haze formation is most efficient at $T_{\rm eq}\sim1000~{\rm K}$ and low atmospheric metallicity ([M/H]$\le2.0$), while soot could be the main haze component only if the atmosphere is hot ($T_{\rm eq}\ga1200~{\rm K}$) and carbon rich (C/O$>1$).
I also compute transmission, emission, and reflected light spectra, thereby suggesting possible observational signatures of diamond hazes, including the $3.53~{\rm {\mu}m}$ feature of hydrogenated diamonds, anomalously faint thermal emission due to thermal scattering, and a drastic increase in geometric albedo.
This study suggests that warm exoplanetary atmospheres may be favorable sites for forming CVD diamonds, which would be testable by future observations by \emph{JWST} and \emph{Ariel} as well as haze synthesis experiments under hot hydrogen-rich conditions.


\end{abstract}
\keywords{Exoplanet atmospheres(487) --- Transmission spectroscopy(2133) --- Atmospheric clouds(2180) --- Hot Jupiters(753) --- Astrochemistry(75)
       }
\section{Introduction}

It has been widely accepted that exoplanetary atmospheres are universally veiled by floating particles---often called clouds or hazes (e.g., \citealt[][]{Bean+10,Kreidberg+14,Knutson+14,Sing+16,Crossfield&Kreidberg17,Libby-Roberts+20,Dymont+22}, see \citet{Zhang20_review} and \citet{Gao+21} for review).
Clouds and hazes affect not only the observable atmospheric spectra but also various planetary properties, including atmospheric thermal structures \citep{Heng+12,Morley+15,Lavvas&Aufaux21,Arfaux&Lavvas22}, chemical compositions \citep{Molaverdikhani+20}, circulation \citep{Steinrueck+23,Teinturier+24}, and thermal evolution of the planet itself \citep{Poser+19,Poser&Redmer24}.
Thus, it is crucial to understand how clouds and hazes form in exoplanetary atmospheres.
\emph{JWST} is now able to provide detailed insights on the cloud/haze properties, such as the mineral composition of condensation clouds \citep{Grant+23,Dyrek+24} and their spatial distributions \citep{Bell+24,Schlawin+24,Murphy+24,Espinoza+24}, which helps to better understand the formation processes of exoplanetary aerosol.

Photochemical haze---particles whose formation is driven by atmospheric photochemistry---is one of the typical aerosols formed in planetary atmospheres.
Previous studies suggested that hazes form efficiently in warm exoplanets with an equilibrium temperature of $\la950~{\rm K}$ due to abundant CH$_4$ \citep{Morley+15,Kawashima&Ikoma19,Gao+20}.
Hazes may still form in hotter exoplanets where CH$_4$ is depleted, as experimental studies have shown that CO and CO$_2$ can serve as haze precursors \citep{He+19_CO,Fleury+19,He+20}.
If it is, the presence of hazes could also explain the muted spectral features and optical spectral slopes that are often seen in the transmission spectra of hot Jupiters \citep{Lavvas&Koskinen17,Ohno&Kawashima20,Steinrueck+21,Steinrueck+23,Arfaux&Lavvas22}. 

Despite their popularity, it remains highly uncertain what exoplanetary hazes are actually made of.
Previous studies used microphysical models to investigate how hazes form in exoplanets \citep[e.g.,][]{Lavvas&Koskinen17,Kawashima&Ikoma18,Kawashima&Ikoma19,Adams+19,Lavvas+19,Ohno&Kawashima20,Gao&Zhang20,Helling+20,Ohno&Tanaka21,Gao+23,Arfaux&Lavvas22,Arfaux&Lavvas24}. 
However, conventional microphysical models were not capable of tracing the substances that make up the haze particles.
Instead, previous studies often assumed that exoplanetary hazes resemble either Titan haze analog \citep[e.g.,][]{Kawashima&Ikoma18}---often called tholin \citep[][]{Sagan&Khare79}---or soot formed by high-temperature flame chemistry \citep[e.g.,][]{Morley+15,Lavvas&Koskinen17}.
Recent experimental studies attempted to synthesize exoplanetary haze analogs from gas mixtures that mimic exoplanetary atmospheres and found that tholin-like substances could be deposited \citep[e.g.,][]{Horst+18,He+18,He+20}, although their compositions are considerably different from that of Titan tholin \citep{Moran+20}.
Most of the current experiments are limited to metal-rich ($\ge100\times$ solar metallicity) warm ($300$--$600~{\rm K}$) gas mixtures \citep[e.g.,][]{Horst+18,He+18}.
Thus, a plausible haze composition for various exoplanetary environments has been largely uncertain in the current exoplanet community.


Recent observations of \emph{JWST} further pose a question about the conventional assumption that exoplanetary hazes resemble tholin or soot.
Previous studies predicted that dark-absorbing hazes, like soot and tholin, produce a temperature inversion in the atmosphere by absorbing stellar light \citep{Morley+15,Lavvas&Aufaux21,Steinrueck+23}, which should be readily detected by emission spectroscopy.
Recent studies used \emph{JWST} to observe the emission spectra of several warm\rev{-to-hot} planets, namely WASP-80b \citep{Bell+23}, WASP-69b \citep{Schlawin+24}, \rev{HD189733b \citep{Inglis+24_HD189} and WASP-17b \citep{Valentine+24}}, for which the transmission spectra \rev{potentially} indicate the presence of photochemical hazes \rev{from optical spectral slopes} \citep{Pont+13_hd189733_slope,Fukui+14,Murgas+20,Zhang20_hd189733b,Khalafinejad+21,Estrela+21,Alderson+22_wasp-17b_slope,Wong+22}.
However, they did not detect any signature of temperature inversion that should be present if tholin or soot hazes exist in the atmosphere.

There are several possibilities to explain the lack of temperature inversion.
One possible explanation is that haze particles have unexpectedly high reflective optical constants to avoid absorbing stellar lights.
Such ``reflective aerosol'' is suggested to be the cause of the unexpectedly faint thermal emission of GJ1214b \citep{Kempton+23}.
Some mineral clouds, such as KCl condensates that are expected to form in warm planets \citep{Morley+13,Mbarek&Kempton16,Ohno&Okuzumi18,Gao&Benneke18,Ohno+20}, have highly reflective optical constants and may serve as the ``reflective aerosol''.
However, current cloud microphysical models struggle to explain the spectral slope \citep{Gao&Benneke18,Powell+19,Gao+20} that is observed in the optical transmission spectrum of WASP-80b and WASP-69b, although the spectral slope may still emerge if the cloud particles are fractal aggregates \citep{Ohno+20}.
It also remains unclear why tholin and soot hazes are absent in the atmosphere.
Another possibility is that tholin or soot hazes do exist, but such hazy regions barely contribute to the emission spectrum due to stronger thermal emission from locally clear regions on the dayside \citep{Schlawin+24}.
One needs a global circulation model that includes the dynamics of haze particles \citep[e.g.,][]{Steinrueck+21,Steinrueck+23,Cohen+24} to test this hypothesis.

In this study, I investigate the first possibility: photochemical hazes in exoplanets may have an unexpectedly reflective composition.
In particular, I propose a novel hypothesis: {\it diamond formation via chemical vapor deposition may be operating in exoplanetary atmospheres}.
Chemical vapor deposition (CVD) is a well-established technology in the industrial community to synthesize diamonds under low pressure conditions (e.g. \citealt[][]{Angus&Hayman88,CVD_diamond_handbook,Butler+09}, see Section \ref{sec:CVD} for review).
In short, this technology deposits diamonds from the gas mixture of H$_2$ and a carbon-containing gas such as CH$_4$.
An efficient formation of CVD diamonds is possible at a hot temperature of $T{\sim}1000~{\rm K}$ and under the presence of an energy source that continuously produces atomic hydrogen.
Intriguingly, the atmospheres of close-in exoplanets satisfy the conditions to drive efficient formation of CVD diamonds, as elaborated in Section \ref{sec:CVD}.

Although diamond is a stable phase at high pressure, it is worthwhile to introduce that diamonds in low-pressure environments are common in the astronomical context.
Several meteorites are known to contain diamonds that are suggested to be of interstellar origin \citep[e.g.,][]{Lewis+87,Blake+88,Anders&Zinner93}.
Spectroscopic observations detected the spectral feature of diamonds with hydrogenated surfaces in molecular clouds \citep{Allamandola+92,Allamandola+93} and protoplanetary disks \citep{Guillois+99,VanKerckhoven02,Habart+04,Goto+09,Greaves+18}.
Several processes have been proposed to explain the origin of interstellar diamonds \citep[for review, see][]{Anders&Zinner93}.
\citet{Tielens+87} suggested that the high-speed collision between two dust grains caused by shock propagation can form interstellar diamonds. 
Meanwhile, several studies suggested that interstellar diamonds formed through the CVD process \citep[e.g.,][]{Lewis+87,Clayton+95,Daulton+96,Mutschke+04}, which I investigate in this study for exoplanetary atmospheres.
I note that this paper focuses on low-pressure diamond formation, which differs from previous studies that discussed the presence of diamond layers in the deep high-pressure interiors of exoplanets \citep{Kuchner&Seager05,Madhusudhan+12_diamond}. 


The organization of this paper is as follows.
In Section \ref{sec:CVD}, I briefly review the CVD diamond growth process and the soot growth process.
In Section \ref{sec:method}, I describe a new microphysical model to investigate the evolution of the haze composition.
In Section \ref{sec:result}, I use the model to investigate how the formation of diamond and soot haze depends on planetary equilibrium temperature, atmospheric metallicity, and atmospheric C/O ratio.
In Section \ref{sec:results_observation}, I investigate the impacts of diamond hazes on transmission, emission, and reflected light spectra of exoplanetary atmospheres to discuss their observational signature. 
In Section \ref{sec:discussion}, I discuss the relevance to existing experimental studies and another potential mechanism that might also form diamonds in exoplanetary atmospheres.
In Section \ref{sec:summary}, I summarize the findings of this paper.

\section{Overview of CVD diamond and Soot Formation}\label{sec:CVD}

\subsection{CVD Diamond}

In the industry community, it has been well known that diamonds can be produced even under low pressure such as $\sim 1~{\rm mbar}$ \citep[e.g.,][]{Spytsyn+81,Angus&Hayman88,Kobashi+88,Angus+94_review}.
Importantly, diamond is a metastable phase in low-pressure environments, as there is a large activation energy barrier \citep[45 kcal/mol,][]{Butenko+00} that prevents immediate transformation from diamond to graphite.
Thus, once it forms, the diamond can be stable for a long time even under low pressure.
The method of synthesis of diamonds at low pressure is called chemical vapor deposition (CVD) \citep[e.g.,][]{Celii&Butler91_CVD_review,Angus+94_review,CVD_diamond_handbook,May00_review,Butler+93_review,Butler+09,Schwander&Partes11_CVD_review}.

The basic procedure for CVD diamond synthesis is to introduce a certain energy source (e.g., a hot filament, microwave plasma) into a carbon-containing hydrogen gas mixture (e.g., H$_2$ + CH$_4$) at high temperature ($\sim1000~{\rm K}$), thereby depositing condensed diamonds on a substrate in the gas chamber \citep[e.g.,][]{Kobashi+88,Angus+94_review,Schwander&Partes11_CVD_review}.
The energy source is required to produce atomic hydrogen that plays a critical role in driving the preferential deposition of diamonds rather than black condensed carbons such as graphites \citep[e.g.,][]{Spytsyn+81,Kobashi+88,Setaka89}.
Non-diamond components such as graphite are also simultaneously deposited during CVD diamond synthesis.
However, atomic hydrogen erodes non-diamond components, the process called etching, much faster than that for diamond \citep[e.g.,][]{Hsu88,Donnelly+97_Etching,Kanai+01,Zhang+23_diamond_etching}, which drives the preferential deposition of diamonds.
Atomic hydrogen also reacts with carbon-containing molecules to produce reactive carbon radicals such as CH$_3$, which then serve as precursors to deposit fresh diamond layers.

Interestingly, the experimental environment of CVD diamond synthesis is considerably similar to the atmospheres of close-in exoplanets.
Gas giants and sub-Neptunes are expected to have hydrogen-dominated atmospheres with a trace amount of carbon-containing molecules, such as CH$_4$, CO and CO$_2$, which is similar to the gas mixture used as an ingredient for CVD diamonds.
The reader is referred to Section \ref{sec:bachmann} for a comparison between typical atmospheric compositions and gas compositions used in CVD experiments.
Depending on orbital distances and heat redistribution efficiency, the atmospheric temperature of close-in exoplanets can reach $\sim1000~{\rm K}$, which is comparable to the temperature suitable for CVD diamond synthesis.
CVD diamond synthesis requires an extra energy source to produce atomic hydrogen \citep[e.g.,][]{Kobashi+88,Angus+94_review}, whereas for exoplanets the central star can play the same role: intense stellar UV photons trigger photochemistry to continuously produce atomic hydrogen at upper atmospheres \citep[e.g.,][]{Zahnle+09_soot,Moses+11,Tsai+21}.
In fact, an experimental study of \citet{Fan+18} demonstrated that UV irradiation promotes CVD diamond growth and also improves diamond quality by suppressing the deposition of nondiamond carbons.
High-energy particles from the central star might also be solely responsible for forming diamonds at low pressure conditions (see Section \ref{sec:radiation_induced}).

The deposition processes of CVD diamonds have been extensively studied since the 1990s from theoretical \citep[e.g.,][]{Frenklach&Wang91,Coltrin&Dandy93,Goodwin93,Harris&Goodwin93,Yu&Girshick94} and experimental studies \citep[e.g.,][]{Martin&Hill90,Chu+90,Chu+91,Evelyn+92,Harris&Weiner92_C2H2,Lee+94_Diamond}.
The surface growth of CVD diamond is initiated by hydrogen absorption from the diamond surface.
Atomic hydrogen mainly drives the hydrogen abstraction in hydrogen-rich environments.
The surface growth processes can be described by the following reaction sequence in surface chemistry \citep[e.g.,][]{Goodwin93,Butler+93_review,Ashfold&Mankelevich23}:
\begin{equation}\label{eq:chem1}
    {\rm C\textrm{-}H+H\rightleftarrows  C\textrm{-}{*}+H_2},
\end{equation}
\begin{equation}\label{eq:chem2}
    {\rm C\textrm{-}{*}+H \rightarrow C\textrm{-}H},
\end{equation}
\begin{equation}\label{eq:chem3}
     \rev{{\rm  C\textrm{-}{*}+CH_{\rm 3} \rightarrow (C\textrm{-}H)_{\rm +1}+H_2}},
\end{equation}
where ${\rm C\textrm{-}H}$ represents a hydrogenated carbon on the surface of diamond, and ${\rm C\textrm{-}{*}}$ represents a carbon with dangling bond without termination by hydrogen, called a surface radical site.
The first and second reactions, as well as the reversed first reaction, mainly control how many reactive radical sites are present on the surface \citep[e.g.,][]{Goodwin93,May+07}. 
The third reaction incorporates the precursor hydrocarbons into a new diamond layer through the surface radical sites.
It has been widely accepted that CH$_3$ serves as the main precursor for CVD diamond growth \citep[e.g.,][]{Harris90,Harris&Weiner91_experiment,Butler+09}, as supported by multiple experimental evidence, such as the isotope labeling of CH$_3$ \citep[e.g.,][]{Martin&Hill90,Chu+90,Chu+91,Evelyn+92,Harris&Weiner92_C2H2,Lee+94_Diamond}.
Other first-order hydrocarbon radicals, such as C, CH, and CH$_2$, could also contribute to the growth of diamonds if their abundances are sufficiently high \citep{Coltrin&Dandy93,May+07}.

\begin{figure*}[t]
\centering
\includegraphics[clip, width=\hsize]{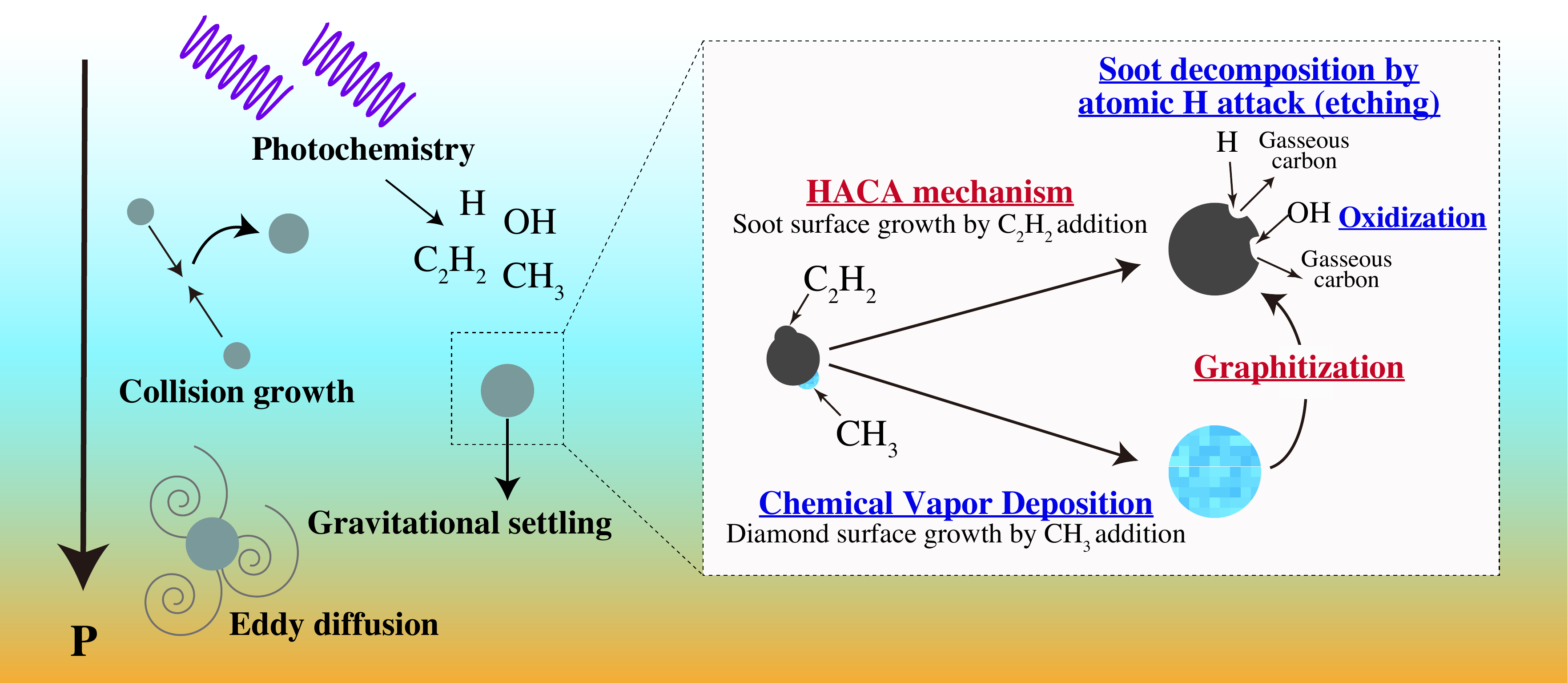}
\caption{Cartoon illustrating microphysical process considered in this study. Several processes, namely the HACA mechanism, CVD process, etching, and oxidization, deposit and decompose soot and CVD diamonds, which would drive the evolution of the haze compositions.
}
\label{fig:Cartoon}
\end{figure*}

\subsection{Soot}
Soot is defined as carbonaceous particles produced by pyrolysis or incomplete combustion of hydrocarbons \citep{Wang&Suk19_soot_review}.
The exoplanet community has speculated that photochemical hazes in exoplanetary atmospheres may resemble soot \citep{Morley+15,Lavvas&Koskinen17}, possibly motivated by the atmospheric temperature being as hot as the soot-forming fuel combustion.
Polycyclic aromatic hydrocarbons (PAHs) have been believed to serve as a precursor to nucleate soot particles \citep{Wang&Suk19_soot_review}. 
Subsequently, nucleated soot grows through the deposition of gas-phase hydrocarbons and PAHs (surface growth) as well as coagulation \citep{Wang&Suk19_soot_review}.

Intriguingly, the chemical process that drives the soot surface growth shares similarities with the surface growth process of CVD diamond. 
In industrial soot models, soot surface growth has been conventionally modeled by the so-called hydrogen-abstraction-C$_2$H$_2$-addition (HACA) mechanism \citep[e.g.,][]{Frenklach&Wang91_soot,Kazakov+95,Appel+00,Wang+15_soot}.
The HACA mechanism describes the soot surface growth through the following reaction sequence \citep[e.g.,][]{Wang&Suk19_soot_review}
\begin{equation}\label{eq:sootchem1}
    {\rm C_{\rm soot}\textrm{-}H+H\rightleftarrows  C\textrm{-}{*}+H_2},
\end{equation}
\begin{equation}\label{eq:sootchem2}
    {\rm C_{\rm soot}\textrm{-}{*}+H \rightarrow C\textrm{-}H},
\end{equation}
\begin{equation}
    {\rm  C_{\rm soot}\textrm{-}{*}+C_{\rm 2}H_{\rm 2} \rightarrow (C_{\rm soot}\textrm{-}H)_{\rm +2}+H},
\end{equation}
where ${\rm C_{\rm soot}\textrm{-}H}$ represents the hydrogenated surface site of soot, and ${\rm C_{\rm soot}\textrm{-}*}$ represents the radical surface site.
The surface growth of the soot is initiated by abstracting surface-terminating hydrogen by atomic hydrogen, allowing the insertion of C$_2$H$_2$ onto surface radical sites to drive further soot growth.
This is essentially the same as the CVD diamond growth process described in \eqref{eq:chem1}--\eqref{eq:chem3}, except that C$_2$H$_2$ serves as a soot precursor. 
The deposition of PAHs can also contribute to soot growth, but it dominates over the HACA mechanism when the temperature is relatively low \citep{Wang+15_soot}.

An important difference between soot formation environments and exoplanetary atmospheres is the difference in background gas compositions.
Soot formation occurs in highly carbon-rich gases produced by fuel combustion, whereas giant exoplanets have much more hydrogen-rich atmospheres. 
Several experimental studies have shown that the addition of H$_2$ acts to reduce soot production \citep[e.g.,][]{Gulder+96,Pandey+07,Xu+20_soot_hydrogen}, which has motivated the use of hydrogen in diesel engines \citep[e.g.,][]{Hosseini+23}.
Therefore, although in the same temperature regime, it is not trivial whether soot really represents exoplanetary hazes because of highly hydrogen-rich environments of exoplanetary atmospheres.

\section{Modeling strategy}\label{sec:method}

It is tempting to consider CVD diamond as a possible candidate of reflective hazes; however, there are many questions to be examined.
It is unclear whether CVD diamond growth is fast enough to affect haze compositions while haze particles stay in the observable upper atmosphere.
One should also investigate whether CVD diamond growth is faster than soot deposition; otherwise, soot would quickly coat haze particles to prohibit diamond deposition.
To assess these open questions, I developed a novel microphysical model that takes into account CVD diamond and soot deposition and decomposition, as well as vertical transport and collisional growth of haze particles (Figure \ref{fig:Cartoon}).
The following sections describe the details of the present model.

\subsection{Atmospheric Thermal and Chemical Structures}\label{sec:method_TP_VULCAN}

\begin{figure}[t]
\centering
\includegraphics[clip, width=\hsize]{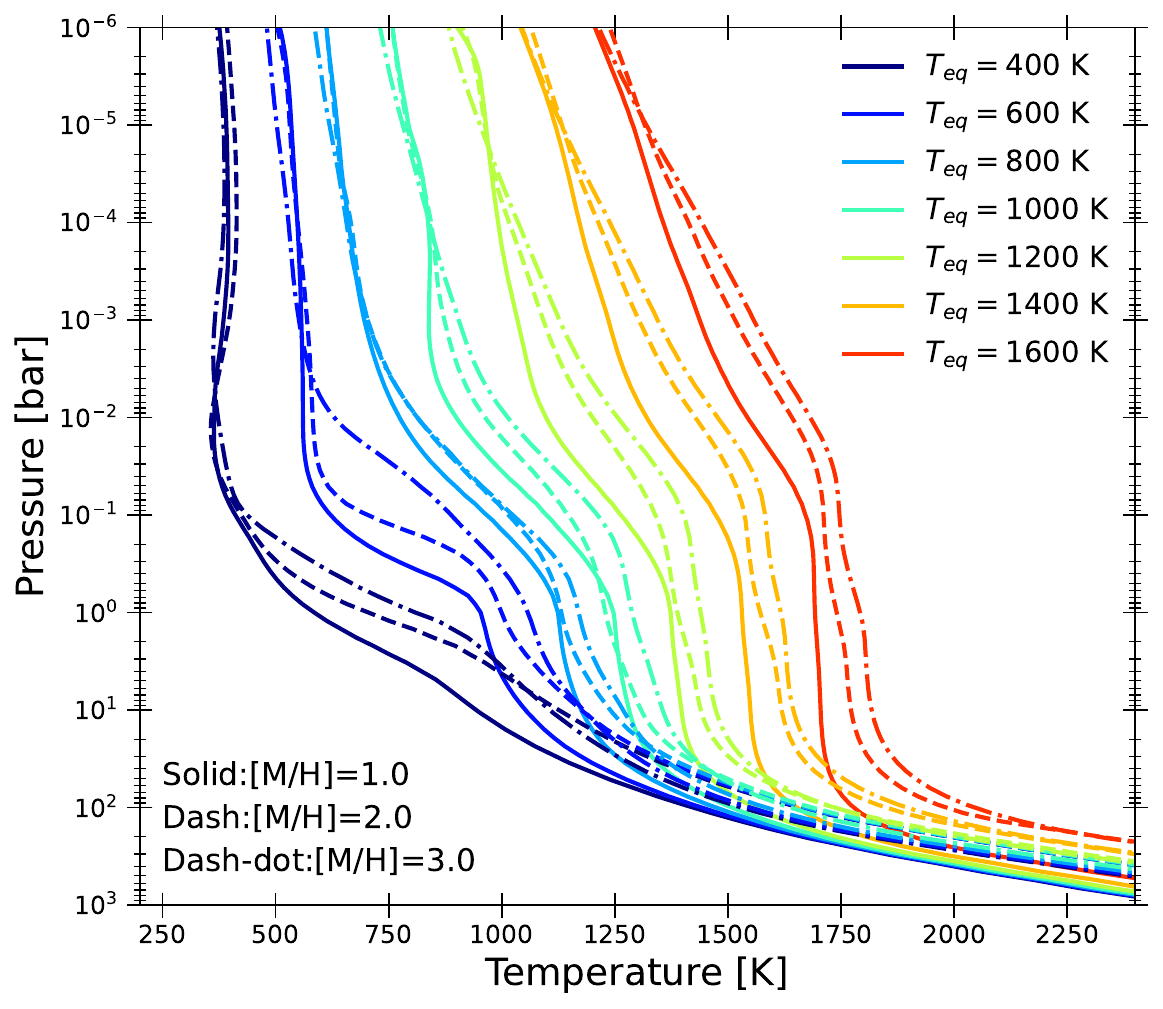}
\caption{Atmospheric TP profiles used in this study. Different colored lines show the profiles for different equilibrium temperature $T_{\rm eq}$. The solid, dashed, and dash-dot lines denote the profiles for atmospheric metallicity of [M/H]=1.0, 2.0, and 3.0, respectively. 
}
\label{fig:TP_haze}
\end{figure}

Atmospheric temperature is one of the key factors that controls the CVD diamond and soot growth rate, as introduced in Section \ref{sec:CVD_method}.
I use the 1D radiative-convective equilibrium model called \texttt{EGP} code \citep[e.g.,][]{McKay+89,Marley&McKay99,Fortney+05,Morley+13,Marley&Robinson15,Thorngren+19,Gao+20,Ohno&Fortney22a} to compute atmospheric temperature-pressure (TP) profiles for a variety of planetary equilibrium temperatures and atmospheric metallicities.
The model performs non-gray radiative transfer with correlated-k coefficients from \citet{Lupu+23}\footnote{The k coefficients for [M/H]$>$2 have not been publicly available and were kindly provided by R. Lupu for 1060 pressure-temperature grid points in personal communication.} under the assumption of thermochemical equilibrium.
I consider a GJ1214-like central star that has a radius of $0.215R_{\rm \odot}$, effective temperature of 3250 K, metallicity of [Fe/H]=0.29, and log surface gravity of $\log{g}=5.026$ \citep{Cloutier+21}.
The stellar spectrum is interpolated from the PHOENIX stellar grid in the Astrolib \texttt{PySynphot} package \citep{STScI}.
I vary a planetary orbital distance as a free parameter to change the equilibrium temperature with zero Bond albedo $T_{\rm eq}$ assuming full heat redistribution, planetary intrinsic temperature of $T_{\rm int}=100~{\rm K}$, and surface gravity of $g=10~{\rm m~s^{-2}}$. 
Figure \ref{fig:TP_haze} shows the computed TP profiles that will be used as input to the photochemical and microphysical models.
Note that the current model assumes clear atmospheres in the vertical 1D configuration for simplification. 
I will examine the effects of the radiative feedback of photochemical hazes in Section \ref{sec:TP_haze}.

The computed TP profiles are postprocessed to the open-source photochemical model \texttt{VULCAN} \citep{Tsai+17,Tsai+21} to compute the vertical distributions of molecular abundances.
I particularly focus on the abundance profiles of H, CH$_3$, C$_2$H$_2$, and OH which enable us to evaluate the growth and loss rates of CVD diamond and soot (Section \ref{sec:CVD_method}).
\texttt{VULCAN} solves the vertical transport equations with the production and loss terms of each chemical species.
I adopt the chemical network of \texttt{SNCHO\_full\_photo\_network} pre-implemented in \texttt{VULCAN} that includes 573 forward reactions and their reversed reactions (1146 reactions in total) as well as the photodissociation of 69 chemical species.
The eddy diffusion coefficient is assumed to obey the following functional form
\begin{equation}\label{eq:Kzz}
    K_{\rm zz}=3\times{10}^{8}~{\rm cm^2~s^{-1}}\left(\frac{P}{\rm bar}\right)^{-1/2}\left(\frac{T_{\rm eq}}{\rm 1000~{\rm K}}\right)^{5}\left(\frac{g}{\rm 10~{\rm m~s^{-2}}}\right)\left(\frac{m_{\rm g}}{\rm 2.3~{\rm amu}}\right)^{-1},
\end{equation}
where $P$ is the pressure, and $g$ is the surface gravity.
This formula is almost equivalent to the empirical $K_{\rm zz}$ profile of \citet{Moses+21}, but I have approximated the pressure scale height at $1~{\rm mbar}$ to $H_{\rm 1 mbar}\approx k_{\rm B}T_{\rm eq}/m_{\rm g}g$, where $k_{\rm B}$ is the Boltzmann constant, and $m_{\rm g}$ is the mean molecular mass.
Although many theoretical studies examined the eddy diffusion coefficients using global circulation models \citep[e.g.,][]{Parmentier+13,Charnay+15,Zhang&Showman18a,Zhang&Showman18b,Komacek+19,Steinrueck+21,Tan22}, the coefficient has been poorly constrained from an observational point of view \citep{Kawashima&Min21}.
Recent \emph{JWST} observations have begun to constrain $K_{\rm zz}$ from disequilibrium chemical species \citep{Welbanks+24,Sing+24}.
Future observations would obtain better insights on how atmospheric mixing takes place in exoplanets.

\subsection{Microphysical Model}
To investigate the growth, transport, and compositional evolution of haze particles, I use a two-moment microphyiscal model of cloud/haze formation developed by \cite{Ohno&Okuzumi18,Ohno+20,Ohno&Kawashima20}.
The model calculates the vertical distributions of the haze mass density and mean particle size by taking into account vertical transport by eddy diffusion and gravitational settling as well as particle growth by collisional sticking and condensation.
In this study, I assumed compact spherical particles and ignored the condensation of saturated mineral vapors for simplicity.

Since the model description can be found in previous publications \citep{Ohno&Okuzumi18,Ohno&Kawashima20}, here I only briefly explain the master equations in the model.
The model solves the advection-diffusion equation of particle number density $n_{\rm p}$ and mass densities. 
The former is given by
\begin{equation}\label{eq:np}
    \frac{\partial n_{\rm p}}{\partial t}=\frac{\partial}{\partial z}\left[ \rho_{\rm g}K_{\rm zz}\frac{\partial}{\partial z}\left( \frac{n_{\rm p}}{\rho_{\rm g}}\right)+v_{\rm t}n_{\rm p}\right]-\left| \frac{\partial n_{\rm p}}{\partial t}\right|_{\rm coll}+\frac{S}{m_{\rm p0}},
\end{equation}
where $\rho_{\rm g}$ is the atmospheric mass density, $v_{\rm t}$ is the particle terminal velocity, and $S$ and $m_{\rm p0}$ are the production rate and mass of initial seed particles.
The seed particles are assumed to have radii of $1~{\rm nm}$, \rev{as in some previous studies of haze modeling \citep[e.g.,][]{Adams+19,Ohno&Kawashima20}}.
The second term on the right-hand side represents the rate of decrease in number density due to collisional sticking. 
The terminal velocity is given by \citep{Ohno&Okuzumi17,Ohno&Okuzumi18}
\begin{equation}
    v_{\rm t}=\frac{2\rho_{\rm p}g}{9\eta}r_{\rm p}^2\beta \left[ 1+\left( \frac{0.45gr_{\rm p}^3\rho_{\rm g}\rho_{\rm p}}{54\eta^2}\right)^{2/5}\right]^{-5/4},
\end{equation}
where $\rho_{\rm p}$ is the particle internal density, $r_{\rm p}$ is the particle radius, $\eta$ is the dynamic viscosity of the atmosphere, and $\beta$ is the slip correction factor \citep[e.g.,][]{Davies45} that accounts for the transition from the viscous flow (Stokes) regime to the kinetic (Epstein) regime.
The bracket part is a correction factor that accounts for the transition from the Stokes regime to the turbulent flow (Newton) regime \citep[see Appendix of][]{Ohno&Okuzumi17}.

Since the formation process of initial seed particles of photochemical hazes are still highly uncertain, I prescribe the seed production rate by a log-normal distribution given by \citep{Ohno&Kawashima20}
\begin{equation}
    S=\rho_{\rm g}g\frac{F_{\rm haze}}{\sigma P\sqrt{2\pi}}\exp{\left[ -\frac{1}{2}\left( \frac{\ln{(P/P_{\rm *})}}{\sigma}\right)^2 \right]},
\end{equation}
where $F_{\rm haze}$ is the column-integrated mass production rate of seed particles, and $P_{\rm *}$ is the peak pressure level of seed production, and $\sigma$ is the width of the production profile.
I adopt $P_{\rm *}={10}^{-6}~{\rm bar}$ and $\sigma=0.5$ as in \citet{Ohno&Kawashima20}.
For fiducial simulations, I assume the \rev{seed mass flux of $F_{\rm haze}={10}^{-14}~{\rm g~{cm}^{-2}~s^{-1}}$}, which is comparable to the haze mass flux estimated for Titan \citep[$3\times{10}^{-14}~{\rm g~{cm}^{-2}~s^{-1}}$; e.g.,][]{McKay+01,Lavvas+11}, Pluto \citep[$1.2\times{10}^{-14}~{\rm g~{cm}^{-2}~s^{-1}}$;][]{Gao+17}, and Triton \citep[$\sim2$--$8\times{10}^{-15}~{\rm g~{cm}^{-2}~s^{-1}}$;][]{Ohno+21}.
I note that the haze mass flux in the present study becomes much higher than the column-integrated seed production rate $F_{\rm haze}$ because the present model explicitly simulates the incorporation of gas-phase hydrocarbons into haze particles, as explained below.

To investigate the fraction of CVD diamonds and soot within haze particles, I have updated the original model to treat the mixed composition of aerosol particles.
To this end, I separately simulate the transport equations of haze mass density for CVD diamonds $\rho_{\rm dia}$ and soot $\rho_{\rm soot}$ as \rev{\citep[this approach is essentially the same as][]{Huang+24}}
\begin{equation}\label{eq:master_diamond}
    \frac{\partial \rho_{\rm dia}}{\partial t}=\frac{\partial}{\partial z}\left[ \rho_{\rm g}K_{\rm zz}\frac{\partial}{\partial z}\left( \frac{\rho_{\rm dia}}{\rho_{\rm g}}\right)+v_{\rm t}\rho_{\rm dia}\right]+f_{\rm dia}\frac{dm_{\rm dia}}{dt}n_{\rm p}+\epsilon_{\rm dia}S
\end{equation}
and
\begin{equation}\label{eq:master_soot}
    \frac{\partial \rho_{\rm soot}}{\partial t}=\frac{\partial}{\partial z}\left[ \rho_{\rm g}K_{\rm zz}\frac{\partial}{\partial z}\left( \frac{\rho_{\rm soot}}{\rho_{\rm g}}\right)+v_{\rm t}\rho_{\rm soot}\right]+(1-f_{\rm dia})\frac{dm_{\rm soot}}{dt}n_{\rm p}+(1-\epsilon_{\rm dia})S,
\end{equation}
where $\epsilon_{\rm dia}$ is the initial mass fraction of CVD diamonds contained in the seed particles, which is set to $\epsilon_{\rm dia}=0.1$. 
The second terms in Equations \ref{eq:master_diamond} and \ref{eq:master_soot} represent the surface growth and erosion rates of CVD diamonds and soot, respectively, which I introduce in the next section.
$f_{\rm dia}$ is the volume fraction of CVD diamond within haze particles, defined as
\begin{equation}
    f_{\rm dia}=\frac{\rho_{\rm dia}/\rho_{\rm d0}}{\rho_{\rm dia}/\rho_{\rm d0}+\rho_{\rm soot}/\rho_{\rm s0}}.
\end{equation}
where $\rho_{\rm d0}=3.5~{\rm g/{cm}^3}$ and $\rho_{\rm s0}=1~{\rm g/{cm}^3}$ are the material density of CVD diamond \citep{Partridge+94} and soot, respectively.
\rerev{Here, I recall the reader that the present model has postulated haze particles within which CVD diamonds and soot are uniformly mixed, and diamond/soot deposition can take place only on the particle surface covered by diamond/soot.
Thus, the prefactors $f_{\rm dia}$ and $(1-f_{\rm dia})$ account for surface fractions available for CVD diamonds and soot deposition.  
}
The total mass density, mean particle radius, and internal density of haze particles can be calculated as
\begin{equation}
    \rho_{\rm haze}=\rho_{\rm dia}+\rho_{\rm soot},
\end{equation}
\begin{equation}
    r_{\rm p}=\left( \frac{3}{4\pi}\frac{\rho_{\rm dia}/\rho_{\rm d0}+\rho_{\rm soot}/\rho_{\rm s0}}{n_{\rm p}}\right)^{1/3},
\end{equation}
and 
\begin{equation}
    \rho_{\rm p}=\frac{\rho_{\rm haze}}{\rho_{\rm dia}/\rho_{\rm d0}+\rho_{\rm soot}/\rho_{\rm s0}}.
\end{equation}


\subsection{Microphysics of CVD diamond and soot formation}\label{sec:CVD_method}

In this study, I newly implemented the surface growth and erosion of CVD diamonds and soot in the previously used microphysical model \citep{Ohno&Okuzumi18,Ohno+20,Ohno&Kawashima20}.
In the industry community, there have been many theoretical studies that investigates the formation process of CVD diamond \citep[e.g.,][]{Frenklach&Wang91,Coltrin&Dandy93,Goodwin93,Harris&Goodwin93,Yu&Girshick94,May+07,May&Mankelevich08,Ashfold&Mankelevich23} and soot \citep[e.g.,][]{Frenklach&Wang91_soot,Kazakov+95,Appel+00,Wang11_soot_review,Wang+15_soot,Wang&Suk19_soot_review}.
This study utilizes the heritage of those works to investigate the CVD diamond and soot formation processes in exoplanetary atmospheres for the first time.

I consider several key processes that control the growth and decomposition of CVD diamonds and soot.
The rates of change in CVD diamond mass for individual haze particles is modeled by
\begin{equation}\label{eq:dmdia_dt_sum}
    \frac{dm_{\rm dia}}{dt}=\left| \frac{dm_{\rm dia}}{dt}\right|_{\rm CVD}-\left| \frac{dm_{\rm dia}}{dt}\right|_{\rm grap},
\end{equation}
where the first term represents the surface growth via chemical vapor deposition, and the second term represents the conversion to soot component (graphitazation).
I also model the rates of change in soot mass as
\begin{equation}\label{eq:dmsoot_dt_sum}
    \frac{dm_{\rm soot}}{dt}=\left| \frac{dm_{\rm soot}}{dt}\right|_{\rm HACA}-\left| \frac{dm_{\rm soot}}{dt}\right|_{\rm etch}-\left| \frac{dm_{\rm soot}}{dt}\right|_{\rm oxid}+\left| \frac{dm_{\rm dia}}{dt}\right|_{\rm grap},
\end{equation}
where the first term expresses the surface growth through the HACA process, the second and third term stand for the soot decomposition through atomic hydrogen attack (etching) and oxidization, and the last term expresses the mass gain by the graphitization of CVD diamonds.
Each rate was often derived as a linear growth/loss rate $dr/dt$ in the literature.
In that case, the linear rate is converted to the mass growth rate using a relation of 
\begin{equation}
    \frac{dm_{\rm dia/soot}}{dt}=4\pi r_{\rm p}^2\rho_{\rm d0/s0}\frac{dr}{dt}.
\end{equation}
In the subsequent subsections, I elaborate each growth and decomposition process of CVD diamonds and soot.

\subsubsection{Surface growth of CVD diamond}
\begin{figure}[t]
\centering
\includegraphics[clip, width=\hsize]{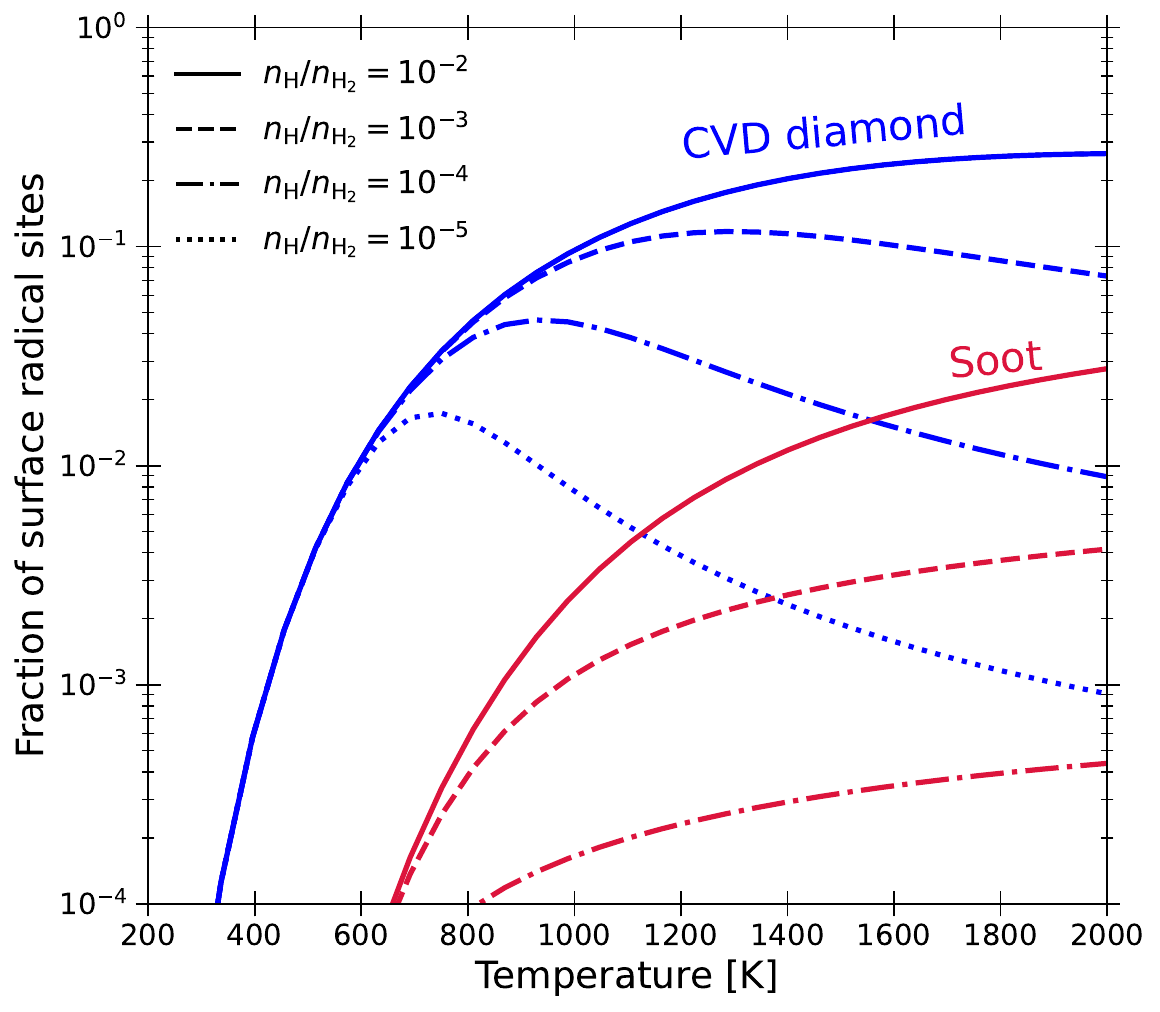}
\caption{Fraction of surface radical sites on diamond $R_{\rm dia}^{*}$ and soot $R_{\rm soot}^{*}$ as a function of temperature. Solid, dashed, dash-dotted and dotted lines show the radical site fraction for the atomic hydrogen abundances of $n_{\rm H}/n_{\rm H2}={10}^{-2}$, ${10}^{-3}$, ${10}^{-4}$, and ${10}^{-5}$ respectively.
}
\label{fig:radical_site}
\end{figure}

The surface growth rate of CVD diamond is known to be approximately proportional to the number density of CH$_3$ in surrounding gases \citep[e.g.,][]{Goodwin93,Harris&Goodwin93,May+07}.
I utilize the linear growth rate of \citet{Ashfold&Mankelevich23} given by
\begin{equation}\label{eq:CVD_diamond}
    \left|\frac{dr}{dt}\right|_{\rm CVD}=2\times{10}^{-13}fT^{1/2} n_{\rm CH_3}R_{\rm dia}^{*}~{\rm {\mu}m~h^{-1}},
\end{equation}
where $T$ is temperature in Kelvin, $n_{\rm CH_3}$ is the number density of CH$_3$ in cgs unit of cm$^{-3}$, and $f$ is the empirical factor accounting for the probability of irreversible incorporation of CH$_3$ into diamond lattice.
\citet{Ashfold&Mankelevich23} estimated $f=0.03$ for nearly nitrogen-free environments. 
It is interesting to note that nitrogen was found to promote CVD diamond growth \citep{Jin&Moustakas94,Dunst+09,Nakano+22} and an optimal amount of nitrogen can increase $f$ by a factor of $\sim3$.
It has been shown that such a simple rate equation \eqref{eq:CVD_diamond} can well explain the experimentally measured rate of CVD diamonds \citep{Harris&Goodwin93,Yu&Girshick94,May+07}.

The factor $R_{\rm dia}^{*}$ is the fraction of surface radical sites.
One can estimate $R_{\rm dia}^{*}$ from the steady state balance between hydrogen abstraction and recombination reactions (i.e., \eqref{eq:chem1} and \eqref{eq:chem2}) \citep[e.g.,][]{Goodwin93,May+07}.
This procedure yields the fraction of surface radical sites given by \citep{May+07}
\begin{eqnarray}\label{eq:radical_site}  
    R_{\rm dia}^{*}&=&\frac{1}{1+k_{\rm 2}/k_{\rm 1}+(k_{\rm -1}/k_{\rm 1})(n_{\rm H_2}/n_{\rm H})}\\
    \nonumber
    &=&\frac{1}{1+0.3\exp{(3430/T)}+0.1\exp{(-4420/T)}(n_{\rm H_2}/n_{\rm H})}
\end{eqnarray}
where $k_{\rm 1}$ and $k_{\rm -1}$ are the forward and reversed rate constants for the Reaction \eqref{eq:chem1}, $k_{\rm 2}$ is the rate constant for Reaction \eqref{eq:chem2}, and $n_{\rm H}$ and $n_{\rm H2}$ are the number density of gas-phase atomic hydrogen and molecular hydrogen.

The temperature dependence of the radical site fraction can explain why CVD diamond growth prefers hot temperature.
Figure \ref{fig:radical_site} shows the fraction of surface radical sites as a function of temperature for several atomic hydrogen abundances.
The fraction of surface radical sites decreases exponentially with decreasing temperature at $T\la 800~{\rm K}$ due to too slow hydrogen abstraction (Reaction \ref{eq:chem1}) compared to recombination (Reaction \ref{eq:chem2}).
The radical site fraction reaches $R_{\rm dia}^{*}\sim0.1$ at high temperature if the atomic hydrogen abundance is high, while it has a maximal value at ${\sim}600$--$1000~{\rm K}$ and decreases with temperature for low abundances of atomic hydrogen (say $n_{\rm H}/n_{\rm H_2}\la{10}^{-3}$).

Since CVD diamond growth takes place under non-equilibrium conditions \citep{Wang+98_Ternay,Wan+98}, I apply the CVD diamond growth (Equation \ref{eq:CVD_diamond}) only when the atomic hydrogen abundance is much higher than the thermochemical equilibrium value.
Assuming that molecular hydrogen abundance is always in equilibrium, the law of mass action yields the equilibrium volume mixing ratio of atomic hydrogen as
\begin{equation}\label{eq:vmr_H_eq}
    q_{\rm H,eq}=\left[ K_{\rm eq}^{-1}\left(\frac{P_{\rm 0}}{P}\right)q_{\rm H_2}\right]^{1/2},
\end{equation}
where $K_{\rm eq}=\exp{[-(\Delta G[H_2]-2\Delta G[H])]}$ is the equilibrium constant, and the standard Gibbs free energy $\Delta G$ is computed from the NASA Polynomials.
Equation \eqref{eq:vmr_H_eq} needs to introduce the standard-state prssure $P_{\rm 0}=1~{\rm bar}$, as the number of reactants is not equal to the number of products in the reaction of H+H$\rightleftarrows {\rm H}_{\rm 2}$ \citep[see e.g.,][]{Vischer&Moses11,Heng+16,Tsai+17}.
I switch on CVD diamond growth (Equation \ref{eq:CVD_diamond}) only when the volume mixing ratio of atomic hydrogen exceeds $3q_{\rm H,eq}$, which allows CVD diamond growth only in upper atmospheres where photochemistry produces super-equilibrium abundance of atomic hydrogen.

\subsubsection{Surface growth of soot}
I implement the soot surface growth through HACA mechanism (see Section \ref{sec:CVD}), as commonly adopted in industrial soot models \citep[e.g.,][]{Frenklach&Wang91_soot,Appel+00,Wang+15_soot}.
Following \citet{Frenklach&Wang91_soot}, the surface growth rate by incorporating C$_2$H$_2$ into soot can be described as
\begin{equation}\label{eq:dm_soot/dt}
    \left|\frac{dm_{\rm soot}}{dt}\right|_{\rm HACA}=4\pi r_{\rm p}^2 k_{\rm soot}n_{\rm C_2H_2}\alpha_{\rm soot}\chi_{\rm s}R_{\rm soot}^{*}\times 2m_{\rm C},
\end{equation}
where $n_{\rm C2H2}$ is the number density of C$_2$H$_2$, $\alpha_{\rm soot}$ is the reduction factor accounting for surface aging of soot which is set to $\alpha_{\rm soot}=0.3$ \citep{Frenklach&Wang94}, $\chi_{\rm s}=2.3\times{10}^{15}~{\rm sites~cm^{-2}}$ is the number density of total surface sites \citep{Frenklach&Wang91_soot}, $m_{\rm C}=12~{\rm amu}$ is the mass of carbon atoms, and $k_{\rm soot}$ is the per-site rate coefficient given by \citep{Appel+00}
\begin{equation}
    k_{\rm soot}=1.3\times{10}^{-16}T^{1.56}\exp{\left(-\frac{E}{RT}\right)} ~{\rm {cm}^{3}~s^{-1}},
\end{equation}
where $R$ is the universal gas constant, and $E=3.8~{\rm kcal~{mol}^{-1}}$.

$R_{\rm soot}^{*}$ is the fraction of radical sites on the soot surface, which can be estimated as similar to that for CVD diamond.
The radical fraction can be expressed by \citep[][]{Frenklach&Wang91_soot}
\begin{eqnarray}\label{eq:radical_site_soot}  
    R_{\rm soot}^{*}&=&\frac{1}{1+k_{\rm s2}/k_{\rm s1}+(k_{\rm s-1}/k_{\rm s1})(n_{\rm H_2}/n_{\rm H})}\\
    \nonumber
    &=&\frac{1}{1+0.47\exp{(6541/T)}+0.09\exp{(1851/T)}(n_{\rm H_2}/n_{\rm H})}
\end{eqnarray}
where $k_{\rm s1}$ and $k_{\rm s-1}$ are the rate constants of forward and reversed reactions of \eqref{eq:sootchem1}, $k_{\rm s2}$ is the rate constant of forward reaction \eqref{eq:sootchem2}, and $T$ is temperature in Kelvin.
For the second formula, I have inserted the rate constants provided by \citet{Wang+15_soot}.
Note that \citet{Frenklach&Wang91_soot} included the loss of radical sites by reactions with C$_2$H$_2$ and O$_2$, while I omit these effects for simplicity.
The surface radical fraction of soot is also shown in Figure \ref{fig:radical_site}, demonstrating that soot growth through the HACA mechanism also prefers a hot temperature.

In this study, I ignore the surface growth of soot through PAH deposition, since the PAH abundance in exoplanetary atmospheres is highly uncertain.
In the soot model of \citet{Wang+15_soot}, pyrene (C$_{16}$H$_{\rm 10}$) is the smallest PAH that can contribute to the surface growth of soot. 
Studies on PAHs in exoplanets have just begun very recently \citep{Ercolano+22,Dubey+23}.
Future studies on PAH chemistry in exoplanetary atmospheres would help to better understand how PAH affects the surface growth of haze particles.
Since the HACA mechanism is designed to model the PAH growth \citep[e.g.,][]{Frenklach&Wang91_soot,Wang&Suk19_soot_review}, it would be interesting to implement the HACA mechanism into the chemical network used for exoplanet modeling.

\subsubsection{Conversion from diamond to soot (graphitization)}\label{eq:graphitization}
Since diamond is a metastable phase under low pressure, diamond is continuously converted to a stable graphite phase (graphitization).
The activation energy of graphitization was measured by \citet{Davies&Evans72} and \citet{Butenko+00}.
\citet{Butenko+00} found that different graphitization mechanisms operate at high and low temperature regimes, for which each mechanism takes a different activation energy. 
To account for this, I model the graphitization as
\begin{equation}\label{eq:dr_graph}
    \left|\frac{dr}{dt}\right|_{\rm grap}=G_{\rm low}\exp{\left(-\frac{E_{\rm low}}{RT}\right)}+G_{\rm high}\exp{\left(-\frac{E_{\rm high}}{RT}\right)},
\end{equation}
where $G_{\rm low}=7.4\times{10}^{-6}~{\rm cm~s^{-1}}$ and $E_{\rm low}=45~{\rm kcal~{\rm mol}^{-1}}$ are the prefactor and activation energy for graphitization in the low-temperature regime \citep{Butenko+00}, and $G_{\rm high}= 3.6\times{10}^{18}~{\rm cm~s^{-1}}$ and $E_{\rm high}=252~{\rm kcal~{\rm mol}^{-1}}$ are those for the high-temperature regime \citep{Davies&Evans72}. 
I have adjusted $G_{\rm high}$ so that the transition from low- to high-temperature regimes takes place at the Debye temperature of diamond ($\approx$1910 K), as suggested by \citet{Butenko+00}.

\begin{figure*}[t]
\centering
\includegraphics[clip, width=\hsize]{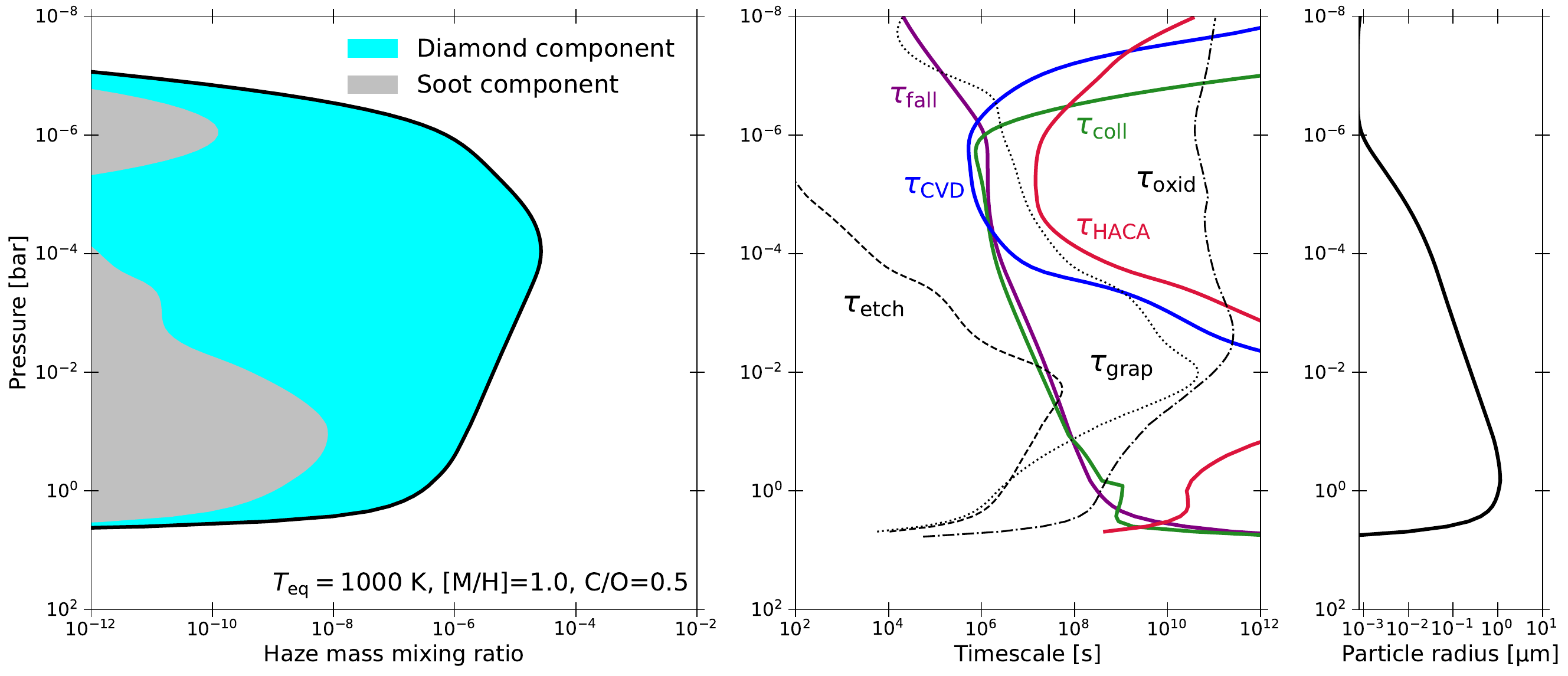}
\caption{Representative result of haze composition evolution. The left panel shows the vertical distribution of the mass mixing ratio of CVD diamonds (cyan) and soot (gray). The solid black line denotes the total mass mixing ratio of the haze particles, demonstrating that the CVD diamond is the dominant component. The middle panel shows timescales relevant for haze composition evolution. The solid red, blue, green, and purple lines denote the timescales of soot growth by the HACA mechanism, diamond growth by chemical vapor deposition, collision growth, and gravitational settling, respectively. The black dotted, dashed, and dash-dot lines show the timescales of graphitization, etching by atomic hydrogen, and oxidization. The right panel shows the vertical distribution of the mean particle radius. I assume $T_{\rm eq}=1000~{\rm K}$, [M/H]=1.0, and C/O=0.5 in this figure.}
\label{fig:result1}

\end{figure*}
\subsubsection{Etching by atomic hydrogen}
Erosion of graphite-like components by atomic hydrogen (etching) plays a critical role in driving the preferential deposition of CVD diamonds \citep[e.g.,][]{Hsu88,Donnelly+97_Etching,Kanai+01,Zhang+23_diamond_etching}.
The etching proceeds through the sequential reactions of atomic hydrogen with hydrogenated carbons on the surface. 
Thus, the etching rate is proportional to the hydrogen surface concentration \citep{Hsu88}. 
In this study, I use the rate constants measured by \citet{Donnelly+97_Etching} who conducted experiments on atomic hydrogen irradiation on graphite\footnote{\rev{Graphite is the sp2-bonded crystaline carbon, while soot has an amorphous structure. In this study, I assume that the etching rate of soot is similar to the rate of graphites.}} under vacuum conditions.
They found that the erosion yield per irradiated H atom can be fitted by
\begin{equation}\label{eq:etch_rate}
    R_{\rm etch}=\frac{A\exp{(-Q_{\rm 1}/RT)}}{J_{\rm H}+B^{-1}\exp{(-Q_{\rm 2}/RT)}},
\end{equation}
where $J_{\rm H}=n_{\rm H}v_{\rm H}/4$ is the number flux of atomic hydrogen, $v_{\rm H}=\sqrt{8k_{\rm B}T/\pi m_{\rm H}}$ is the mean thermal velocity of atomic hydrogen, $A=5.8\times{10}^{16}~{\rm cm^{-2}~s^{-1}}$ and $Q_{\rm 1}=19~{\rm kJ/mol}$ are constants for describing the erosion reaction, and $B=4.3\times{10}^{-28}~{\rm cm^2~s^{1}}$ and $Q_{\rm 2}=160~{\rm kJ/mol}$ are constants introduced to describe the desorption rate of hydrogen from the surface \footnote{\citet{Donnelly+97_Etching} has several typos in the unit of those constants.}.
I use the erosion yield to express the etching rate as
\begin{equation}
    \left| \frac{dm_{\rm soot}}{dt}\right|_{\rm etching}=4\pi r_{\rm p}^2 J_{\rm H} m_{\rm C}\times \min{(0.1R_{\rm etch},0.01)},
\end{equation}
where I have reduced the etching rate of Equation \eqref{eq:etch_rate} by an order of magnitude, as the erosion yield of \citet{Donnelly+97_Etching} is approximately an order of magnitude higher than those obtained by other experiments.
I also restrict the etching yield to not exceed $R_{\rm etch}=0.01$ which is the highest value reported in \citet{Donnelly+97_Etching}.
Note that I neglect the etching for CVD diamonds, as the etching rate for diamonds is several orders of magnitude slower than that for non-diamond carbons \citep[e.g.,][]{Donnelly+97_Etching,Zhang+23_diamond_etching}.

\subsubsection{Soot oxidation}
Oxidized gasses such as OH decompose soot to form gas-phase carbon molecules (e.g., CO).
I take into account the oxidation due to OH following the soot model of \citet{Appel+00}.
The oxidation rate can be expressed by
\begin{equation}\label{eq:oxid}
    \left|\frac{dm_{\rm soot}}{dt}\right|_{\rm oxid}=4\pi r_{\rm p}^2 (1-\rev{R_{\rm soot}^{*}})J_{\rm OH}m_{\rm C}\alpha_{\rm OH},
\end{equation}
where $J_{\rm OH}=v_{\rm OH}n_{\rm OH}/4$ is the flux of OH radicals, $v_{\rm OH}=\sqrt{8k_{\rm B}T/\pi m_{\rm OH}}$, $m_{\rm OH}$, and $n_{\rm OH}$ are the mean thermal velocity, mass, and number density of OH radicals, and $\alpha_{\rm OH}$ is the reaction probability.
\rev{Equation \eqref{eq:oxid} includes a prefactor of $(1-R^{*}_{\rm soot})$, as \citet{Appel+00} modeled the oxidization reaction that takes place on the hydrogenated surface.}
I adopt $\alpha_{\rm OH}=0.13$ following soot modeling studies \citep{Frenklach&Wang91_soot,Appel+00}, which is based on the experiments of \citet{Neoh+81_Soot_oxidization}.

\rev{
\subsection{Calculation Procedure}
I solve the Equations \eqref{eq:np}, \eqref{eq:master_diamond}, and \eqref{eq:master_soot} until the system reach a steady state.
To calculate the CVD diamond growth rate (Equation \ref{eq:CVD_diamond}), graphitization rate (Equation \ref{eq:dr_graph}), soot growth rates (Equation \ref{eq:dm_soot/dt}), the rate of atomic hydrogen etching (Equation \ref{eq:etch_rate}), and oxidization rates (Equation \ref{eq:oxid}), I use the vertical distributions of temperature and number densities of H$_2$, H, CH$_3$, C$_2$H$_2$ and OH calculated by the \texttt{EGP} code and \texttt{VULCAN} (Section \ref{sec:method_TP_VULCAN}).
For this proof-of-concept study, I have ignored any feedbacks from the diamond/soot haze formation on TP and chemical profiles, such as the consumption of gas-phase C$_2$H$_2$ by rapid HACA mechanism, and leave these effects to future studies.
}

\section{Results}\label{sec:result}
\begin{figure*}[t]
\centering
\includegraphics[clip, width=\hsize]{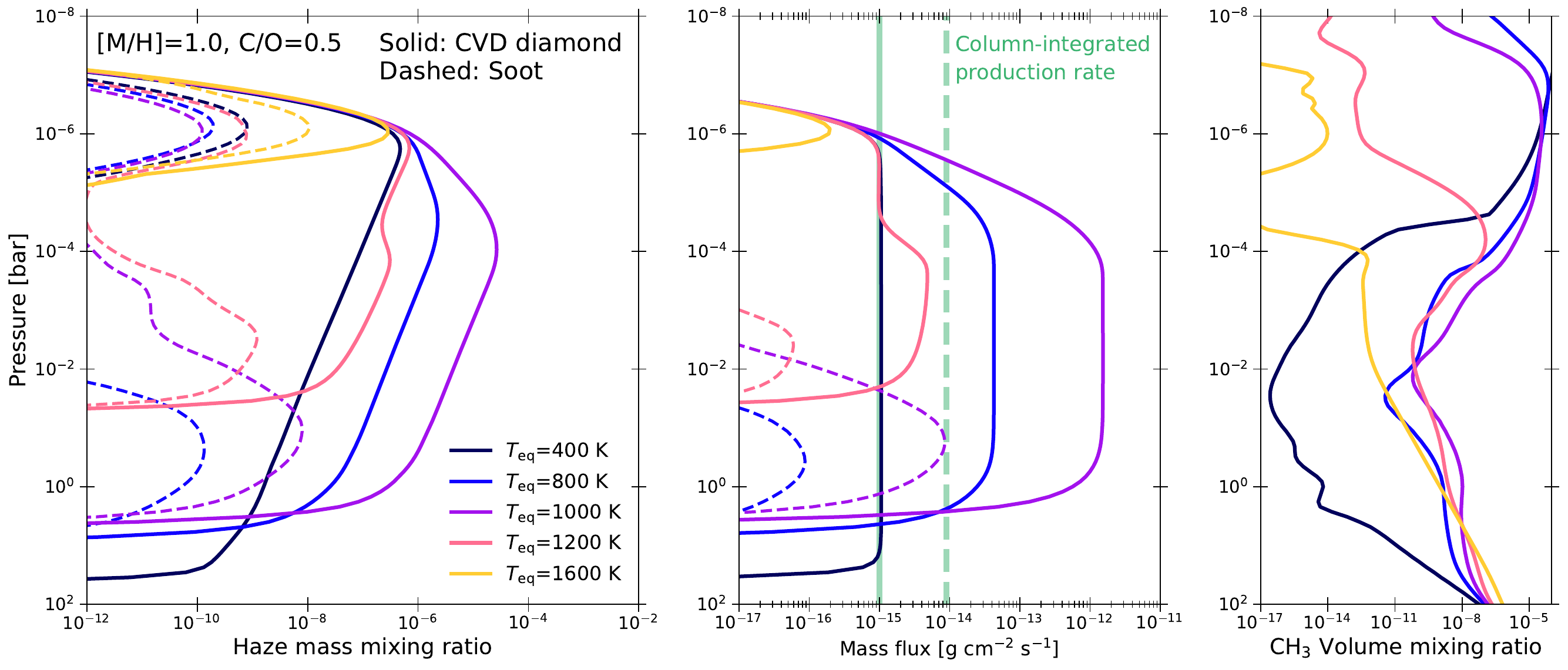}
\caption{Vertical distributions of haze mass mixing ratio (left), mass flux (middle), and volume mixing ratio of CH$_3$ (right).
Each colored line shows the profile for different equilibrium temperature.
Solid and dashed lines show the profiles of CVD diamonds and soot, respectively.
The solid and dashed green lines in the middle panel denote the column-integrated production rate of diamond and soot seeds. 
The haze mass flux should converge to the column-integrated rate if diamond and soot growth/decomposition do not occur, as seen in the diamond mass flux for $T_{\rm eq}=400~{\rm K}$.
Atmospheric compositions are set to [M/H]=1.0 and C/O=0.5 in this figure.
}
\label{fig:result_Teq}
\end{figure*}

In this section, I investigate how the compositions of photochemical hazes evolve in exoplanetary atmospheres.
Since the eddy diffusion coefficient has been highly uncertain for exoplanets, I switch off the particle transport by eddy diffusion for simplicity.
This assumption can be regarded as a setup that maximizes the amount of CVD diamond and soot produced, since eddy diffusion acts to remove haze particles from the atmosphere \citep[e.g.,][]{Lavvas&Koskinen17,Kawashima&Ikoma19,Ohno&Kawashima20}.

\subsection{Haze Composition Evolution}
\begin{figure*}[t]
\centering
\includegraphics[clip, width=\hsize]{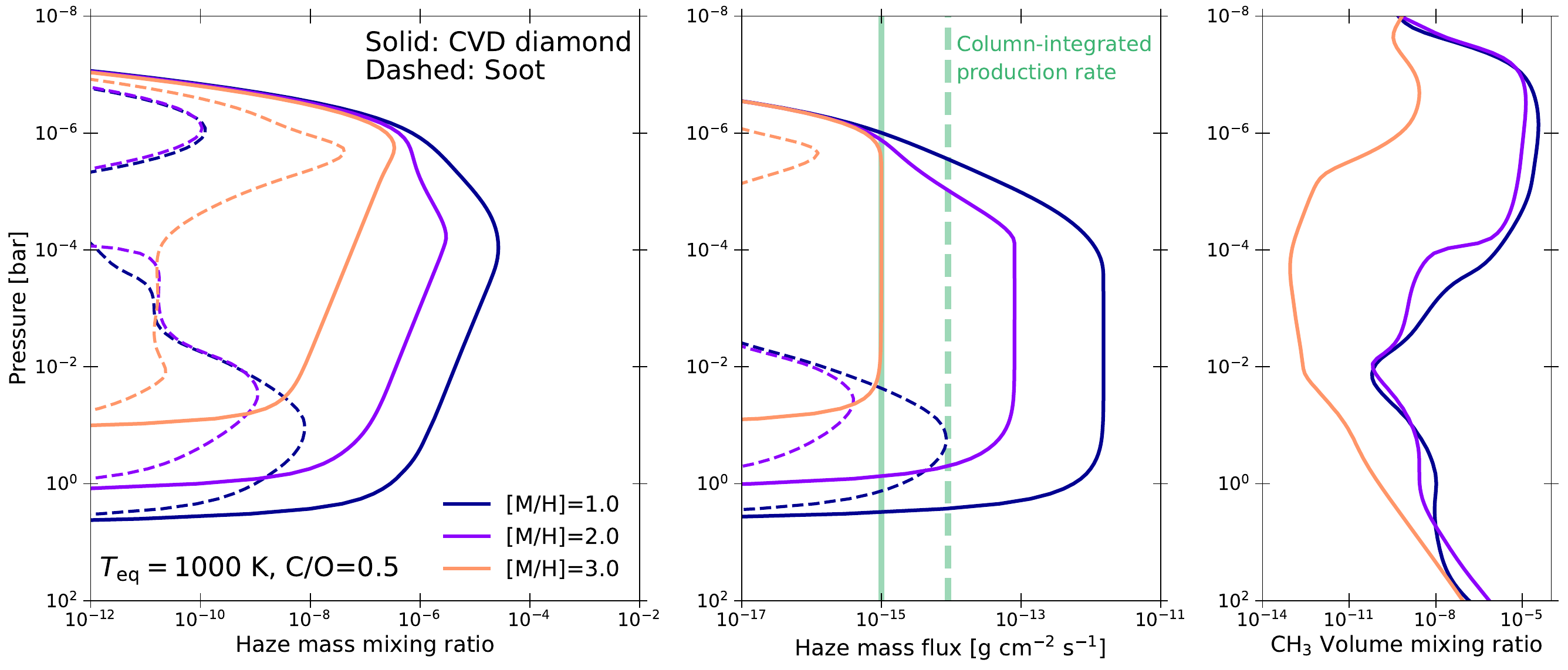}
\includegraphics[clip, width=\hsize]{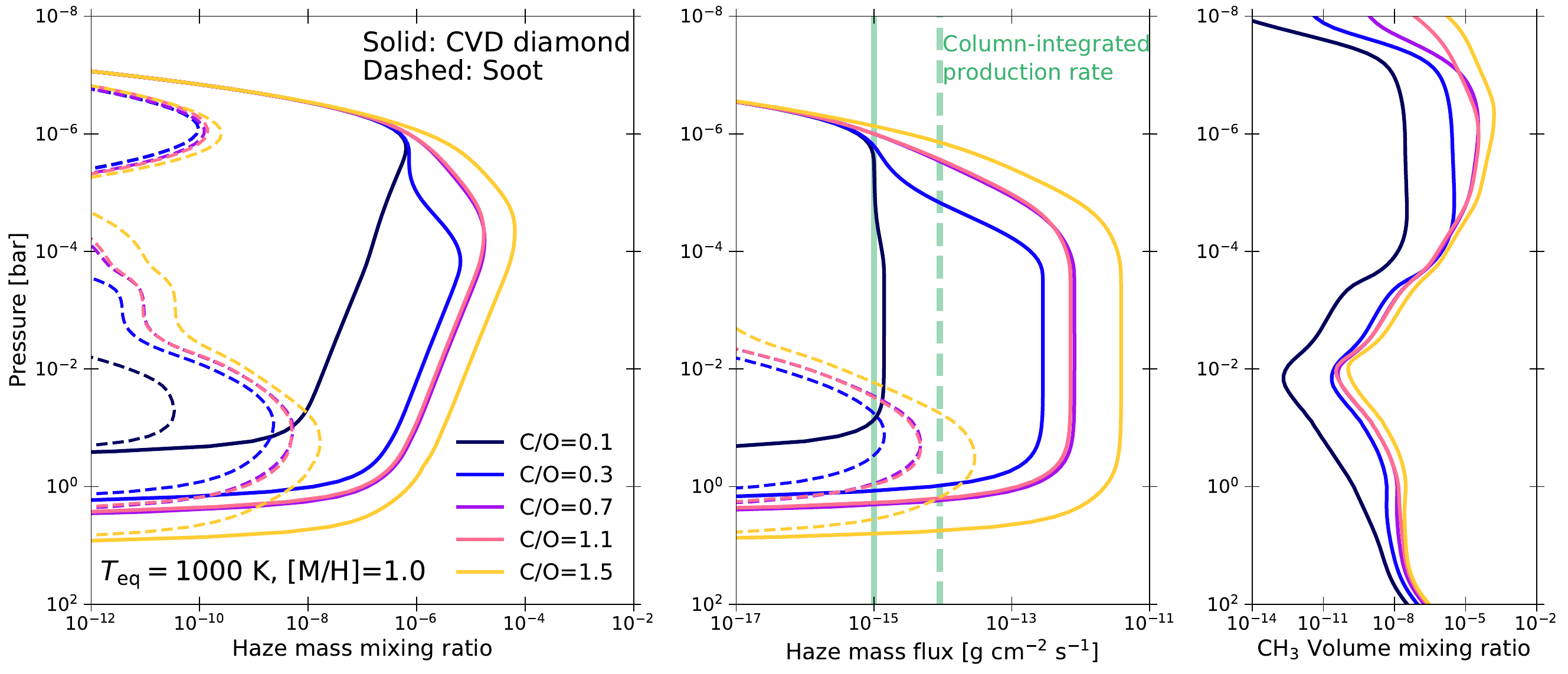}
\caption{Same as Figure \ref{fig:result_Teq}, but varying atmospheric metallicity (top row) or C/O ratio (bottom row). The equilibrium temperature is fixed to $T_{\rm eq}=1000~{\rm K}$. I set the atmospheric C/O and metallicity to be C/O = 0.5 and [M/H] = 1.0 as fiducial values when the parameter is not varied.
}
\label{fig:result_mtoh_ctoo}
\end{figure*}

\begin{figure*}[t]
\centering
\includegraphics[clip, width=\hsize]{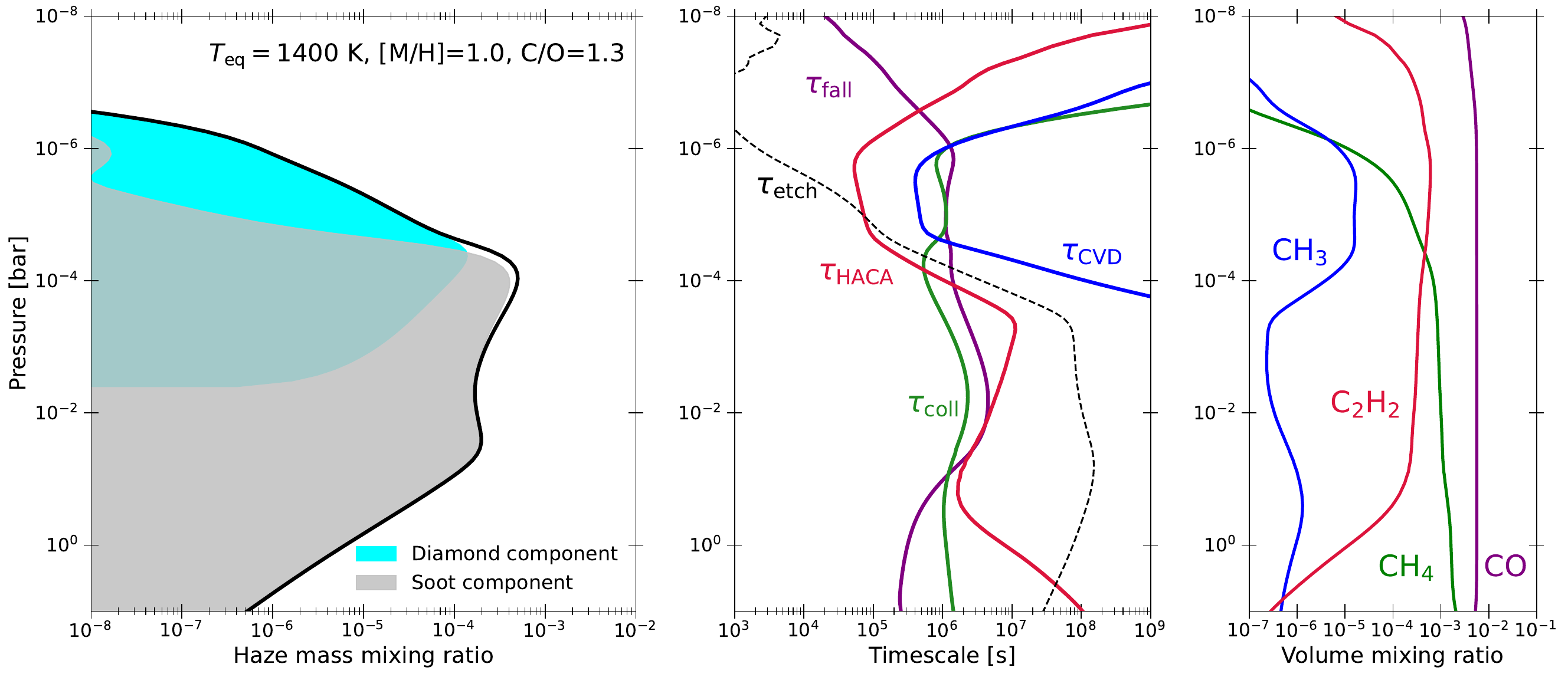}
\caption{Same as Figure \ref{fig:result1}, but for $T_{\rm eq}=1400~{\rm K}$, [M/H]=1.0, and C/O=1.3. The right panel shows the volume mixing ratios of CO, CH$_4$, CH$_3$, and C$_2$H$_2$ at each pressure level instead of mean particle size.
This carbon-rich atmosphere with hot temperature leads to soot-dominated hazes at $P\ga{10^{-4}}~{\rm bar}$.
}
\label{fig:result_soot}
\end{figure*}

In the present model, the haze compositions evolve during particle sedimentation in a complex manner.
The left panel of Figure \ref{fig:result1} shows the vertical distribution of mass mixing ratio of CVD diamond and soot for $T_{\rm eq}=1000~{\rm K}$, [M/H]=1.0, and C/O=0.5.
Once the seed haze particles are injected into the upper atmosphere, the soot component quickly decays, while the diamond component increases with depth.
The diamond mass mixing ratio reaches a maximum at $\sim{10}^{-4}~{\rm bar}$ and then decreases with increasing depth.
In contrast, the soot component gradually increases with depth at $P>{10}^{-4}~{\rm bar}$.
At $P\ga1~{\rm bar}$, mass mixing ratios of soot and diamond rapidly decrease with depth, resulting in haze-free regions in deep atmospheres.
The haze particle size increases with depth in general, as seen in previous haze models \citep[e.g.,][]{Lavvas&Koskinen17,Kawashima&Ikoma18,Ohno&Kawashima20,Gao+23}, while the size rapidly decreases with depth at $P\ga 1~{\rm bar}$ because of particle decomposition.

To better understand the model results, the middle panel of Figure \ref{fig:result1} shows the growth and loss timescales of CVD diamonds and soot.
Here, I have newly defined the timescales of soot growth through the HACA mechanism, CVD diamond growth, graphitization, etching by atomic hydrogen attack, and oxidization.
\rev{Calculating $\rho_{\rm dia}|d\rho_{\rm dia}/dt|^{-1}$ and $\rho_{\rm soot}|d\rho_{\rm soot}/dt|^{-1}$ with Equations \eqref{eq:master_diamond}, \eqref{eq:master_soot} for each term of Equations \eqref{eq:dmdia_dt_sum} and \eqref{eq:dm_soot/dt}, I define}
\begin{equation}
    \tau_{\rm HACA}\equiv \rev{\rho_{\rm soot}\left|\frac{d\rho_{\rm soot}}{dt}\right|^{-1}_{\rm HACA}}=\frac{\rho_{\rm soot}}{n_{\rm p}(1-f_{\rm dia})}\left( \frac{dm_{\rm soot}}{dt}\right)_{\rm HACA}^{-1},
\end{equation}
\begin{equation}
    \tau_{\rm CVD}\equiv \rev{\rho_{\rm dia}\left|\frac{d\rho_{\rm dia}}{dt}\right|^{-1}_{\rm CVD}} =\frac{r_{\rm p}}{3f_{\rm dia}}\left( \frac{dr}{dt}\right)_{\rm CVD}^{-1},
\end{equation}
\begin{equation}
    \tau_{\rm grap}\equiv \frac{r_{\rm p}}{3}
    \left| \frac{dr}{dt}\right|_{\rm grap}^{-1},
\end{equation}
\begin{equation}
    \tau_{\rm etch}\equiv \rev{\rho_{\rm soot}\left|\frac{d\rho_{\rm soot}}{dt}\right|^{-1}_{\rm etch}} = \frac{\rho_{\rm soot}}{n_{\rm p}(1-f_{\rm dia})}\left| \frac{dm_{\rm soot}}{dt}\right|_{\rm etch}^{-1},
\end{equation}
and
\begin{equation}
    \tau_{\rm oxid}\equiv \rev{\rho_{\rm soot}\left|\frac{d\rho_{\rm soot}}{dt}\right|^{-1}_{\rm oxid}} = \frac{\rho_{\rm soot}}{n_{\rm p}(1-f_{\rm dia})}\left| \frac{dm_{\rm soot}}{dt}\right|_{\rm oxid}^{-1},
\end{equation}

respectively.
\rev{I have ignored the factor of $f_{\rm dia}^{-1}$ in the graphitization timescale for simplicity, as it is mostly controlled by the exponential dependence of temperature.}
The definition of collision ($\tau_{\rm coll}$) and settling timescales ($\tau_{\rm fall}$) can be found elsewhere \citep[e.g.,][]{Rossow78,Ohno&Okuzumi18}.
In the case of Figure \ref{fig:result1}, the soot decomposition timescale through etching ($\tau_{\rm etch}$) is much faster than the soot growth timescale through the HACA mechanism ($\tau_{\rm HACA}$) at $\la{10}^{-2}~{\rm bar}$.
This explains why the soot component quickly decays in the upper atmosphere.

The CVD diamond growth timescale ($\tau_{\rm CVD}$) becomes shorter than other relevant timescales at $P\sim {10}^{-4}$--${10}^{-6}~{\rm bar}$ where CH$_4$ photodissociation produces the diamond precursor CH$_3$.
Thus, the CVD diamond grows until the particle settling timescale ($\tau_{\rm fall}$) becomes comparable to the CVD growth timescale.
After CVD diamond growth ceases, haze particles continue to grow during sedimentation through collisional sticking, as seen in conventional haze models \citep[e.g.,][]{Kawashima&Ikoma18,Ohno&Kawashima20,Gao&Zhang20}.
Note that the graphitization timescale ($\tau_{\rm grap}$) is longer than the settling timescale at $P\la{10}^{-2}~{\rm bar}$, which allows haze particles to retain CVD diamonds during sedimentation.
At $P\ga{10}^{-2}~{\rm bar}$ where the atmospheric temperature starts to increase, graphitization becomes faster at deeper atmospheres.
Thus, CVD diamonds are eventually converted to soot in deep hot regions.
However, the etching timescale is also fast in a deep atmosphere as a result of the increased abundance of thermochemically produced atomic hydrogen. 
Thus, CVD diamond and soot eventually convert back to the gas phase in $P\ga 1~{\rm bar}$.

One of the most intriguing results is the nearly complete absence of soot in the upper atmosphere.
This is a consequence of the abundant atomic hydrogen produced by photochemistry, which efficiently decomposes the soot components.
Previous studies speculated that soot represents hazes in hot exoplanets due to its strong resistance to hot temperature \citep[e.g.,][]{Lavvas&Koskinen17,Steinrueck+21}.
However, our result rather suggests that the hot hydrogen-rich atmosphere may not be suitable for soot formation because of the abundant atomic hydrogen produced by photochemistry.

\subsection{Role of Planetary Parameters}
\begin{figure}[t]
\centering
\includegraphics[clip, width=\hsize]{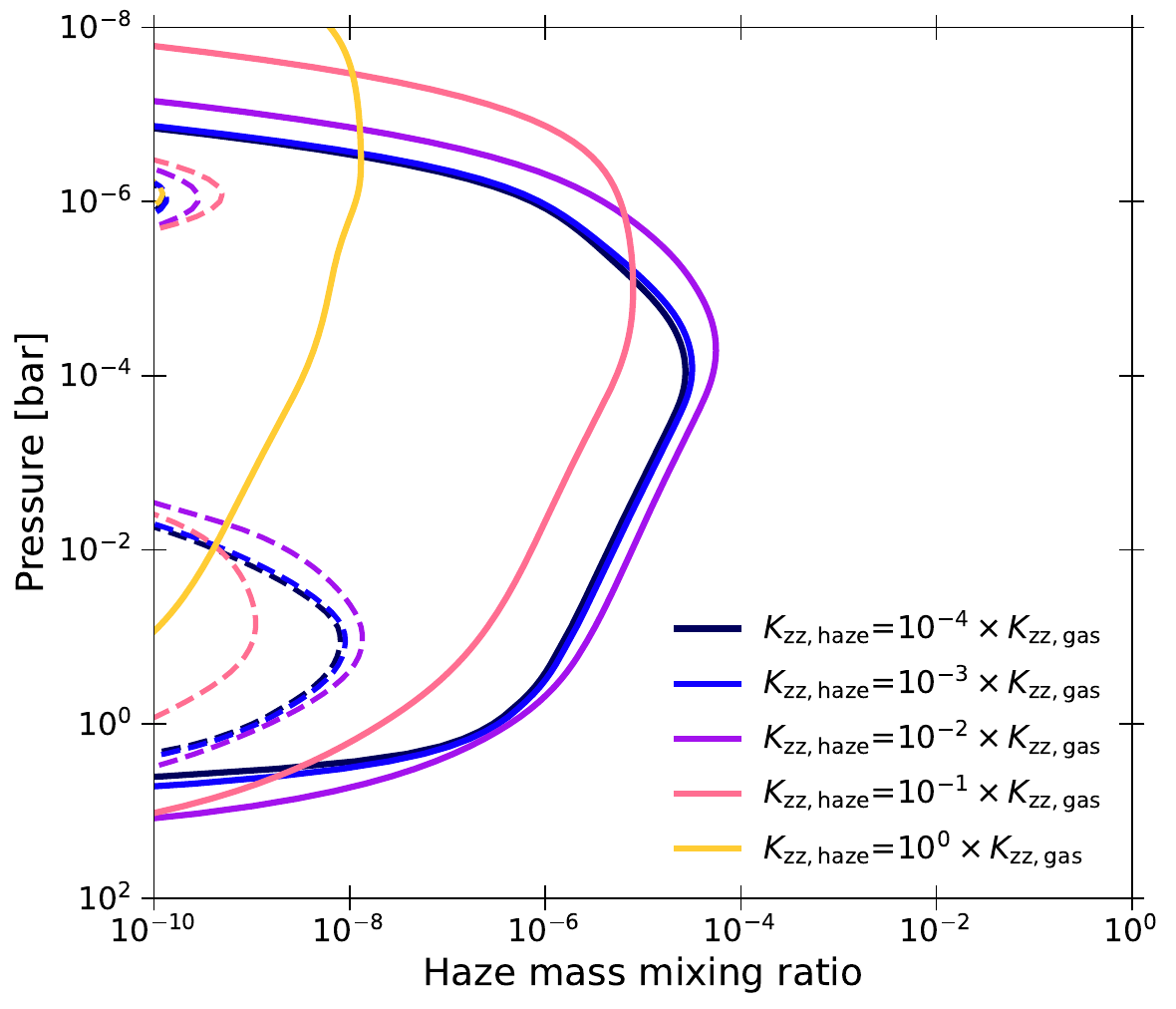}
\includegraphics[clip, width=\hsize]{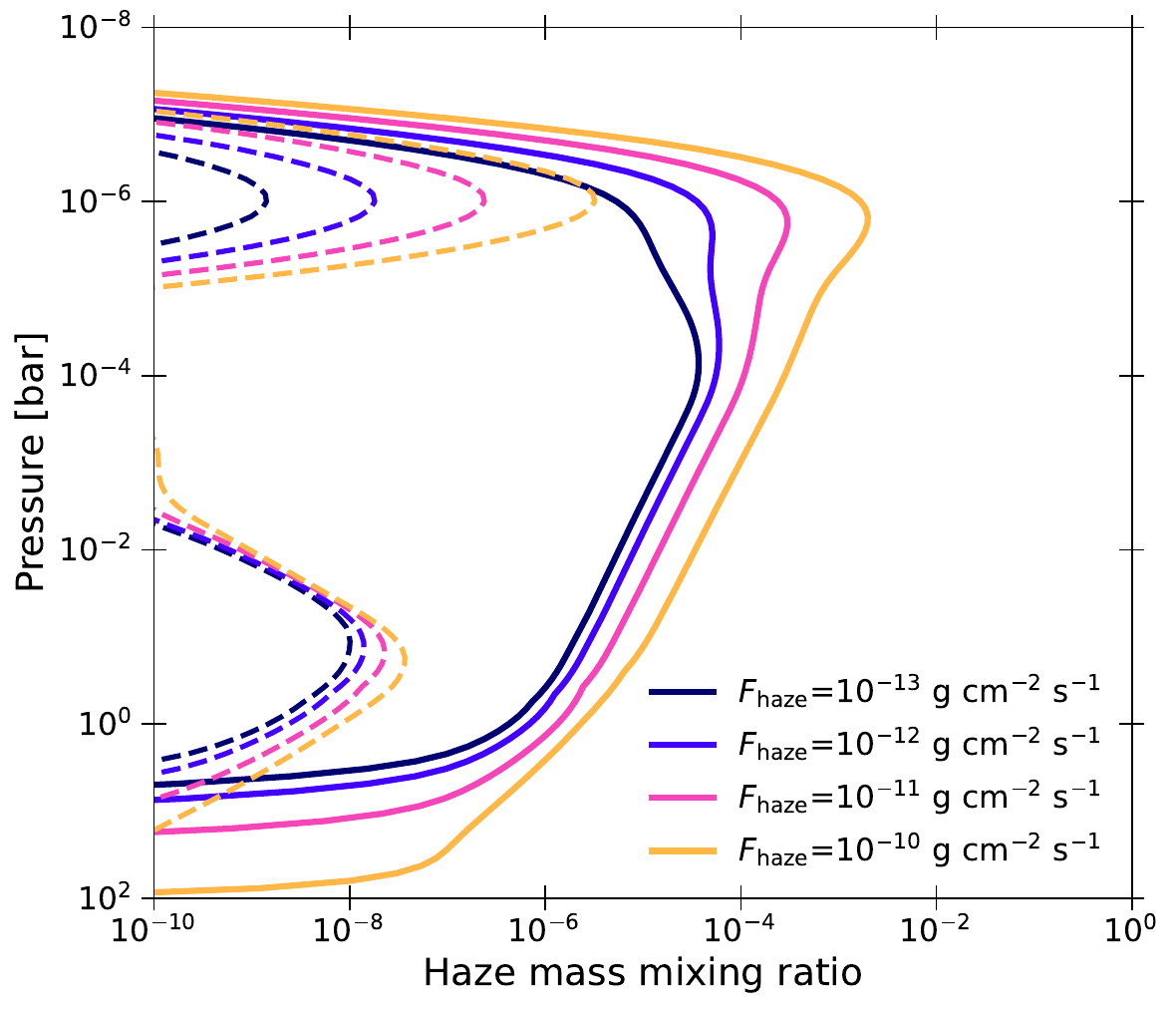}
\caption{Vertical distributions of haze mass mixing ratio for different values of particle eddy diffusion coefficient (top) and column-integrated seed production rate (bottom). The solid and dashed lines show the vertical profiles of CVD diamonds and soot, respectively.
}
\label{fig:result_Kzz_Fhaze}
\end{figure}

The expected haze compositions depend on various planetary properties, such as equilibrium temperature and atmospheric compositions.
Here, I investigate how haze compositions depend on those planetary properties.

\subsubsection{Effects of Equilibrium Temperature}

I first investigated the role of the planetary equilibrium temperature, as previous studies speculated that haze compositions tend to be similar to soot at hotter planets.
The left panel of Figure \ref{fig:result_Teq} shows the mass mixing ratios of the CVD diamond and soot components for various equilibrium temperatures. 
For the assumed atmospheric compositions of [M/H]=1.0 and C/O=0.5, the CVD diamond always dominates over soot at $T_{\rm eq}=400$--$1600~{\rm K}$.
The relative fraction of soot tends to increase with increasing equilibrium temperature at $T_{\rm eq}\la1200~{\rm K}$ due to the faster graphitization of CVD diamonds.
However, soot converted from CVD diamond is vulnerable to the ething by atomic hydrogen.
At a high equilibrium temperature of $T_{\rm eq}=1600~{\rm K}$, CVD diamond is immediately converted to soot through graphitization, but soot is also quickly decomposed to gas phase.
Consequently, the present model predicts that very hot planets have haze-free atmospheres.

To illustrate the CVD diamonds produced by chemical vapor deposition, the middle panel of Figure \ref{fig:result_Teq} shows the mass flux of the CVD diamond and soot components.
CVD diamond growth greatly enhances the haze mass flux in warm to hot exoplanets with $T_{\rm eq}\ga800~{\rm K}$.
The column-integrated production rate of diamond seed is ${\sim}{10}^{-15}~{\rm g~cm^{-2}~s^{-1}}$, while chemical vapor deposition enhances the diamond mass flux to ${\sim}3\times{10}^{-14}~{\rm g~cm^{-2}~s^{-1}}$ for $T_{\rm eq}=800~{\rm K}$ and ${\sim}2\times{10}^{-12}~{\rm g~cm^{-2}~s^{-1}}$ for $T_{\rm eq}=1000~{\rm K}$.
Note that the final quantitative value of the diamond mass flux depends on the column-integrated seed production rate because it determines the number of seed particles on which CVD diamond can be deposited.

\rev{I here provide more in depth understanding on the behavior of the diamond mass flux in Figure \ref{fig:result_Teq}.
In the case of $T_{\rm eq}=1000~{\rm K}$, for example, the diamond mass flux rapidly increases to $\sim2\times{10}^{-12}~{\rm g~cm^{-2}~s^{-1}}$ at $P\la{10}^{-4}~{\rm bar}$. 
This occurs because the CVD growth timescale is comparable to the gravitational settling timescale there (see the middle panel of Figure \ref{fig:result1}).
In other words, CVD diamond is continuously deposited on haze particles during particle sedimentation to lower atmosphere at $P\la{10}^{-4}~{\rm bar}$, resulting in a continuous increase in diamond mass flux.
At $P\ga{10}^{-4}~{\rm bar}$, in contrast, the CVD growth timescale becomes much longer than the sedimentation timescale because the abundance of CH$_3$, the diamond precursor, decreases with depth (see the middle panel of Figure \ref{fig:result1}).
As a result, diamond hazes descend to the lower atmosphere without depositing new CVD diamonds, leading to a vertically constant diamond mass flux.
At the even deeper atmosphere of $P\ga1~{\rm bar}$, the diamond mass flux quickly drops because of  fast timescales of graphitization and subsequent atomic hydrogen etching compared to gravitational settling.
}

The present model suggests that CVD diamond growth is the most efficient at $T_{\rm eq}\sim1000~{\rm K}$.
CVD diamond growth tends to be inefficient at cooler planets because hydrogen abstraction from the diamond surface is inefficient (see Figure \ref{fig:radical_site}).
This can be seen in the case of $T_{\rm eq}=400~{\rm K}$ where the diamond mass flux converges to the column-integrated seed production rate, indicating that chemical vapor deposition barely forms \rev{new CVD} diamonds.
Meanwhile, CVD diamond growth is also inefficient at hotter planets due to depletion of CH$_4$, which provides diamond precurosor CH$_3$.
Consequently, warm planets can achieve both hot temperature and abundant CH$_3$ and serve as the sweet spot for the formation of diamond hazes.

\subsubsection{Effects of Atmospheric Metallicity}\label{sec:metallicity}

Atmospheric metallicity affects the efficiency of CVD diamond growth.
The upper row of Figure \ref{fig:result_mtoh_ctoo} is the same as Figure \ref{fig:result_Teq} but for various atmospheric metallicities, where other parameters are set to $T_{\rm eq}=1000~{\rm K}$ and C/O=0.5.
CVD diamond is still a dominant component of haze particles, whereas CVD diamond growth tends to be inefficient at high atmospheric metallicity. 
This trend stems from the depletion of CH$_4$ in high-metallicity atmospheres, which reduces the abundance of the diamond precursor CH$_3$.
This is particularly noticeable at [M/H]=3.0, where the diamond mass flux converges to the column-integrated seed production rate, implying that chemical vapor deposition barely produces new CVD diamonds.

\subsubsection{Effects of Atmospheric C/O Ratio}
In general, carbon-rich atmospheres with higher C/O ratios promote the formation of both soot and CVD diamonds.
The lower row of Figure \ref{fig:result_mtoh_ctoo} shows the haze profiles for a variety of atmospheric C/O ratios, where other parameters are set to $T_{\rm eq}=1000~{\rm K}$ and [M/H]=1.0.
Although chemical vapor deposition barely enhances the diamond mass flux at C/O=0.1, the haze mass flux increases with increasing C/O ratio at C/O$\ga$0.3.
The efficiency of CVD diamond growth is scaled by the abundance of CH$_3$.

CVD diamond is the dominant haze component for C/O=0.1--1.5 in Figure \ref{fig:result_mtoh_ctoo}, while a high C/O ratio potentially leads to the formation of soot hazes in some cases.
Figure \ref{fig:result_soot} shows the vertical haze profiles for [M/H]=1.0, C/O=1.3, and $T_{\rm eq}=1400~{\rm K}$.
In this carbon-rich hot atmosphere, the soot component rapidly grows and dominates over CVD diamonds at $\la{10}^{-5}~{\rm bar}$.
This occurs because C$_2$H$_2$ becomes the second most dominant carbon reservoir after CO at high temperature and C/O$\gg1$ \citep{Moses+13}, which makes C$_2$H$_2$ more abundant than photochemically produced CH$_3$ (see the right panel of Figure \ref{fig:result_soot}).
Consequently, the HACA mechanism becomes faster than CVD diamond growth (see middle panel), resulting in soot-dominated hazes.

\subsubsection{Effect of Eddy Diffusion and Seed Production Rate}\label{sec:result_Kzz}
I omit the eddy diffusion transport of haze particles in the previous sections for simplicity.
The top panel of Figure \ref{fig:result_Kzz_Fhaze} shows how eddy diffusion affects the model results, where the eddy diffusion coefficient for haze particles is scaled from $K_{\rm zz}$ adopted in photochemical calculations (Equation \ref{eq:Kzz}).
In all cases, CVD diamonds dominate over soot components.
The moderate value of $K_{\rm zz,haze}={10}^{-2}\times{K_{\rm zz,gas}}$ slightly enhances the mass mixing ratio of diamond hazes by redistributing the hazes from the region where CVD diamond deposition is the most efficient.
For stronger eddy diffusion, haze mass mixing ratio decreases with increasing $K_{\rm zz}$ because eddy diffusion quickly removes haze particles from the upper atmosphere \citep[e.g.,][]{Kawashima&Ikoma19}.

The present model still needs to assume the column-integrated seed production rate, which is highly uncertain.
The bottom panel of Figure \ref{fig:result_Kzz_Fhaze} shows the haze profiles for different values of the column-integrated seed production rate $F_{\rm haze}$.
CVD diamonds always dominate over soot regardless of the seed production rate.
Meanwhile, the mass mixing ratio of diamond hazes increases with increasing the seed production rate, as a higher value of $F_{\rm haze}$ supplies more seed particles on which CVD diamonds can be deposited.
Therefore, one should keep in mind that the quantitative value of the haze mass flux computed by the present model depends on the uncertain seed production rate $F_{\rm haze}$.

\subsection{Regime of Haze Composition}
\begin{figure*}[t]
\centering
\includegraphics[clip, width=\hsize]{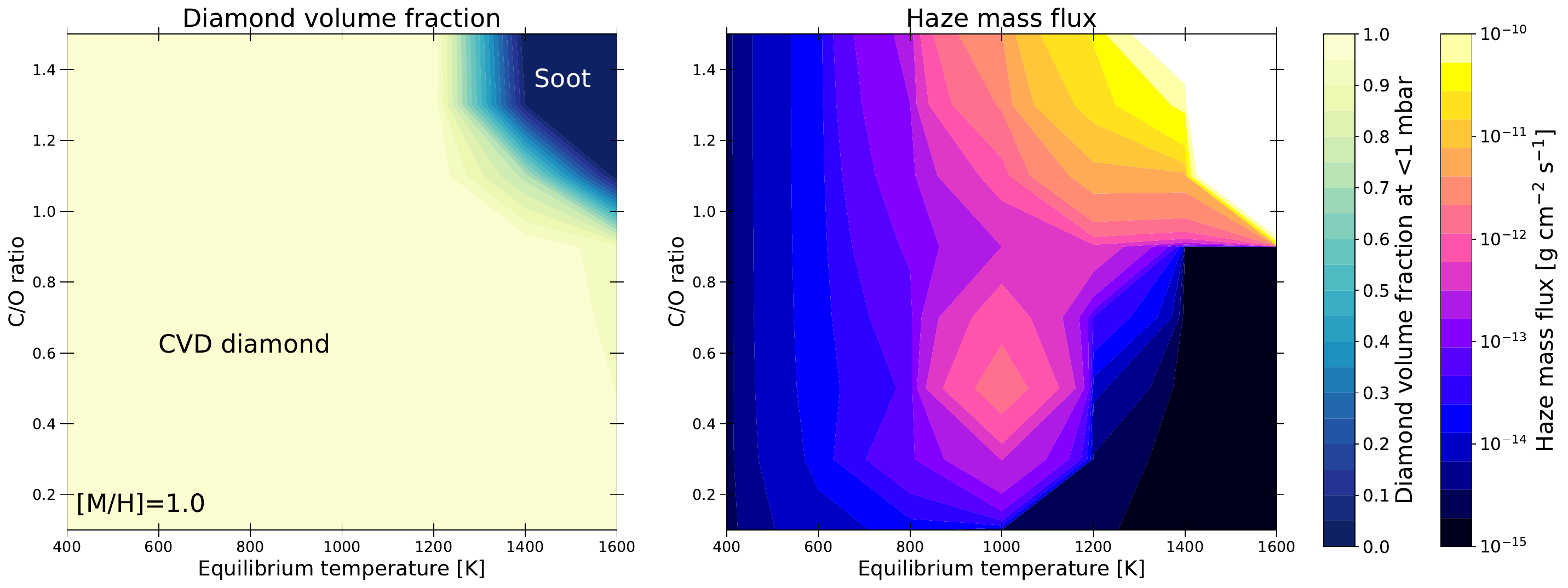}
\includegraphics[clip, width=\hsize]{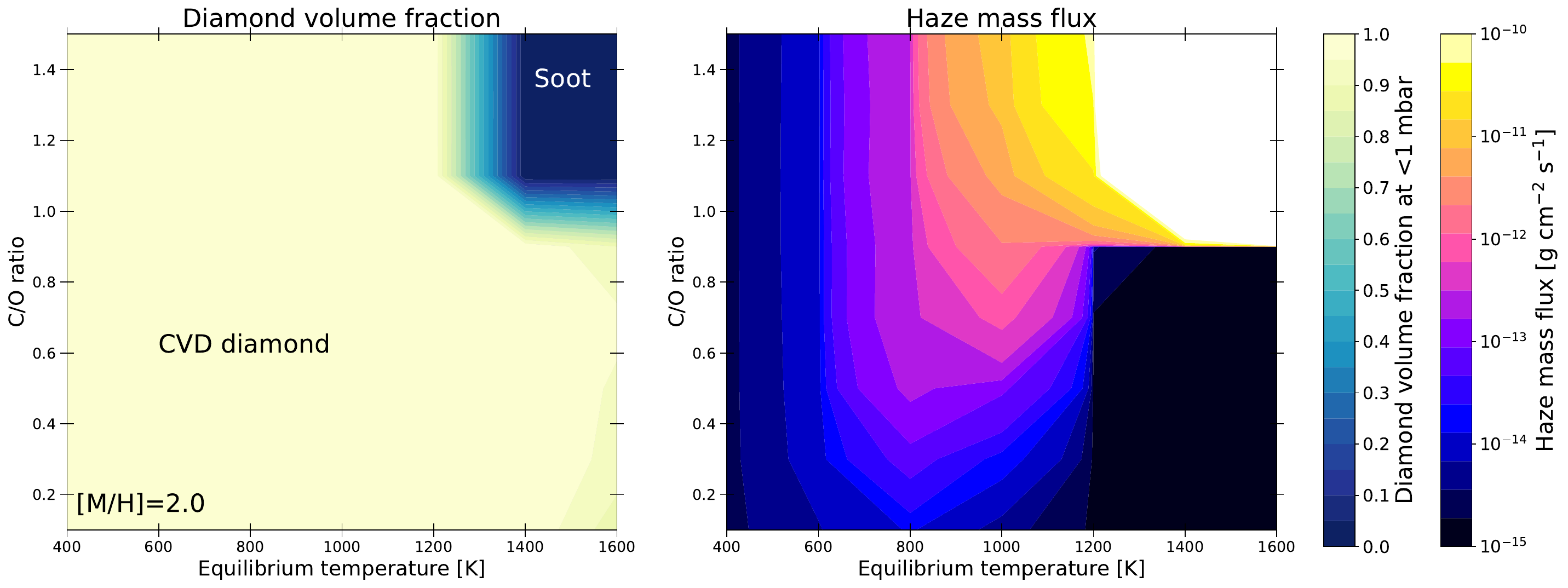}
\includegraphics[clip, width=\hsize]{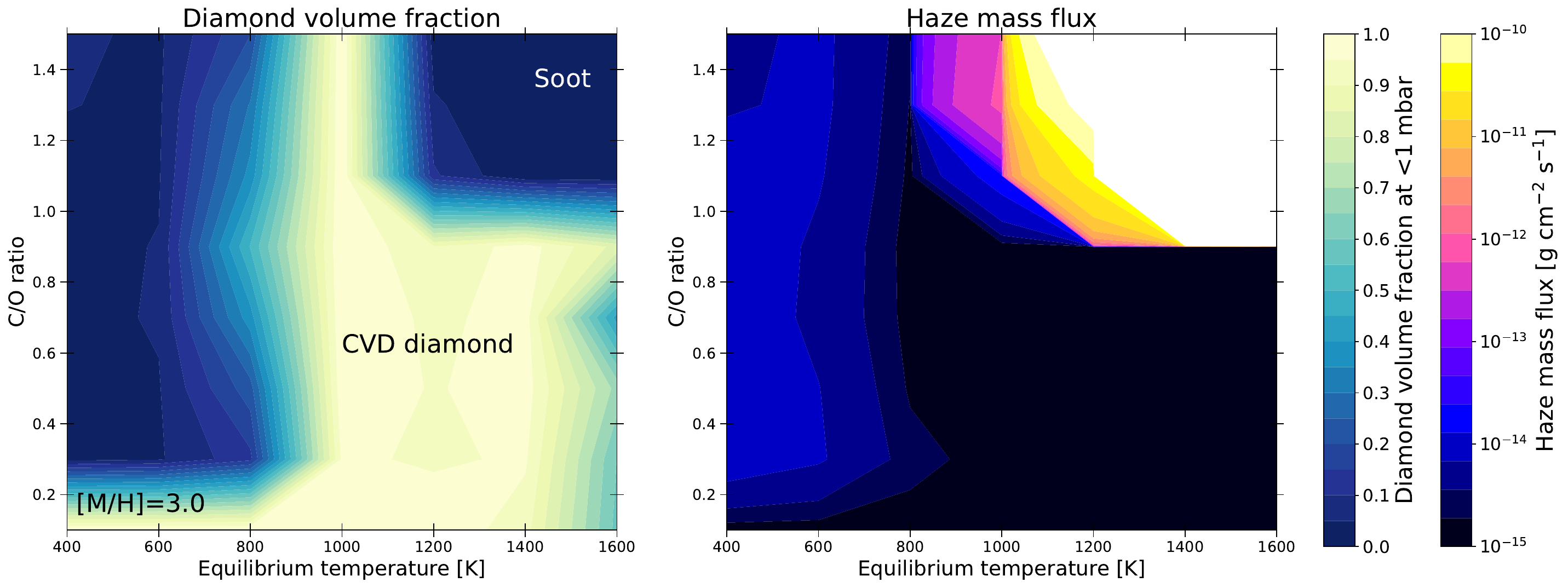}
\caption{2D diagram of haze compositions and mass flux. Left column shows the volume fraction of CVD diamond $\mathcal{F}_{\rm dia}$ in atmospheric column at $P<1~{\rm mbar}$ as a function of atmospheric C/O ratio and equilibrium temperature. The right column shows the mass flux of total (soot+diamond) haze particles measured at 1 mbar.
From top to bottom, each row shows the diagram for atmospheric metallicity of [M/H]=1.0, 2.0, and 3.0, respectively. 
}
\label{fig:result_map}
\end{figure*}

Here I summarize the main haze composition as a function of equilibrium temperature, atmospheric metallicity, and C/O ratio.
To quantify the main haze compositions relevant to atmospheric observations, I define the volume fraction of CVD diamonds in atmospheric column, given by
\begin{equation}
    \mathcal{F}_{\rm dia}=\frac{\int_{\rm 0}^{P_{\rm obs}} (\rho_{\rm dia}/\rho_{\rm d0})Hd\ln{P}}{\int_{\rm 0}^{P_{\rm obs}} (\rho_{\rm dia}/\rho_{\rm d0}+\rho_{\rm soot}/\rho_{\rm s0})Hd\ln{P}},
\end{equation}
where $P_{\rm obs}$ is the pressure level relevant to observations, which is set to $P_{\rm obs}=1~{\rm mbar}$ in this study.

The present model suggests that, for atmospheric metallicity of [M/H]$\le 2.0$, CVD diamond always dominates over soot at ${\rm C/O}<1$ regardless of the equilibrium temperature.
The left column of Figure \ref{fig:result_map} shows the column volume fraction of CVD diamonds as a function of the equilibrium temperature and the C/O ratio for [M/H]=1.0, 2.0, and 3.0.
For [M/H]$\le2.0$, CVD diamond is the main haze component even at ${\rm C/O}=1$--$1.5$ for warm planets with $T_{\rm eq}\la1200~{\rm K}$.
Soot dominates over CVD diamond only in $T_{\rm eq}\ga1200~{\rm K}$ and ${\rm C/O}>1$, although soot can be dominant at slightly cooler planets with $T_{\rm eq}\sim1000~{\rm K}$ if the metallicity is [M/H]=3.0.
This result poses a question for the traditional assumption that soot represents exoplanetary hazes.

For the atmospheric metallicity of [M/H]$=3.0$, haze compositions tend to be soot-like rather than CVD diamonds.
The growth of CVD diamonds is inefficient due to the depletion of the diamond precursor CH$_3$ at very high metallicity (see Section \ref{sec:metallicity}).
Soot decomposition through atomic hydrogen etching is also relatively inefficient under high-metallicity conditions.  
This is why the haze composition is soot-like at $T_{\rm eq}\le800~{\rm K}$ where the atomic hydrogen abundance is insufficient to fully decompose soot, although the predominant soot composition is due to the assumption of seed particles mainly made of soot in the present model ($\epsilon_{\rm dia}=0.1$, see Section \ref{sec:method}).
However, it should be noted that the present model assumes hydrogen-rich environments and may underestimate the growth rate of CVD diamonds, as discussed in Section \ref{sec:bachmann}.

The present model is also capable of predicting how the efficiency of haze formation may depend on planetary parameters.
The right column of Figure \ref{fig:result_map} shows the haze mass flux measured at $P=1~{\rm mbar}$, demonstrating that the formation of diamond and soot hazes tends to be inefficient at hot planets of $T_{\rm eq}\ga1200~{\rm K}$ as long as C/O$<$1.
This supports the hypothesis that main aerosol composition is condensed silicate rather than photochemical organic hazes in hot Jupiters \citep{Gao+20}, although hazes may still contribute to the atmospheric opacity of hot Jupiters with $T_{\rm eq}\la1200~{\rm K}$.
Meanwhile, in case of C/O$>$1, haze mass flux further increases with increasing equilibrium temperature even at $T_{\rm eq}\ga1200~{\rm K}$ because of efficient soot formation through the HACA mechanism.
Therefore, less hazy atmospheres at hotter planets might imply that many hot Jupiter atmospheres have C/O$<$1.
Figure \ref{fig:result_map} also indicates that haze formation tends to be inefficient at cool planets with $T_{\rm eq}\la600~{\rm K}$ and/or high-metallicity atmospheres where CH$_3$ is depleted.
However, it should be noted that the current model does not take into account tholin formation that can promote tholin haze formation on cool metal-rich planets.
Another caveat is that the present study assumes a planetary intrinsic temperature of $T_{\rm int}=100~{\rm K}$, which is relevant to gas giants. 
Sub-Neptunes with cooler $T_{\rm int}$ may extend the parameter space for efficient diamond haze formation by enhancing the abundances of CH$_4$ and thus CH$_3$.


\section{Observational Implications}\label{sec:results_observation}
\begin{figure}[t]
\centering
\includegraphics[clip, width=\hsize]{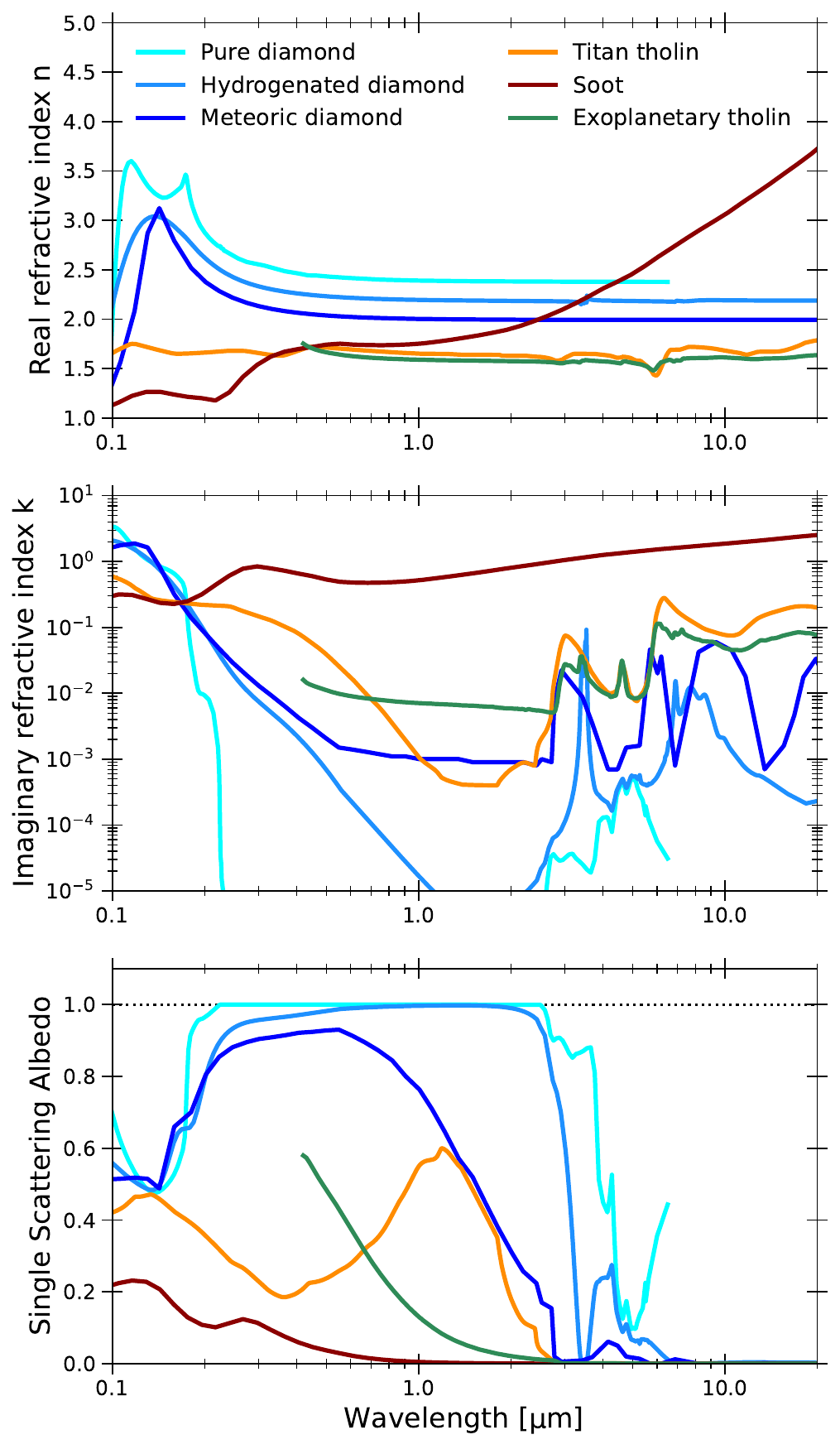}
\caption{Real (top) and imaginary (middle) parts of refractive indices for several materials that potentially represent exoplanetary hazes. The bottom panel shows the single scattering albedo of a particle with a radius of $0.03~{\rm {\mu}m}$ for each substance, as computed by the Mie theory code \texttt{PyMieScatt} \citep{Sumlin+18}. Refractive indices are taken from \citet{Palik85} for pure diamond, \citet{Jones&Ysard22_diamond_optics} for hydrogenated diamonds, \citet{Mutschke+04} for meteoric diamonds, \citet{Khare+84} for Titan tholin, \citet{Lavvas&Koskinen17} for soot, and \citet{He+24} for exoplanetary tholin.
}
\label{fig:optic}
\end{figure}




In this section, I investigate the possible consequence of diamond hazes on various types of atmospheric observations.
Here, I consider how diamond hazes influence atmospheric transmission, emission, and reflected light spectra \rev{\citep[for review of each technic, see e.g.,][]{Kreidberg18}}.

\subsection{Methods}
For transmission spectroscopy, I use the open source radiative transfer code \texttt{CHIMERA} \citep{Line+13,Mai&Line19} to compute the transmission spectrum, while I use another open source code \texttt{PICASO} \citep{Batalha+19,Mukherjee+22a} to calculate the emission and reflected light spectra.
For transmission and emission spectra, I take into account the molecular opacity of H$_2$O, CH$_4$, CO, CO$_2$, NH$_3$, Na, and K in addition to gasseous Rayleigh scattering and collision-induced absorption opacity.
For the reflected light spectrum, for simplicity, I include only Rayleigh scattering, collision-induced absorption, and absorption by H$_2$O molecules as a source of gas opacity.
The atmospheric TP and chemical abundance profiles are extracted from the profiles used in Section \ref{sec:result}. 
I have adopted the atmospheric TP and chemical profiles for $T_{\rm eq}=1000~{\rm K}$, [M/H]=$1.0$, and C/O=$0.5$ as a representative case.
I recall that our generic warm planet has a surface gravity of $g=10~{\rm m~s^{-2}}$.

To clarify the effects of diamond hazes, I computed vertical haze profiles assuming $\epsilon_{\rm dia}=1$ and $dm_{\rm dia}/dt=dm_{\rm soot}/dt=0$ so that the haze particles grow only through collisional sticking.
The diamond mass flux becomes equivalent to the column-integrated seed production rate $F_{\rm haze}$ in this setup. 
The model adopted here is equivalent to \citet{Ohno&Kawashima20} except for the substance that represents haze compositions.
Vertical haze profiles are calculated for various values of the column-integrated haze production rate with a vertically constant eddy diffusion coefficient of $K_{\rm zz}={10}^{7}~{\rm cm^2~s^{-1}}$.
Then I compute the extinction cross section, the single scattering albedo, and the asymmetry parameter of the haze particles at each pressure level using the publicly available Mie theory code \texttt{PyMieScatt} \citep{Sumlin+18} assuming a spherical particle.
In reality, haze particles may be represented by fractal aggregates that have optical properties distinct from those of perfect spheres \citep[e.g.,][]{Tazaki&Tanaka18,Adams+19,Ohno+20,Lodge+24,Vahidinia+24}, but I leave the investigation of non-spherical particles to future studies.
To clarify the observational signatures of diamond hazes, I assume the haze particles purely made of CVD diamonds in the subsequent sections \rerev{(see Appendix \ref{appendix:mix} for discussions on the effects of soot contamination)}.
The next section introduces the refractive index of diamond hazes used in this study.

\subsection{Optical constants}
Pure diamonds have an extremely low imaginary refractive index of $k<{10}^{-5}$ at $\ga0.2~{\rm {\mu}m}$ accompanied by some weak absorption features at $\sim3~{\rm {\mu}m}$ and $\sim5~{\rm {\mu}m}$.
The imaginary index is remarkably lower than those of previously considered haze analogs, namely soot and tholin. 
Thus, exoplanetary hazes would be more reflective than previous thoughts if diamond hazes indeed form.

Pure diamonds are extremely reflective, as shown in Figure \ref{fig:optic}; however, pure diamonds may not represent the optical properties of diamond hazes formed in exoplanetary atmospheres.
For instance, a diamond with a hydrogenated surface is known to produce the absorption features of the C-H bond at $3.43$ and $3.53~{\rm {\mu}m}$ (e.g., \citealt[][]{Sheu+02,Chen+02,Jones+04}, see Figure \ref{fig:optic}), which was used to identify diamonds in protoplanetary disks around HD 97048, Elias 1 and HR 4049 \citep[e.g.,][]{Guillois+99,VanKerckhoven+02,Habart+04,Goto+09}.
Diamonds can also contain impurities, depending on the formation process, which can also affect optical constants.
The effect of impurities is notable in presolar meteoritic diamonds that exhibit several oxygen-related features (e.g., OH stretching at $2.95~{\rm {\mu}m}$, C=O stretching at $5.70~{\rm {\mu}m}$, see also Figure \ref{fig:optic}), indicating the incorporation of oxygen during meteoric diamond formation \citep{Mutschke+04}.

In this study, I adopt the optical constants of diamonds with a hydrogenated surface \citep{Jones&Ysard22_diamond_optics} \footnote{The complex refractive index of hydrogenated diamonds is planned to be available from \url{https://www.ias.u-psud.fr/themis/index.html}.} when computing the optical properties of diamond hazes unless otherwise indicated.
\citet{Jones&Ysard22_diamond_optics} obtained the complex refractive indices of hydrogenated diamonds using the optEC(s)(a) framework \citep{Jones12a,Jones12b,Jones12c} with calibration by existing laboratory data.
The optical constants of hydrogenated diamonds depend on the particle size, as the ratio of the hydrogenated surface to the bulk volume varies with the size \citep[e.g.,][]{Jones+04,Jones12c}.
I interpolate the size-dependent optical constants of hydrogenated diamonds from \citet{Jones&Ysard22_diamond_optics} at each pressure level assuming that the fractional surface coverage of hydrogen is unity.

\subsection{Transmission Spectrum}
\begin{figure*}[t]
\centering
\includegraphics[clip, width=\hsize]{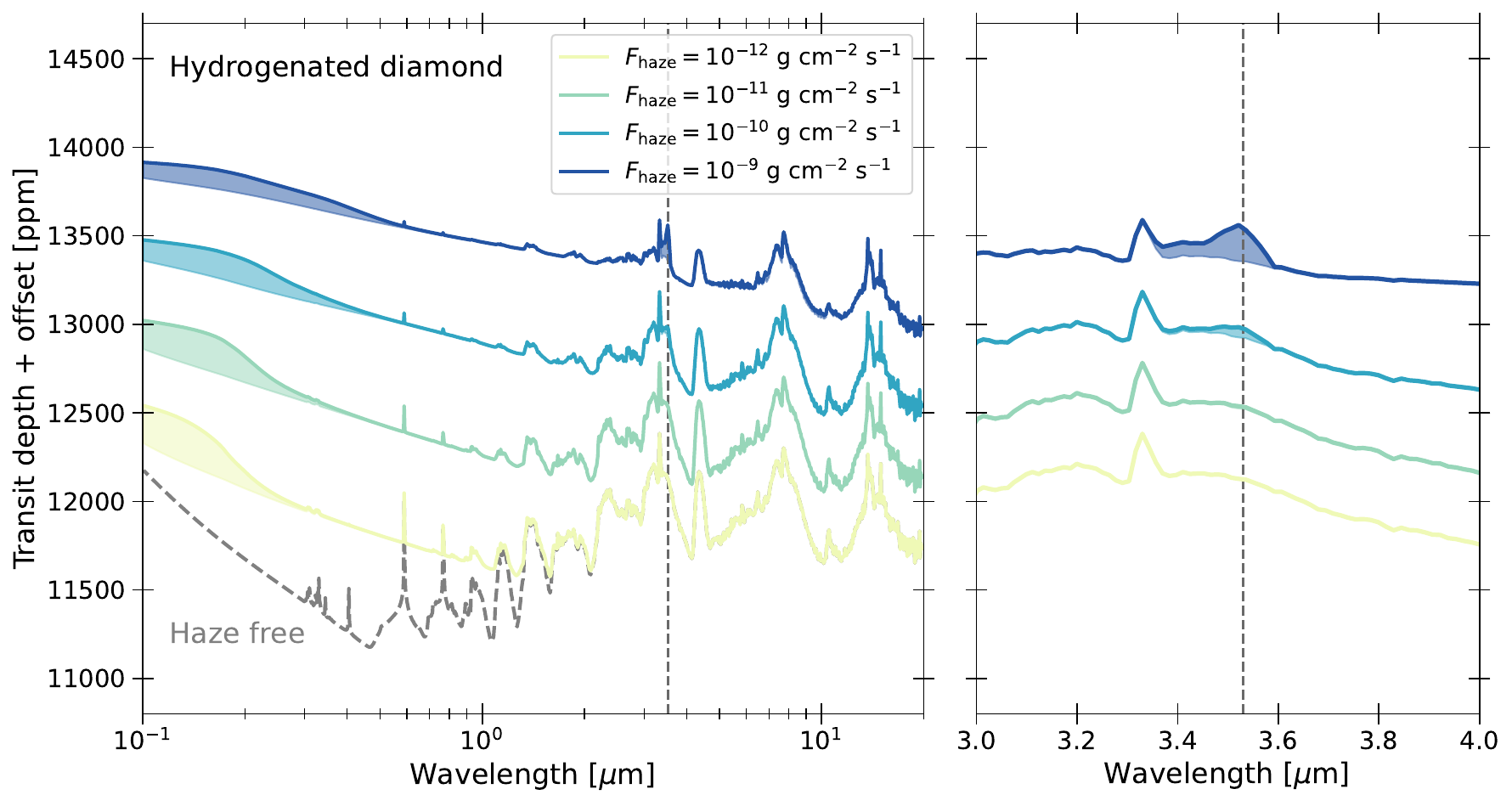}
\caption{Transmission spectra of Saturn-like planets ($M_{\rm P}=0.3M_{\rm jup}$, $R_{\rm 10bar}=R_{\rm jup}$, where $R_{\rm 10bar}$ is the planetary radius at $P=10~{\rm bar}$) with diamond hazes around a Sun-like star. The different colored lines show the spectra for different values of the column-integrated \rev{seed} production rate. The gray dashed line denotes the spectrum with a clear atmosphere. The colored shaded regions denote the contribution of absorption by diamond hazes, \rev{defined as the arising difference if we remove the absorption opacity of diamond hazes from total (absorption+scattering) diamond haze opacity.} The right panel zooms in the wavelength range of $3$--$4~{\rm {\mu}m}$, where the hydrogenated diamonds produce a spectral feature at $3.53~{\rm {\mu}m}$, as indicated by the black dashed line. The spectra for $F_{\rm haze}={10}^{-11}$--${10}^{-9}~{\rm g~cm^{-2}~s^{-1}}$ have been shifted for the sake of clarity.
In this figure, I used the atmospheric TP and chemical profiles for $T_{\rm eq}=1000~{\rm K}$, [M/H]=1.0, and C/O=0.5.
This figure adopts the optical constants of hydrogenated diamonds \citep{Jones&Ysard22_diamond_optics}.
The spectra with diamond hazes containing impurities are shown in Appendix \ref{appendix:meteo}.
\rev{Note that this figure vertically shifts each spectrum by adding constant offsets for clarification purpose.}
}
\label{fig:trans}
\end{figure*}



Diamond hazes affect the transmission spectra mainly at short wavelengths.
Figure \ref{fig:trans} shows the transmission spectrum for several choices of the column-integrated haze production rate.
Diamond hazes obscure the near-infrared H$_2$O feature and produce a spectral slope at visible-to-UV wavelengths. 
This effect is essentially the same as how tholin and soot haze influence the transmission spectrum \citep[e.g.,][]{Lavvas&Koskinen17,Kawashima&Ikoma19,Ohno&Kawashima20,Arfaux&Lavvas22,Steinrueck+23}.
At UV wavelengths of $\la0.3~{\rm {\mu}m}$, the diamond hazes enhance the spectral slope due to the contribution of absorption in addition to scattering.

It is worth noting that diamond hazes produce a scattering slope with a sub-Rayleigh slope.
The onset of a shallow spectral slope originates from the vertical size gradient of haze particles.
\citet{Ohno&Kawashima20} demonstrated that the optical spectral slope becomes steeper/shallower as the atmospheric opacity increases/decreases with increasing altitude.
Because the Rayleigh scattering cross section is larger for larger particles, the spectral slope becomes shallower if the particle size is larger at lower altitudes, which makes atmospheric opacity smaller at higher altitudes. 
It should be noted that Figure \ref{fig:trans} shows the spectrum for which the vertical transport of haze particles is dominated by gravitational settling.
Diamond hazes could produce a super-Rayleigh slope if strong eddy diffusion hinders particle growth and creates a strong vertical gradient in the haze mass mixing ratio, as found in \citet{Ohno&Kawashima20}.

Diamond haze produces its own absorption feature if the haze production rate is very high.
The right panel of Figure \ref{fig:trans} shows the spectrum at wavelength of $3$--$4~{\rm {\mu}m}$, where the hydrogenated diamond has a absorption feature at $3.53~{\rm {\mu}m}$.
The $3.53~{\rm {\mu}m}$ absorption feature overlaps with the absorption feature of CH$_4$, which would make the detection of the diamond feature nontrivial.
Nonethless, the $3.53~{\rm {\mu}m}$ feature may serve as the smoking gun of diamond hazes in exoplanetary atmospheres if the atmosphere is extremely hazy.
It should be noted that the absorption features of diamonds depend on the impurities incorporated in the diamonds (see Appendix \ref{appendix:meteo}).
Further experimental studies are highly warranted to investigate the possible absorption features of diamond hazes.

\subsection{Atmospheric TP Profiles}\label{sec:TP_haze}
\begin{figure*}[t]
\centering
\includegraphics[clip, width=\hsize]{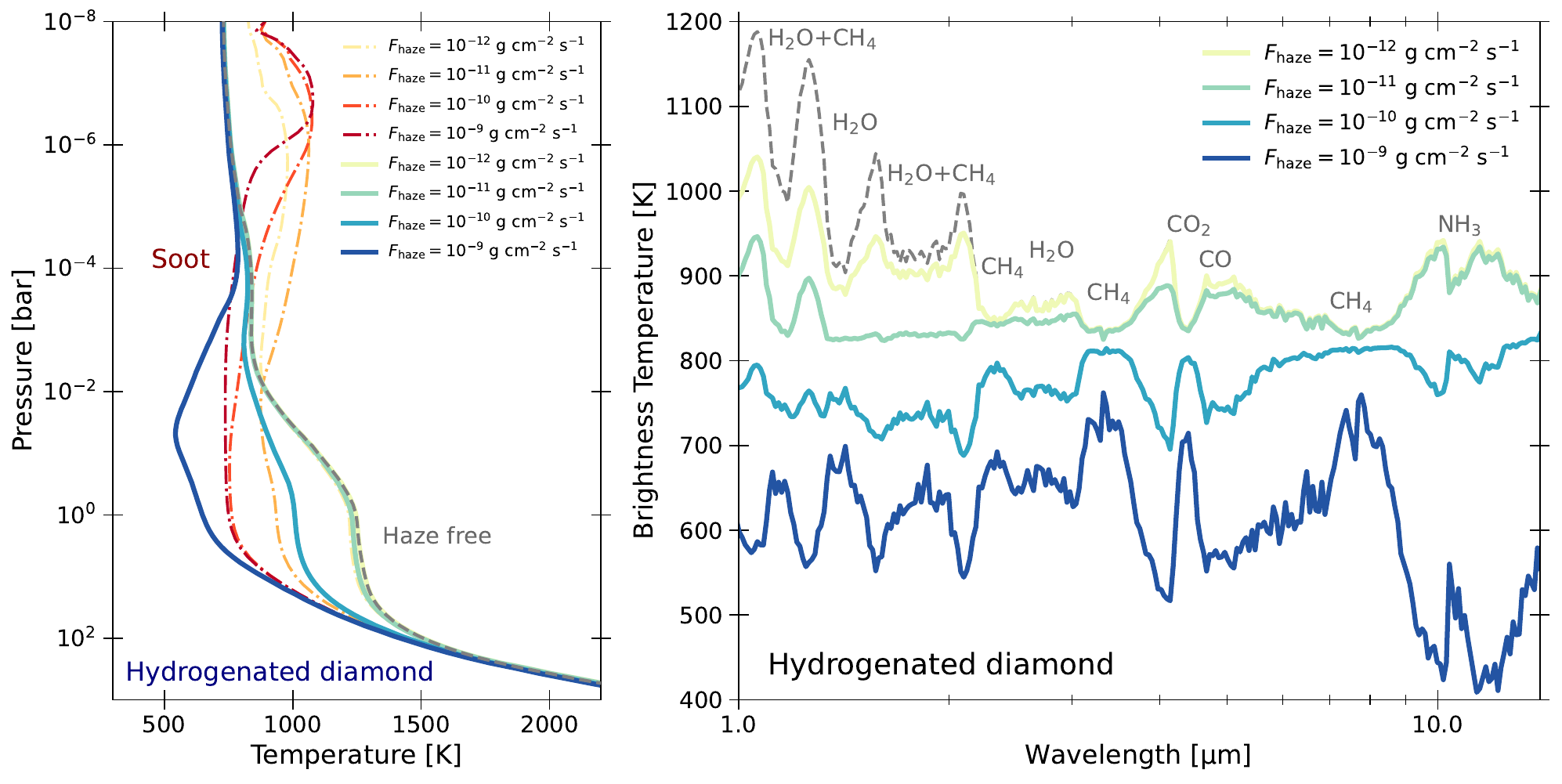}
\caption{Atmospheric TP profiles (left panel) and brightness temperature spectrum (right panel) of Saturn-like planets with $T_{\rm eq}=1000~{\rm K}$ and surface graivity of $g=10~{\rm m~s^{-2}}$ around a generic M-dwarf represented by GJ1214 ($T_{\rm eff}=3250~{\rm K}$). The different colored lines show the profiles for different values of the column-integrated haze production rate. 
I have used the optical constants of hydrogenated diamonds \citep{Jones&Ysard22_diamond_optics}.
The gray dashed lines denote the profile of clear atmospheres, while the reddish colored dash-dotted lines in the right panel show the TP profiles with soot optical constants \citep{Lavvas&Koskinen17} for reference.}
\label{fig:TP_diamond}
\end{figure*}

Diamond hazes likely affect atmospheric thermal structure in a different manner from tholin and soot hazes studied previously \citep{Morley+15,Lavvas&Aufaux21,Steinrueck+23}.
To investigate the impact of diamond hazes, I iteratively run the \texttt{EGP} radiative-convective equilibrium model \citep[e.g.,][]{McKay+89,Marley&McKay99,Fortney+05,Marley&Robinson15,Ohno&Fortney22a} and the microphysical model to obtain a self-consistent set of atmospheric TP and haze profiles. 
For simplicity, I have ignored the temperature decoupling of haze particles and ambient gases \citep{Lavvas&Koskinen17,Lavvas&Aufaux21}.

Diamond hazes tend to suppress, but could still produce temperature inversion if the haze production rate is extremely high.
Figure \ref{fig:TP_diamond} shows the TP profiles that include hydrogenated diamond hazes with various haze production rates.
The figure also shows the TP profiles calculated with soot optical constants as a reference.
Diamond hazes suppress the onset of temperature inversion at $F_{\rm haze}\la{10}^{-10}~{\rm g~{cm}^{-2}~s^{-1}}$ due to inefficient absorption at short wavelengths. 
This trend is remarkably different from tholin and soot hazes that cause temperature inversion by absorbing incident stellar lights (\citealt{Morley+15,Lavvas&Aufaux21,Steinrueck+23}, also see reddish dash-dotted lines in the left panel of Figure \ref{fig:TP_diamond}).
However, when the haze production rate is extremely high such as $F_{\rm haze}={10}^{-9}~{\rm g~cm^{-2}~s^{-1}}$, diamond hazes turn out to produce a temperature inversion at $\sim{10}^{-4}~{\rm bar}$. 
This inversion is caused by cooling in the lower atmosphere: Diamond hazes efficiently scatter incoming stellar lights back to space, making the lower atmosphere of $P\ga{10}^{-4}~{\rm bar}$ much colder than the upper atmosphere of $P\la{10}^{-4}~{\rm bar}$.
\subsection{Emission Spectrum}
Diamond hazes weaken the molecular features when the haze production is modest.
The right panel of Figure \ref{fig:TP_diamond} shows the emission spectrum in terms of brightness temperature for various haze production rates.
Note that I used the TP profile in the left panel of Figure \ref{fig:TP_diamond} to calculate the emission spectra.
When the haze production is moderate ($F_{\rm haze}\le{10}^{-11}~{\rm g~{cm}^{-2}~s^{-1}}$), diamond hazes hide thermal emission from deep hot atmospheres, resulting in a reduction of the brightness temperature at wavelengths where the atmosphere is relatively transparent, such as ${\sim}1.25~{\rm {\mu}m}$ and ${\sim}3.9~{\rm {\mu}m}$.
Thus, diamond hazes act to mute molecular features.
Note that diamond hazes less affect the emission spectrum at a longer wavelength of $\ga5~{\rm {\mu}m}$ due to the lower scattering opacity.

As the haze production rate further increases, the brightness temperature goes down at entire wavelengths, making faint thermal emission.
Intriguingly, at such a very high haze production rate, diamond hazes rather enhance the molecular features and make them emission features. 
This trend originates from the reduction of atmospheric emissivity---the effect also called thermal scattering \citep[e.g.,][]{Taylor+21}.
It has been known that the emergent flux from an isothermal semi-infinite atmosphere can be expressed by \citep[e.g.,][]{Ribicki&Lightman79,Taylor+21}
\begin{equation}\label{eq:thermal_scattering}
    F_{\rm \nu}=\frac{4\pi}{\sqrt{3}}\frac{\sqrt{\epsilon}}{1+\sqrt{\epsilon}}B_{\rm \nu}(T),
\end{equation}
where $\epsilon=1-\omega$ and $\omega$ are the emissivity and the single scattering albedo of the atmosphere.
The emergent flux approaches the black body radiation $\approx \pi B_{\rm \nu}(T)$ in the limit of a purely absorbing atmosphere ($\epsilon=1$), whereas atmospheric scattering makes the flux faint compared to the anticipated black body radiation.
Diamond hazes reduce $\epsilon$ while molecular absorption increases $\epsilon$.
As a result, the spectrum shows emission features at wavelengths where molecular absorption increases atmospheric emissivity \citep{Taylor+21}.
This thermal scattering effect acts to enhance the molecular features.
If the haze production rate is extremely high, diamond hazes create the temperature inversion and further boost the molecular emission features.

\rev{To better understand the behavior at extremely hazy cases, let us consider a simplified case of an isothermal atmosphere where only the line absorption by gasseous molecules and scattering by diamond hazes contribute the atmospheric opacity.
In this case, the atmospheric emissivity is given by $\epsilon=\kappa_{\rm gas}/(\kappa_{\rm gas}+w_{\rm dia}\kappa_{\rm dia})$, where $\kappa_{\rm gas}$ is the line absorption opacity of gasseous molecules, and $w_{\rm dia}$ and $\kappa_{\rm dia}$ are the mass mixing ratio and scattering opacity of diamond hazes.
For extremely hazy atmosphere, since $\epsilon \approx \kappa_{\rm gas}/w_{\rm dia}\kappa_{\rm dia}\ll1$, Equation \eqref{eq:thermal_scattering} is simplified to
\begin{equation}\label{eq:thermal_scattering_simple}
    F_{\rm \nu}\approx \frac{4\pi}{\sqrt{3}}\left( \frac{\kappa_{\rm gas}}{w_{\rm dia}\kappa_{\rm dia}}\right)^{1/2}B_{\rm \nu}(T).
\end{equation}
Since the scattering opacity of diamond haze is a continuous function, higher haze production rate $F_{\rm haze}$ yields a higher haze mass mixing ratio $w_{\rm dia}$ and thus reduces the entire brightness temperature.
On the other hand, molecular line absorption provides a sudden increase in $\kappa_{\rm gas}$ at specific wavelengths, such as $\sim3.2~{\rm {\mu}m}$ for CH$_4$ and  $\sim4.3~{\rm {\mu}m}$ for CO$_2$, which also suddenly increases the brightness temperature according to Equation \eqref{eq:thermal_scattering_simple}.
In summary, although diamond hazes tend to suppress the onset of temperature inversion, they could still produce emission features for molecular line absorption if diamond haze production is very efficient.
}


\subsection{Reflected Light Spectrum}
\begin{figure*}[t]
\centering
\includegraphics[clip, width=\hsize]{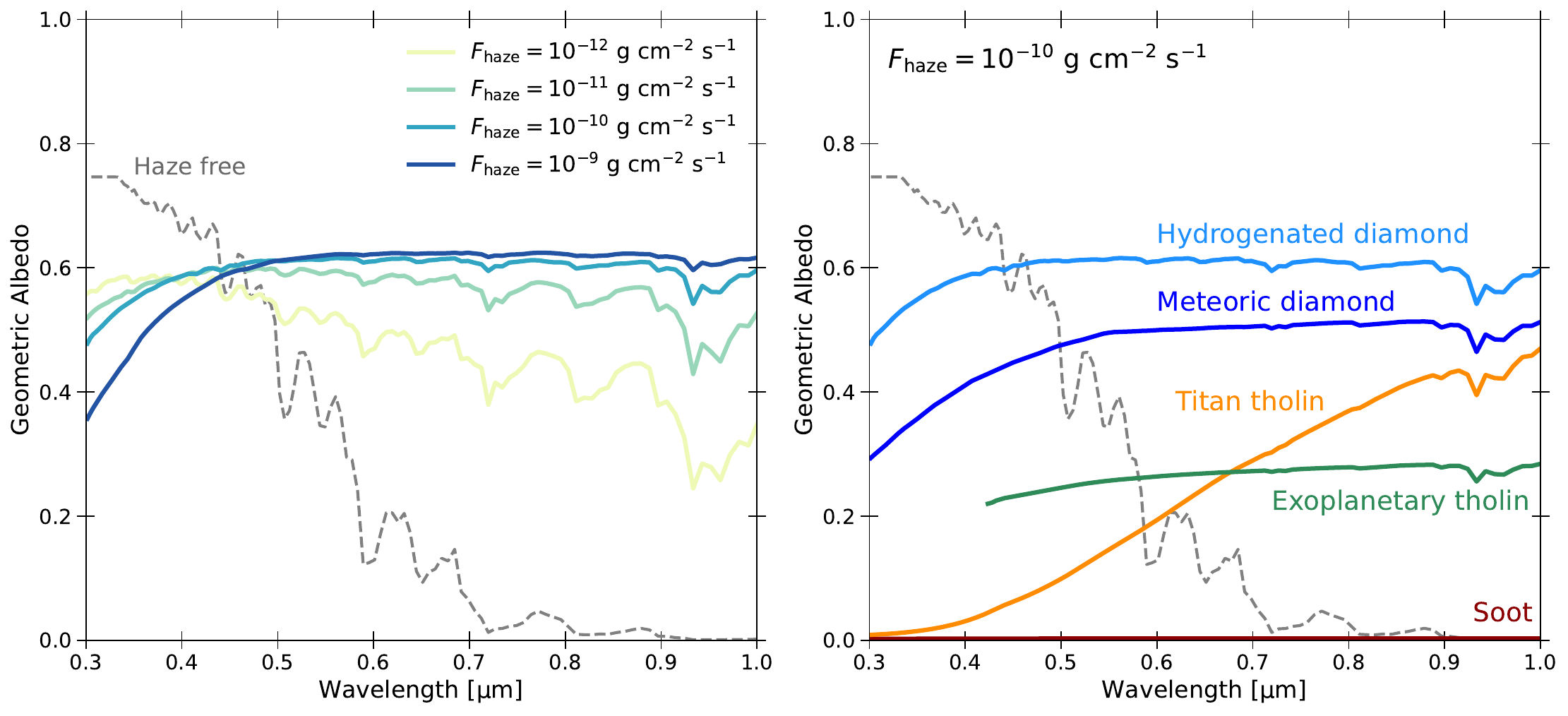}
\caption{Geometric Albedo spectrum with photochemical hazes. Different colored lines in the left panel show the albedo spectra for various values of the column-integrated haze production rate, where I have adopted the optical constants of hydrogenated diamonds. The right panel shows the albedo spectra for various choices of representative haze optical constants. All albedo spectra are calculated for the atmospheric profile of $T_{\rm eq}=1000~{\rm K}$, [M/H]=1.0, and C/O=0.5. The gray dashed lines show the albedo spectrum for a clear atmosphere for reference.}
\label{fig:Ag_diamond}
\end{figure*}

Future space-based facilities, such as the Habitable World Observatory, will leverage reflected light spectroscopy to characterize the scattering properties of exoplanetary atmospheres.
The left panel of Figure \ref{fig:Ag_diamond} shows the albedo spectra with diamond hazes for various values of the column-integrated haze production rate. 
Diamond haze greatly enhances the geometric albedo at $\ga0.5~{\rm {\mu}m}$ and decreases the albedo at $\la0.5~{\rm {\mu}m}$ as the haze production rate increases.
This trend is similar to the impacts of sulfur hazes on reflected light spectra, although sulfur hazes cause a more sharp drop in geometric albedo at $<0.45~{\rm {\mu}m}$ \citep{Hu+13,Gao+17_sulfur}.

Future reflected light spectroscopy would open a novel window for directly diagnosing the actual composition of exoplanetary hazes.
The right panel of Figure \ref{fig:Ag_diamond} shows the geometric albedo spectrum for various choice of optical constants that might represent exoplanetary hazes.
Soot hazes cause very low geometric albedo at all visible to UV wavelengths, whereas Titan tholin causes a gradual increase in geometric albedo over a wide range of UV to near-infrared wavelengths, as found by \citet[][]{Morley+15}.
Hydrogenated diamond haze causes the greatest increase in geometric albedo among the haze analog tested here, though the presence of impurities in diamond hazes may lower the geometric albedo as represented by the meteoric diamond.
Exoplanetary tholin, a haze analog produced from high metallicity (1000$\times$ solar value) warm ($400~{\rm K}$) gas \citep{He+24}, yields the intermediate albedo between hydrogenated diamonds and soot, although the shape of the spectrum at $<0.4~{\rm {\mu}m}$ is currently unknown due to the lack of optical constant data.

\section{Discussion}\label{sec:discussion}
\subsection{Relation to Existing Experimental Studies}\label{sec:bachmann}

The industry community has established an experimental understanding on the gas compositions required to synthesize CVD diamonds, which is known as Bachmann's diamond domain \citep{Bachmann+91}.
The Bachmann's diamond domain is the empirical criterion of gas-phase C-H-O ratios that allows CVD diamond deposition, as revealed by the compilation of CVD diamond experiments over 30 years \citep{Bachmann+91}.
Several studies later constructed a theoretical basis for the diamond domain \citep[e.g.,][]{Ford96,Wang+98_Ternay,Wan+98}. 
For example, \citet{Petherbridge+01_Ternary} provided an intuitive interpretation that the diamond domain corresponds to the phase space that produces a sufficiently high abundance ($\ga{10}^{-6}$) of CH$_3$---a main precursor of CVD diamond \citep{Harris90,Harris&Weiner91_experiment}---and high abundance of atomic H relative to C$_2$H$_2$---a main precursor of graphitic condensed carbons such as soot.

\begin{figure*}[t]
\centering
\includegraphics[clip, width=\hsize]{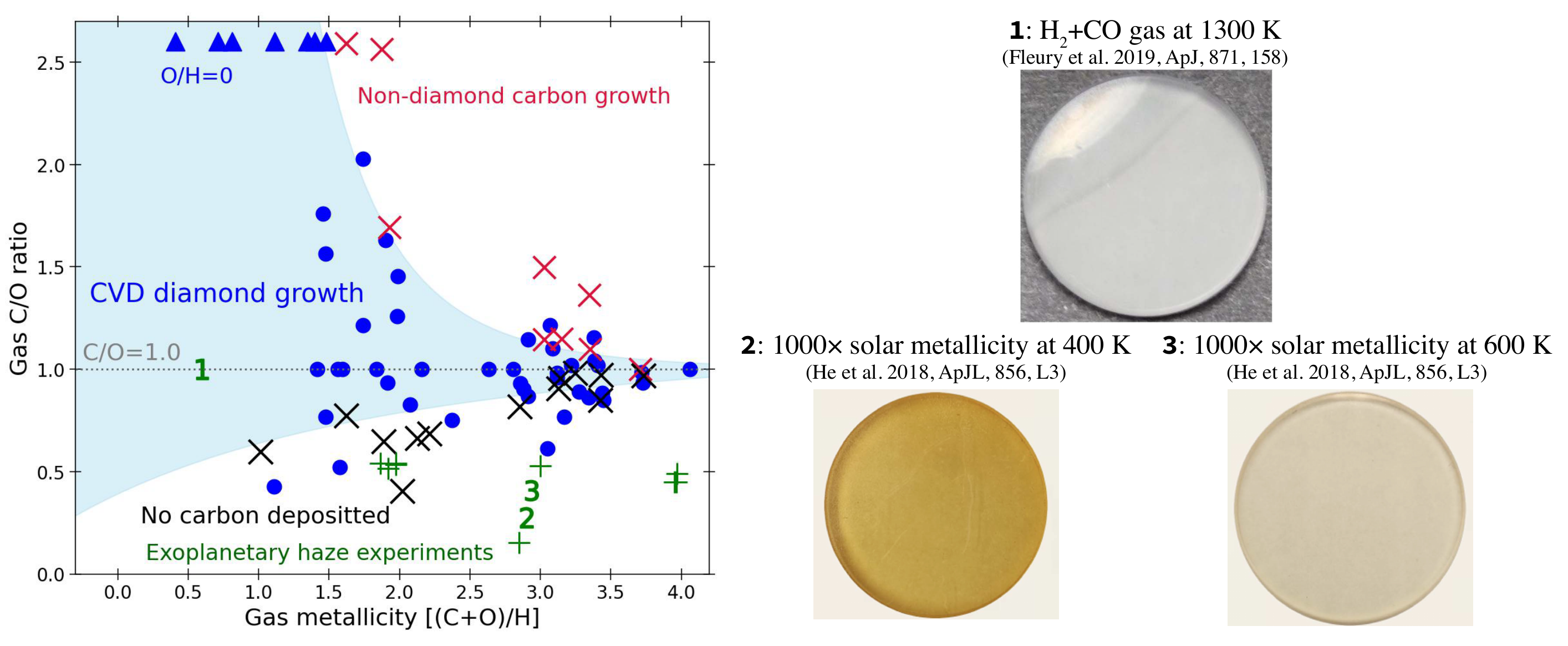}
\caption{Compilation of CVD experiments, known as the Bachmann's diamond diagram \citep{Bachmann+91}. Blue symbols denote the gas-phase metallicity (expressed by the [(C+O)/H]) and C/O ratio relation that succeeded in producing CVD diamonds, whereas black and red symbols denote the phase space where no carbons or non-diamond carbons (e.g., graphites) are deposited. These data are taken from Table 1 of \citet{Bachmann+91} with adding the results of \citet{Marinelli+94}.
For illustrative purposes, the blue shaded region roughly denotes the diamond domain that allows CVD diamond formation.
The green symbols denote the metallicity-C/O relation of exoplanetary haze experiments \citep{Horst+18,Fleury+19,He+20}.
The right panels show the pictures of the haze analogues synthesized from the gas compositions labeled in the left panel, \rerev{taken from Figure 9 of \citet{Fleury+19} (\href{https://iopscience.iop.org/article/10.3847/1538-4357/aaf79f/meta}{DOI:10.3847/1538-4357/aaf79f}) and Figure 2 of \citet{He+18} (\href{https://iopscience.iop.org/article/10.3847/2041-8213/aab42b}{DOI:10.3847/2041-8213/aab42b}). {\copyright}AAS. Reproduced with permission. }
}
\label{fig:diamond_domain}
\end{figure*}

Figure \ref{fig:diamond_domain} shows the Bachmann's diamond domain in terms of gas metallicity and C/O ratio, which is relevant to exoplanetary science.
Figure \ref{fig:diamond_domain} demonstrates that CVD diamond synthesis is successful for a wide range of C/O ratio.
In particular, CVD diamonds could be synthesized even without oxygen (i.e., ${\rm C/O}=\infty$) if the gas metallicity is $\la30\times$ solar value.
On the other hand, at a high metallicity of $\ga100~{\times}$ solar value, CVD experiments tend to produce non-diamond carbon such as graphites especially at C/O$>1$, although CVD diamonds could still be synthesized at ${\rm C/O}\sim$1.

It is interesting to note that the haze composition predicted by this study is qualitatively consistent with the Bachman's diamond domain.
The present model suggests that diamond haze production tends to be more efficient at a lower atmospheric metallicity of $\la100\times$ solar value and a higher C/O ratio (\rev{see Figure \ref{fig:result_mtoh_ctoo}}), which is in line with CVD experiments in Figure \ref{fig:diamond_domain}.
Meanwhile, this study barely produces diamond hazes at [M/H]=3.0, whereas some CVD experiments could succeed in synthesizing CVD diamonds at a very high metallicity of [(C+O)/H]${\ge}3.0$.
This discrepancy may indicate some missing processes which lead to an underestimate of the diamond growth rate in the present model under such high-metallicity conditions.


Recent laboratory studies have begun to synthesize exoplanetary haze analogs and examine their physical and chemical nature \citep{Horst+18,He+18,He+18b,He+20,He+20_sulfur,He+24,Fleury+19,Moran+20,Yu+21}, while no experiments have reported the formation of diamonds thus far.
One of the reasons is presumably the gas compositions used in those experiments, which are denoted by green symbols in Figure \ref{fig:diamond_domain}.
Most previous haze experiments used high-metallicity gas mixtures that are a bit off from the compositions yielding CVD diamonds in the Bachman's diagram.
Another potential reason is that most haze experiments adopt the temperature of $\le600~{\rm K}$ \citep[e.g.,][]{Horst+18,He+18}, at which the growth of CVD diamonds is inefficient due to inefficient hydrogen abstraction from the diamond surface.

It is worth mentioning that the experiment of \citet{Fleury+19} actually satisfied both the hot temperature and the gas compositions within the Bachman diamond domain.
\citet{Fleury+19} conducted Ly$\alpha$ (121.6 nm) UV irradiation of the gas mixture of H$_2$ and CO (99.7\%:0.3\%) at $\sim1300~{\rm K}$ and reported the deposition of solid phase condensates.
Intriguingly, they reported the emergence of ``white deposits'' in their experiments, which is reminiscent of CVD diamonds.
Moreover, haze experiments of \citet{He+18} showed that, for $1000\times$ and $10000\times$ solar metallicity gas, haze analogs tend to be colorless when they are synthesized at hotter temperatures, at which CVD diamonds can grow more efficiently.
It would be definitely interesting to conduct haze synthesis experiments using gas compositions within Bachman's diamond domain at sufficiently hot temperatures, say $\sim1000~{\rm K}$ as in typical CVD diamond experiments.

\subsection{Radiation-induced Graphite-Diamond Conversion}\label{sec:radiation_induced}
Although this paper has focused on CVD diamond formation, intense high energy photons from the central star might be solely responsible for the formation of diamond hazes.
An experimental study of \citet{Lyutovich&Banhart99} observed the transformation of graphite into diamond under a low pressure condition ($2\times{10}^{-6}~{\rm Pa}$) when the graphite is irradiated by high energy ($1.25~{\rm MeV}$) electrons at temperature higher than $\approx 600~{\rm K}$.
Another experimental study of \citet{Kouchi+05} similarly observed the formation of diamonds during UV irradiation onto the interstellar ice analog to synthesize the interstellar organic analog.
\citet{Zaiser&Banhart97} suggested that the irradiation-induced diamond formation is caused by a difference in the cross sections of the irradiation-induced displacements of carbon atoms in diamond and graphite.
Namely, incident high energy particles preferentially displace the position of graphitic carbon due to a lower threshold energy than that for diamond carbon. 
\citet{Zaiser&Banhart97} introduced a nonequilibrium ``effective free energy'' to diagnose the relative stability of graphite and diamond under irradiation, which well explains the experimental results \citep{Zaiser+00}.
This irradiation-induced diamond formation has been proposed as a possible origin of diamonds detected in protoplanetary disks around Herbig Ae/Be stars showing active X-ray flares \citep{Goto+09}.
Since a close-in planet orbits close to the star, radiation-induced diamond formation might form diamond hazes as well if the star has sufficiently strong activity, which would be worthwhile to investigate in future studies.

\section{Summary}\label{sec:summary}
Reflective hazes hinted by recent \emph{JWST} observations may suggest diamond formation in exoplanetary atmospheres.
I have utilized the theory of CVD diamond formation and soot formation through the HACA mechanism to revisit the plausible composition of exoplanetary hazes.
I have demonstrated that the CVD diamond formation can dominate over soot formation at wide ranges of equilibrium temperature, atmospheric metallicity, and C/O ratio.
The findings of this paper are summarized below.

\begin{enumerate}
\item The growth of CVD diamond can be efficient at $P\sim{10}^{-4}$--${10}^{-6}~{\rm bar}$ where photochemistry produces the diamond precursor CH$_3$ and atomic hydrogen H, as similar to the experimental environments of CVD diamond synthesis.

\item Soot growth is also relatively efficient at $P\sim{10}^{-4}$--${10}^{-6}~{\rm bar}$ where the CH$_4$ dissociation triggers the formation of C$_2$H$_2$ that is the main soot precursor. However, soot components are vulnerable to the decomposition due to photochemically-produced atomic hydrogen. 

\item The present model suggests that CVD diamond is always the main haze component at C/O$<1$. This is mainly due to the efficient atomic hydrogen etching of soot components. This process is the analogy of how to deposit diamonds with suppressing the deposition of graphite components in CVD diamond synthesis. 

\item Soot could be the main haze component only at hot exoplanets ($T_{\rm eq}\ga1200~{\rm K}$) with high C/O ratio of C/O$>$1, where the soot growth through the HACA mechanism overcomes the atomic hydrogen etching and CVD diamond growth.

\item Diamond haze formation is the most efficient on planets with an equilibrium temperature of $T_{\rm eq}{\sim}1000~{\rm K}$ due to abundant CH$_3$ and a sufficiently hot temperature that promotes the hydrogen abstraction from diamond surfaces.
Diamond hazes tend to be depleted at hotter planets because of faster graphitization followed by the atomic hydrogen etching, whereas CVD diamond growth also becomes slower at cooler planets because of inefficient hydrogen abstraction from the diamond surface. 

\item Diamond hazes produce a scattering slope in optical transmission spectra, as is the case with conventional tholin and soot hazes. Furthermore, diamond hazes may produce an absorption feature at $3.53~{\rm {\mu}m}$ due to the hydrogenated surface, although CH$_4$ feature tends to obscure the hydrogenated diamond feature.
Infrared observations by \emph{JWST} and future atmospheric survey by Ariel may be able to detect the diamond feature if they observe an extremely hazy planet.

\item Diamond hazes tend to suppress the onset of a temperature inversion because of their reflective nature.
However, if the haze production is too efficient, diamond hazes cause a temperature inversion by scattering incident stellar lights back to space, leading to the middle atmosphere being colder than the upper atmosphere.

\item Diamond hazes mute the molecular features in emission spectra when the haze production rate is modest. 
However, if the haze production rate is very high, diamond hazes turn out to enhance the molecular features and make them emission rather than absorption features.
This trend can happen because the shape of the emission spectrum is controlled by wavelength dependence of atmospheric emissivity rather than temperature at wavelength-dependent photospheres.

\item Diamond hazes greatly enhance the geometric albedo in the visible wavelength, similar to sulfur hazes.
The geometric albedo spectra are remarkably different from each other among various substances that were considered as exoplanetary haze analogs.
Thus, reflected light spectroscopy by future facilities, such as the Habitable World Observatory, would be able to observationally diagnose what exoplanetary hazes are made of.

\item The results of the present model are roughly consistent with the gas compositions that are found to deposit diamonds in CVD experiments, known as the Bachman diamond domain.
However, CVD experiments could deposit diamonds with a very high metallicity of [M/H]$\ge$3.0 for which the present model barely forms CVD diamonds.
The discrepancy might suggest that diamond haze formation is even easier than what the present model predicts.
Future laboratory studies specific to exoplanetary environments are highly desired to test the hypothesis proposed in this study.

\end{enumerate}

Atmospheric observations of GJ1214b, WASP-80b, and WASP-69b have hinted at the presence of reflective hazes \citep{Kempton+23,Bell+23,Schlawin+24}, as introduced earlier.
Diamond hazes can partly reconcile the presence of a scattering slope with the lack of temperature inversion, although too thick diamond hazes would be still inconsistent with observations due to the emergence of emission features (see Section \ref{sec:results_observation}).
In particular, diamond hazes would be able to reconcile the spectral slope \citep{Fukui+14,Wong+22} and the lack of temperature inversion in WASP-80b, as the emission spectrum seems to be less affected by clouds and hazes \citep{Bell+23}. 
WASP-80b is an interesting target for searching for diamond hazes, as it shows the hint of reflected light in the near-infrared wavelength \citep{Jacobs+23} and also shows the spectral feature of CH$_4$ \citep{Bell+23} which provides the diamond precursors CH$_3$.
Putative reflective aerosols in WASP-69b \citep{Schlawin+24} may have a different origin, as CH$_4$ appears to be depleted on the planet.
Meanwhile, WASP-69b shows strong horizontal inhomogeneity \citep{Schlawin+24}, which might lead to localized diamond hazes.
GJ1214b is suggested to have an extremely metal-rich atmosphere \citep{Kempton+23,Gao+23}, which makes the formation of diamond haze inefficient in the present framework.
However, some CVD experiments succeeded in forming CVD diamonds even at [M/H]$=4.0$ (\citealt{Bachmann+91}, see also Section \ref{sec:bachmann}).
Further observations of GJ1214b would help investigate the possibility of diamond hazes by searching for the CH$_4$ feature as well as the spectral feature of the diamond haze itself.


\section*{Acknowledgements}
I thank anonymous reviewer for helpful comments that improved the quality of this paper.
I also thank Anthony Jones for sharing the optical constants of hydrogenated nanodiamonds, Sarah Moran for helping the use of \texttt{PICASO}, and Chao He and Benjamin Fleury for giving permision to reuse Figures of \citet{He+18} and \citet{Fleury+19}.
I also thank Yuichi Ito and Tadahiro Kimura for the early conversation from which K.O. conceived the idea of CVD diamond, and Masahiro Ikoma, Jonathan Fortney, Xinting Yu and Xi Zhang for encouraging discussion and constructive comments. 
This work benefited from the 2024 Exoplanet Summer Program in the Other
Worlds Laboratory (OWL) at the University of California, Santa Cruz, a program funded by the Heising-Simons Foundation.
A part of numerical computations were carried out on PC cluster at Center for Computational Astrophysics, National Astronomical Observatory of Japan.
This work is supported by the JSPS KAKENHI Grant Number JP23K19072.

\software{\texttt{Matplotlib} \citep{Hunter2007matplotlib},
    \texttt{NumPy} \citep{harris2020NumPy},
    \texttt{SciPy} \citep{2020SciPy-NMeth},
    \texttt{astropy} \citep{astropy:2018},
    \texttt{PySynphot} \citep{STScI},
    \texttt{EGP} \citep{Marley&Robinson15},
    \texttt{VULCAN} \citep{Tsai+17,Tsai+21},
    \texttt{PyMieScatt} \citep{Sumlin+18},
    \texttt{CHIMERA} \citep{Line+13},
    \texttt{PICASO} \citep{Batalha+17,Mukherjee+22a}
}

\appendix

\section{Transmission Spectra with Diamond Hazes Containing Impurities}\label{appendix:meteo}
\begin{figure*}[t]
\centering
\includegraphics[clip, width=0.7\hsize]{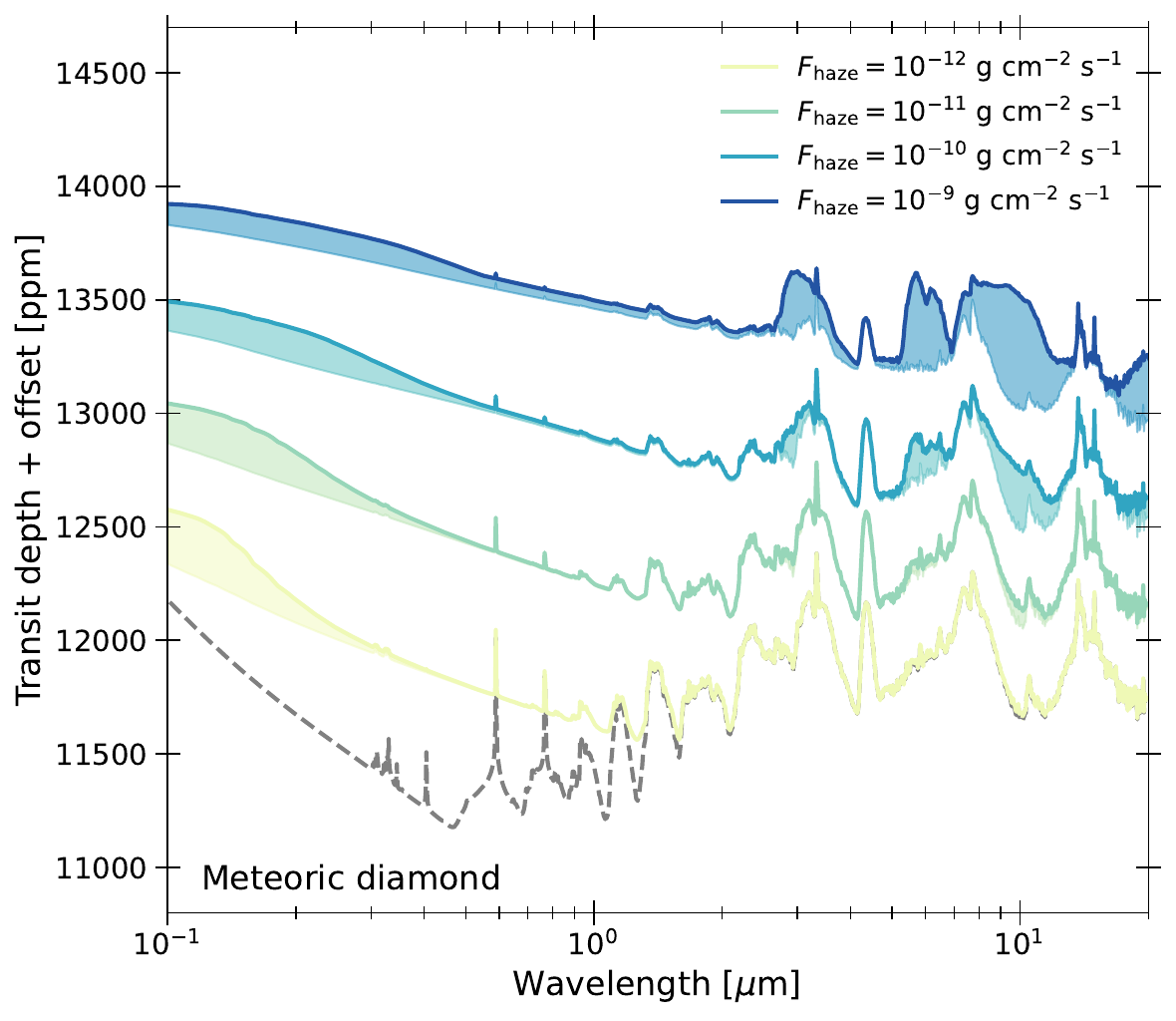}
\caption{Same as Figure \ref{fig:trans}, but I used the optical constants of meteoric diamonds \citep{Mutschke+04} as an example of diamond hazes with impurities.
}
\label{fig:trans_meteo}
\end{figure*}

The main text introduces the transmission spectra, including hydrogenated diamonds (Figure \ref{fig:trans}).
However, impurities within diamond hazes can alter the optical constants and thus shape of transmission spectra.
Figure \ref{fig:trans_meteo} shows the transmission spectra, where I now use the optical constants of meteoric diamonds \citep{Mutschke+04} as an example of diamonds with impurities.
Figure \ref{fig:trans_meteo} demonstrates that impurities indeed change the spectral features of diamond hazes, suggesting the importance of further experimental studies to measure the optical constants of diamonds that are synthesized from a gas mixture mimicking exoplanetary atmospheres.

\rerev{
\section{Optical Properties of Diamond-Soot Mixture}\label{appendix:mix}
\begin{figure*}[t]
\centering
\includegraphics[clip, width=\hsize]{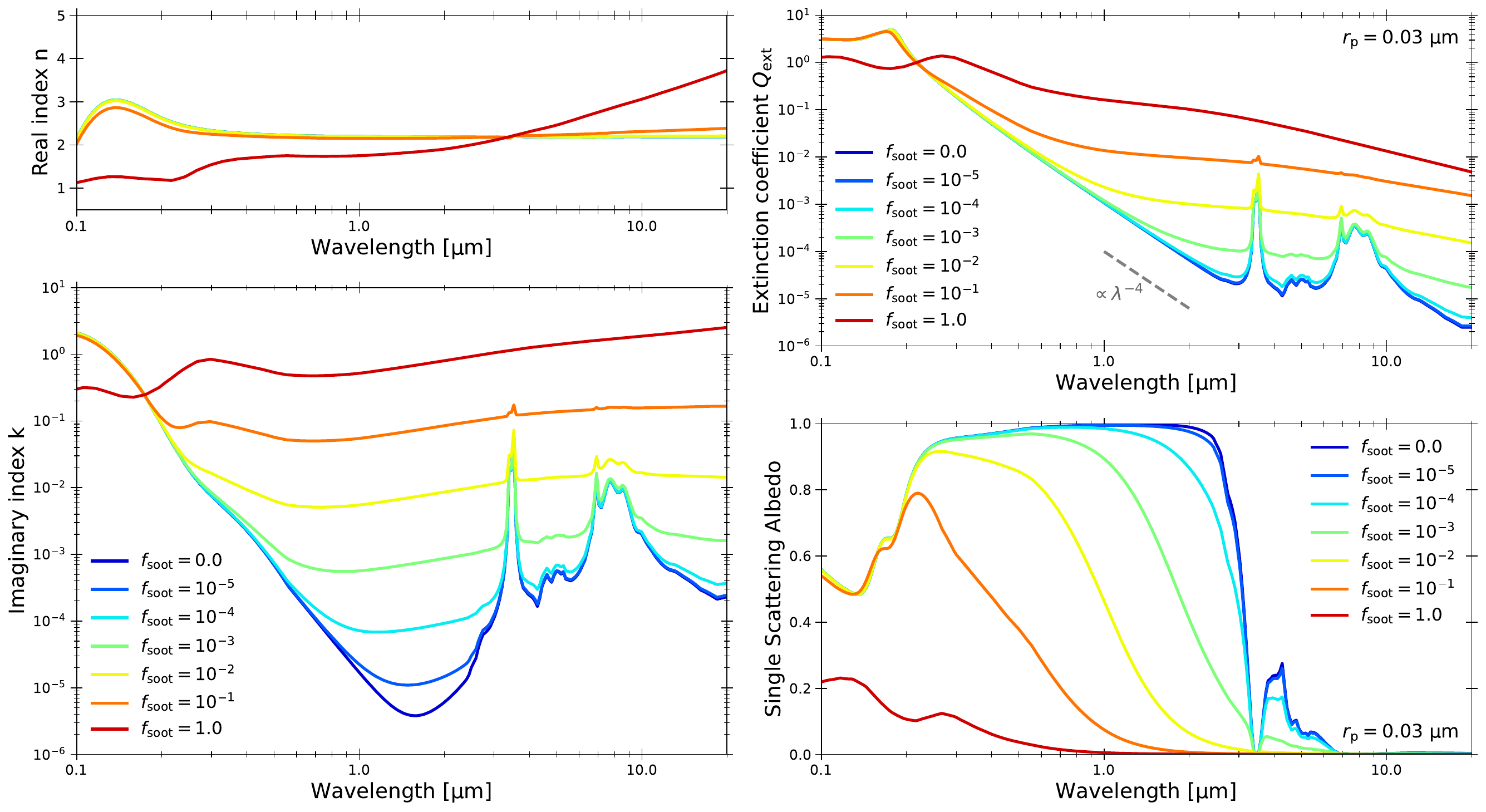}
\caption{(Left) Real and imaginary refractive indices, $n$ and $k$, of hydrogenated diamonds mixed with soot as a function of wavelengths. Different colored lines show the indices for different soot volume fraction $f_{\rm soot}=1-f_{\rm dia}$. (Right) Extinction coefficient $Q_{\rm ext}$ (top) and single scattering albedo (bottom) of a spherical particle made of a diamond-soot mixture for various soot volume fraction. The particle radius is set to $r_{\rm p}=0.03~{\rm {\mu}m}$.
}
\label{fig:appendix_mix}
\end{figure*}
In Section \ref{sec:results_observation}, I have assumed that haze particles are purely made of hydrogenated diamonds to clarify its observational signatures. In reality, haze particles could be the mixtures of CVD diamonds and other carbon deposits, such as soot. Here, I evaluate how contamination of soot components may affect the optical properties of diamond hazes. To this end, I calculated the effective refractive index of hydrogenated diamonds containing soot based on the Effective Medium theory with the Bruggeman mixing rule \citep[e.g.,][]{Bohren&Huffman83,Kiefer+24}.
The left panels of Figure \ref{fig:appendix_mix} show the effective refractive indices of the diamond-soot mixture. The real part of the refractive index $n$ is almost independent of the soot contamination, whereas the soot contamination could affect the imaginary part $k$ even with a soot volume fraction as low as $f_{\rm soot}\sim{10}^{-5}$, since hydrogenated diamonds intrinsically have an extremely low $k$ value. The right panels of Figure \ref{fig:appendix_mix} show the extinction coefficient $Q_{\rm ext}$ and the single scattering albedo of a spherical particle with a radius of $r_{\rm p}=0.03~{\rm {\mu}m}$.
Soot contamination starts to impact these optical properties once the soot volume fraction becomes $f_{\rm soot}\ga10^{-4}$--$10^{-3}$, depending on the particle size. 
Within the present model, the mass mixing ratio of soot is typically $\ga4$ orders of magnitude lower than that of CVD diamonds at a range of pressure levels relevant to atmospheric observations (Section \ref{sec:result}).
Thus, soot contamination would not greatly alter the qualitative conclusion for the observational signatures of diamond hazes predicted in Section \ref{sec:results_observation}.
}

\bibliography{references}

\begin{thebibliography}{}
\expandafter\ifx\csname natexlab\endcsname\relax\def\natexlab#1{#1}\fi
\providecommand{\url}[1]{\href{#1}{#1}}
\providecommand{\dodoi}[1]{doi:~\href{http://doi.org/#1}{\nolinkurl{#1}}}
\providecommand{\doeprint}[1]{\href{http://ascl.net/#1}{\nolinkurl{http://ascl.net/#1}}}
\providecommand{\doarXiv}[1]{\href{https://arxiv.org/abs/#1}{\nolinkurl{https://arxiv.org/abs/#1}}}

\bibitem[{{Adams} {et~al.}(2019){Adams}, {Gao}, {de Pater}, \& {Morley}}]{Adams+19}
{Adams}, D., {Gao}, P., {de Pater}, I., \& {Morley}, C.~V. 2019, \apj, 874, 61, \dodoi{10.3847/1538-4357/ab074c}

\bibitem[{{Alderson} {et~al.}(2022){Alderson}, {Wakeford}, {MacDonald}, {Lewis}, {May}, {Grant}, {Sing}, {Stevenson}, {Fowler}, {Goyal}, {Batalha}, \& {Kataria}}]{Alderson+22_wasp-17b_slope}
{Alderson}, L., {Wakeford}, H.~R., {MacDonald}, R.~J., {et~al.} 2022, \mnras, 512, 4185, \dodoi{10.1093/mnras/stac661}

\bibitem[{{Allamandola} {et~al.}(1992){Allamandola}, {Sandford}, {Tielens}, \& {Herbst}}]{Allamandola+92}
{Allamandola}, L.~J., {Sandford}, S.~A., {Tielens}, A.~G.~G.~M., \& {Herbst}, T.~M. 1992, \apj, 399, 134, \dodoi{10.1086/171909}

\bibitem[{{Allamandola} {et~al.}(1993){Allamandola}, {Sandford}, {Tielens}, \& {Herbst}}]{Allamandola+93}
---. 1993, Science, 260, 64, \dodoi{10.1126/science.11538059}

\bibitem[{{Anders} \& {Zinner}(1993)}]{Anders&Zinner93}
{Anders}, E., \& {Zinner}, E. 1993, Meteoritics, 28, 490, \dodoi{10.1111/j.1945-5100.1993.tb00274.x}

\bibitem[{Angus {et~al.}(1994)Angus, Argoitia, Gat, Li, Sunkara, Wang, \& Wang}]{Angus+94_review}
Angus, J.~C., Argoitia, A., Gat, R., {et~al.} 1994, Chemical vapour deposition of diamond, ed. A.~H. Lettington \& J.~W. Steeds (Dordrecht: Springer Netherlands), 1--14, \dodoi{10.1007/978-94-011-0725-9_1}

\bibitem[{Angus \& Hayman(1988)}]{Angus&Hayman88}
Angus, J.~C., \& Hayman, C.~C. 1988, Science, 241, 913, \dodoi{10.1126/science.241.4868.913}

\bibitem[{Appel {et~al.}(2000)Appel, Bockhorn, \& Frenklach}]{Appel+00}
Appel, J., Bockhorn, H., \& Frenklach, M. 2000, Combustion and Flame, 121, 122, \dodoi{https://doi.org/10.1016/S0010-2180(99)00135-2}

\bibitem[{{Arfaux} \& {Lavvas}(2022)}]{Arfaux&Lavvas22}
{Arfaux}, A., \& {Lavvas}, P. 2022, \mnras, \dodoi{10.1093/mnras/stac1772}

\bibitem[{{Arfaux} \& {Lavvas}(2024)}]{Arfaux&Lavvas24}
---. 2024, \mnras, 530, 482, \dodoi{10.1093/mnras/stae826}

\bibitem[{Ashfold \& Mankelevich(2023)}]{Ashfold&Mankelevich23}
Ashfold, M.~N., \& Mankelevich, Y.~A. 2023, Diamond and Related Materials, 137, 110097, \dodoi{https://doi.org/10.1016/j.diamond.2023.110097}

\bibitem[{{Bachmann} {et~al.}(1991){Bachmann}, {Leers}, \& {Lydtin}}]{Bachmann+91}
{Bachmann}, P.~K., {Leers}, D., \& {Lydtin}, H. 1991, Diamond and Related Materials, 1, 1, \dodoi{10.1016/0925-9635(91)90005-U}

\bibitem[{{Batalha} {et~al.}(2019){Batalha}, {Marley}, {Lewis}, \& {Fortney}}]{Batalha+19}
{Batalha}, N.~E., {Marley}, M.~S., {Lewis}, N.~K., \& {Fortney}, J.~J. 2019, \apj, 878, 70, \dodoi{10.3847/1538-4357/ab1b51}

\bibitem[{{Batalha} {et~al.}(2017){Batalha}, {Mandell}, {Pontoppidan}, {Stevenson}, {Lewis}, {Kalirai}, {Earl}, {Greene}, {Albert}, \& {Nielsen}}]{Batalha+17}
{Batalha}, N.~E., {Mandell}, A., {Pontoppidan}, K., {et~al.} 2017, \pasp, 129, 064501, \dodoi{10.1088/1538-3873/aa65b0}

\bibitem[{{Bean} {et~al.}(2010){Bean}, {Miller-Ricci Kempton}, \& {Homeier}}]{Bean+10}
{Bean}, J.~L., {Miller-Ricci Kempton}, E., \& {Homeier}, D. 2010, \nat, 468, 669, \dodoi{10.1038/nature09596}

\bibitem[{{Bell} {et~al.}(2023){Bell}, {Welbanks}, {Schlawin}, {Line}, {Fortney}, {Greene}, {Ohno}, {Parmentier}, {Rauscher}, {Beatty}, {Mukherjee}, {Wiser}, {Boyer}, {Rieke}, \& {Stansberry}}]{Bell+23}
{Bell}, T.~J., {Welbanks}, L., {Schlawin}, E., {et~al.} 2023, \nat, 623, 709, \dodoi{10.1038/s41586-023-06687-0}

\bibitem[{{Bell} {et~al.}(2024){Bell}, {Crouzet}, {Cubillos}, {Kreidberg}, {Piette}, {Roman}, {Barstow}, {Blecic}, {Carone}, {Coulombe}, {Ducrot}, {Hammond}, {Mendon{\c{c}}a}, {Moses}, {Parmentier}, {Stevenson}, {Teinturier}, {Zhang}, {Batalha}, {Bean}, {Benneke}, {Charnay}, {Chubb}, {Demory}, {Gao}, {Lee}, {L{\'o}pez-Morales}, {Morello}, {Rauscher}, {Sing}, {Tan}, {Venot}, {Wakeford}, {Aggarwal}, {Ahrer}, {Alam}, {Baeyens}, {Barrado}, {Caceres}, {Carter}, {Casewell}, {Challener}, {Crossfield}, {Decin}, {D{\'e}sert}, {Dobbs-Dixon}, {Dyrek}, {Espinoza}, {Feinstein}, {Gibson}, {Harrington}, {Helling}, {Hu}, {Iro}, {Kempton}, {Kendrew}, {Komacek}, {Krick}, {Lagage}, {Leconte}, {Lendl}, {Lewis}, {Lothringer}, {Malsky}, {Mancini}, {Mansfield}, {Mayne}, {Evans-Soma}, {Molaverdikhani}, {Nikolov}, {Nixon}, {Palle}, {Petit dit de la Roche}, {Piaulet}, {Powell}, {Rackham}, {Schneider}, {Steinrueck}, {Taylor}, {Welbanks}, {Yurchenko}, {Zhang}, \& {Zieba}}]{Bell+24}
{Bell}, T.~J., {Crouzet}, N., {Cubillos}, P.~E., {et~al.} 2024, Nature Astronomy, \dodoi{10.1038/s41550-024-02230-x}

\bibitem[{{Blake} {et~al.}(1988){Blake}, {Freund}, {Krishnan}, {Echer}, {Shipp}, {Bunch}, {Tielens}, {Lipari}, {Hetherington}, \& {Chang}}]{Blake+88}
{Blake}, D.~F., {Freund}, F., {Krishnan}, K.~F.~M., {et~al.} 1988, \nat, 332, 611, \dodoi{10.1038/332611a0}

\bibitem[{{Bohren} \& {Huffman}(1983)}]{Bohren&Huffman83}
{Bohren}, C.~F., \& {Huffman}, D.~R. 1983, {Absorption and scattering of light by small particles}

\bibitem[{Butenko {et~al.}(2000)Butenko, Kuznetsov, Chuvilin, Kolomiichuk, Stankus, Khairulin, \& Segall}]{Butenko+00}
Butenko, Y.~V., Kuznetsov, V.~L., Chuvilin, A.~L., {et~al.} 2000, Journal of Applied Physics, 88, 4380, \dodoi{10.1063/1.1289791}

\bibitem[{Butler {et~al.}(2009)Butler, Mankelevich, Cheesman, Ma, \& Ashfold}]{Butler+09}
Butler, J.~E., Mankelevich, Y.~A., Cheesman, A., Ma, J., \& Ashfold, M. N.~R. 2009, Journal of Physics: Condensed Matter, 21, 364201, \dodoi{10.1088/0953-8984/21/36/364201}

\bibitem[{Butler {et~al.}(1993)Butler, Woodin, Brown, \& Fallon}]{Butler+93_review}
Butler, J.~E., Woodin, R.~L., Brown, L.~M., \& Fallon, P. 1993, Philosophical Transactions: Physical Sciences and Engineering, 342, 209.
\newblock \url{http://www.jstor.org/stable/54081}

\bibitem[{Celii \& Butler(1991)}]{Celii&Butler91_CVD_review}
Celii, F.~G., \& Butler, J.~E. 1991, Annual Review of Physical Chemistry, 42, 643, \dodoi{10.1146/annurev.pc.42.100191.003235}

\bibitem[{{Charnay} {et~al.}(2015){Charnay}, {Meadows}, \& {Leconte}}]{Charnay+15}
{Charnay}, B., {Meadows}, V., \& {Leconte}, J. 2015, \apj, 813, 15, \dodoi{10.1088/0004-637X/813/1/15}

\bibitem[{{Chen} {et~al.}(2002){Chen}, {Wu}, {Cheng}, {Sheu}, \& {Chang}}]{Chen+02}
{Chen}, C.~F., {Wu}, C.~C., {Cheng}, C.~L., {Sheu}, S.~Y., \& {Chang}, H.~C. 2002, \jcp, 116, 1211, \dodoi{10.1063/1.1434947}

\bibitem[{Chu {et~al.}(1990)Chu, D'Evelyn, Hauge, \& Margrave}]{Chu+90}
Chu, C.~J., D'Evelyn, M.~P., Hauge, R.~H., \& Margrave, J.~L. 1990, Journal of Materials Research, 5, 2405

\bibitem[{{Chu} {et~al.}(1991){Chu}, {D'Evelyn}, {Hauge}, \& {Margrave}}]{Chu+91}
{Chu}, C.~J., {D'Evelyn}, M.~P., {Hauge}, R.~H., \& {Margrave}, J.~L. 1991, Journal of Applied Physics, 70, 1695, \dodoi{10.1063/1.349539}

\bibitem[{{Clayton} {et~al.}(1995){Clayton}, {Meyer}, {Sanderson}, {Russell}, \& {Pillinger}}]{Clayton+95}
{Clayton}, D.~D., {Meyer}, B.~S., {Sanderson}, C.~I., {Russell}, S.~S., \& {Pillinger}, C.~T. 1995, \apj, 447, 894, \dodoi{10.1086/175927}

\bibitem[{{Cloutier} {et~al.}(2021){Cloutier}, {Charbonneau}, {Deming}, {Bonfils}, \& {Astudillo-Defru}}]{Cloutier+21}
{Cloutier}, R., {Charbonneau}, D., {Deming}, D., {Bonfils}, X., \& {Astudillo-Defru}, N. 2021, \aj, 162, 174, \dodoi{10.3847/1538-3881/ac1584}

\bibitem[{{Cohen} {et~al.}(2024){Cohen}, {Palmer}, {Paradise}, {Bollasina}, \& {Tiranti}}]{Cohen+24}
{Cohen}, M., {Palmer}, P.~I., {Paradise}, A., {Bollasina}, M.~A., \& {Tiranti}, P.~I. 2024, \aj, 167, 97, \dodoi{10.3847/1538-3881/ad1ab9}

\bibitem[{Coltrin \& Dandy(1993)}]{Coltrin&Dandy93}
Coltrin, M.~E., \& Dandy, D.~S. 1993, Journal of Applied Physics, 74, 5803

\bibitem[{{Crossfield} \& {Kreidberg}(2017)}]{Crossfield&Kreidberg17}
{Crossfield}, I. J.~M., \& {Kreidberg}, L. 2017, \aj, 154, 261, \dodoi{10.3847/1538-3881/aa9279}

\bibitem[{{Daulton} {et~al.}(1996){Daulton}, {Eisenhour}, {Bernatowicz}, {Lewis}, \& {Buseck}}]{Daulton+96}
{Daulton}, T.~L., {Eisenhour}, D.~D., {Bernatowicz}, T.~J., {Lewis}, R.~S., \& {Buseck}, P.~R. 1996, \gca, 60, 4853, \dodoi{10.1016/S0016-7037(96)00223-2}

\bibitem[{{Davies}(1945)}]{Davies45}
{Davies}, C.~N. 1945, Proceedings of the Physical Society, 57, 259, \dodoi{10.1088/0959-5309/57/4/301}

\bibitem[{{Davies} \& {Evans}(1972)}]{Davies&Evans72}
{Davies}, G., \& {Evans}, T. 1972, Proceedings of the Royal Society of London Series A, 328, 413, \dodoi{10.1098/rspa.1972.0086}

\bibitem[{Donnelly {et~al.}(1997)Donnelly, McCullough, \& Geddes}]{Donnelly+97_Etching}
Donnelly, C., McCullough, R., \& Geddes, J. 1997, Diamond and Related Materials, 6, 787, \dodoi{https://doi.org/10.1016/S0925-9635(96)00606-1}

\bibitem[{{Dubey} {et~al.}(2023){Dubey}, {Gr{\"u}bel}, {Arenales-Lope}, {Molaverdikhani}, {Ercolano}, {Rab}, \& {Trapp}}]{Dubey+23}
{Dubey}, D., {Gr{\"u}bel}, F., {Arenales-Lope}, R., {et~al.} 2023, \aap, 678, A53, \dodoi{10.1051/0004-6361/202346958}

\bibitem[{Dunst \& Sternschulte(2009)}]{Dunst+09}
Dunst, S., \& Sternschulte, H. 2009, Applied physics letters, 94

\bibitem[{{Dymont} {et~al.}(2022){Dymont}, {Yu}, {Ohno}, {Zhang}, {Fortney}, {Thorngren}, \& {Dickinson}}]{Dymont+22}
{Dymont}, A.~H., {Yu}, X., {Ohno}, K., {et~al.} 2022, \apj, 937, 90, \dodoi{10.3847/1538-4357/ac7f40}

\bibitem[{{Dyrek} {et~al.}(2024){Dyrek}, {Min}, {Decin}, {Bouwman}, {Crouzet}, {Molli{\`e}re}, {Lagage}, {Konings}, {Tremblin}, {G{\"u}del}, {Pye}, {Waters}, {Henning}, {Vandenbussche}, {Ardevol Martinez}, {Argyriou}, {Ducrot}, {Heinke}, {van Looveren}, {Absil}, {Barrado}, {Baudoz}, {Boccaletti}, {Cossou}, {Coulais}, {Edwards}, {Gastaud}, {Glasse}, {Glauser}, {Greene}, {Kendrew}, {Krause}, {Lahuis}, {Mueller}, {Olofsson}, {Patapis}, {Rouan}, {Royer}, {Scheithauer}, {Waldmann}, {Whiteford}, {Colina}, {van Dishoeck}, {{\"O}stlin}, {Ray}, \& {Wright}}]{Dyrek+24}
{Dyrek}, A., {Min}, M., {Decin}, L., {et~al.} 2024, \nat, 625, 51, \dodoi{10.1038/s41586-023-06849-0}

\bibitem[{D’Evelyn {et~al.}(1992)D’Evelyn, Chu, Hange, \& Margrave}]{Evelyn+92}
D’Evelyn, M.~P., Chu, C.~J., Hange, R.~H., \& Margrave, J.~L. 1992, Journal of Applied Physics, 71, 1528, \dodoi{10.1063/1.351223}

\bibitem[{{Ercolano} {et~al.}(2022){Ercolano}, {Rab}, {Molaverdikhani}, {Edwards}, {Preibisch}, {Testi}, {Kamp}, \& {Thi}}]{Ercolano+22}
{Ercolano}, B., {Rab}, C., {Molaverdikhani}, K., {et~al.} 2022, \mnras, 512, 430, \dodoi{10.1093/mnras/stac505}

\bibitem[{{Espinoza} {et~al.}(2024){Espinoza}, {Steinrueck}, {Kirk}, {MacDonald}, {Savel}, {Arnold}, {Kempton}, {Murphy}, {Carone}, {Zamyatina}, {Lewis}, {Samra}, {Kiefer}, {Rauscher}, {Christie}, {Mayne}, {Helling}, {Rustamkulov}, {Parmentier}, {May}, {Carter}, {Zhang}, {L{\'o}pez-Morales}, {Allen}, {Blecic}, {Decin}, {Mancini}, {Molaverdikhani}, {Rackham}, {Palle}, {Tsai}, {Ahrer}, {Bean}, {Crossfield}, {Haegele}, {H{\'e}brard}, {Kreidberg}, {Powell}, {Schneider}, {Welbanks}, {Wheatley}, {Brahm}, \& {Crouzet}}]{Espinoza+24}
{Espinoza}, N., {Steinrueck}, M.~E., {Kirk}, J., {et~al.} 2024, arXiv e-prints, arXiv:2407.10294, \dodoi{10.48550/arXiv.2407.10294}

\bibitem[{{Estrela} {et~al.}(2021){Estrela}, {Swain}, {Roudier}, {West}, {Sedaghati}, \& {Valio}}]{Estrela+21}
{Estrela}, R., {Swain}, M.~R., {Roudier}, G.~M., {et~al.} 2021, \aj, 162, 91, \dodoi{10.3847/1538-3881/ac0c7c}

\bibitem[{Fan {et~al.}(2018)Fan, Constantin, Li, Liu, Keramatnejad, Azina, Huang, Golgir, Lu, Ahmadi, Wang, Shield, Cui, Silvain, \& Lu}]{Fan+18}
Fan, L.-S., Constantin, L., Li, D.-w., {et~al.} 2018, Light: Science \& Applications, 7, 17177

\bibitem[{{Fleury} {et~al.}(2019){Fleury}, {Gudipati}, {Henderson}, \& {Swain}}]{Fleury+19}
{Fleury}, B., {Gudipati}, M.~S., {Henderson}, B.~L., \& {Swain}, M. 2019, \apj, 871, 158, \dodoi{10.3847/1538-4357/aaf79f}

\bibitem[{Ford(1996)}]{Ford96}
Ford, I.~J. 1996, Journal of Physics D: Applied Physics, 29, 2229, \dodoi{10.1088/0022-3727/29/9/002}

\bibitem[{{Fortney} {et~al.}(2005){Fortney}, {Marley}, {Lodders}, {Saumon}, \& {Freedman}}]{Fortney+05}
{Fortney}, J.~J., {Marley}, M.~S., {Lodders}, K., {Saumon}, D., \& {Freedman}, R. 2005, \apjl, 627, L69, \dodoi{10.1086/431952}

\bibitem[{Frenklach \& Wang(1991{\natexlab{a}})}]{Frenklach&Wang91}
Frenklach, M., \& Wang, H. 1991{\natexlab{a}}, Phys. Rev. B, 43, 1520, \dodoi{10.1103/PhysRevB.43.1520}

\bibitem[{Frenklach \& Wang(1991{\natexlab{b}})}]{Frenklach&Wang91_soot}
---. 1991{\natexlab{b}}, Symposium (International) on Combustion, 23, 1559, \dodoi{https://doi.org/10.1016/S0082-0784(06)80426-1}

\bibitem[{Frenklach \& Wang(1994)}]{Frenklach&Wang94}
---. 1994, Detailed Mechanism and Modeling of Soot Particle Formation, ed. H.~Bockhorn (Berlin, Heidelberg: Springer Berlin Heidelberg), 165--192, \dodoi{10.1007/978-3-642-85167-4_10}

\bibitem[{{Fukui} {et~al.}(2014){Fukui}, {Kawashima}, {Ikoma}, {Narita}, {Onitsuka}, {Ita}, {Onozato}, {Nishiyama}, {Baba}, {Ryu}, {Hirano}, {Hori}, {Kurosaki}, {Kawauchi}, {Takahashi}, {Nagayama}, {Tamura}, {Kawai}, {Kuroda}, {Nagayama}, {Ohta}, {Shimizu}, {Yanagisawa}, {Yoshida}, \& {Izumiura}}]{Fukui+14}
{Fukui}, A., {Kawashima}, Y., {Ikoma}, M., {et~al.} 2014, \apj, 790, 108, \dodoi{10.1088/0004-637X/790/2/108}

\bibitem[{{Gao} \& {Benneke}(2018)}]{Gao&Benneke18}
{Gao}, P., \& {Benneke}, B. 2018, \apj, 863, 165, \dodoi{10.3847/1538-4357/aad461}

\bibitem[{{Gao} {et~al.}(2017{\natexlab{a}}){Gao}, {Marley}, {Zahnle}, {Robinson}, \& {Lewis}}]{Gao+17_sulfur}
{Gao}, P., {Marley}, M.~S., {Zahnle}, K., {Robinson}, T.~D., \& {Lewis}, N.~K. 2017{\natexlab{a}}, \aj, 153, 139, \dodoi{10.3847/1538-3881/aa5fab}

\bibitem[{{Gao} {et~al.}(2021){Gao}, {Wakeford}, {Moran}, \& {Parmentier}}]{Gao+21}
{Gao}, P., {Wakeford}, H.~R., {Moran}, S.~E., \& {Parmentier}, V. 2021, Journal of Geophysical Research (Planets), 126, e06655, \dodoi{10.1029/2020JE006655}

\bibitem[{{Gao} \& {Zhang}(2020)}]{Gao&Zhang20}
{Gao}, P., \& {Zhang}, X. 2020, \apj, 890, 93, \dodoi{10.3847/1538-4357/ab6a9b}

\bibitem[{{Gao} {et~al.}(2017{\natexlab{b}}){Gao}, {Fan}, {Wong}, {Liang}, {Shia}, {Kammer}, {Yung}, {Summers}, {Gladstone}, {Young}, {Olkin}, {Ennico}, {Weaver}, {Stern}, \& {New Horizons Science Team}}]{Gao+17}
{Gao}, P., {Fan}, S., {Wong}, M.~L., {et~al.} 2017{\natexlab{b}}, \icarus, 287, 116, \dodoi{10.1016/j.icarus.2016.09.030}

\bibitem[{{Gao} {et~al.}(2020){Gao}, {Thorngren}, {Lee}, {Fortney}, {Morley}, {Wakeford}, {Powell}, {Stevenson}, \& {Zhang}}]{Gao+20}
{Gao}, P., {Thorngren}, D.~P., {Lee}, G. K.~H., {et~al.} 2020, Nature Astronomy, \dodoi{10.1038/s41550-020-1114-3}

\bibitem[{{Gao} {et~al.}(2023){Gao}, {Piette}, {Steinrueck}, {Nixon}, {Zhang}, {Kempton}, {Bean}, {Rauscher}, {Parmentier}, {Batalha}, {Savel}, {Arnold}, {Roman}, {Malsky}, \& {Taylor}}]{Gao+23}
{Gao}, P., {Piette}, A. A.~A., {Steinrueck}, M.~E., {et~al.} 2023, arXiv e-prints, arXiv:2305.05697, \dodoi{10.48550/arXiv.2305.05697}

\bibitem[{Goodwin(1993)}]{Goodwin93}
Goodwin, D.~G. 1993, Journal of Applied Physics, 74, 6888

\bibitem[{{Goto} {et~al.}(2009){Goto}, {Henning}, {Kouchi}, {Takami}, {Hayano}, {Usuda}, {Takato}, {Terada}, {Oya}, {J{\"a}ger}, \& {Andersen}}]{Goto+09}
{Goto}, M., {Henning}, T., {Kouchi}, A., {et~al.} 2009, \apj, 693, 610, \dodoi{10.1088/0004-637X/693/1/610}

\bibitem[{{Grant} {et~al.}(2023){Grant}, {Lewis}, {Wakeford}, {Batalha}, {Glidden}, {Goyal}, {Mullens}, {MacDonald}, {May}, {Seager}, {Stevenson}, {Valenti}, {Visscher}, {Alderson}, {Allen}, {Ca{\~n}as}, {Col{\'o}n}, {Clampin}, {Espinoza}, {Gressier}, {Huang}, {Lin}, {Long}, {Louie}, {Pe{\~n}a-Guerrero}, {Ranjan}, {Sotzen}, {Valentine}, {Anderson}, {Balmer}, {Bellini}, {Hoch}, {Kammerer}, {Libralato}, {Mountain}, {Perrin}, {Pueyo}, {Rickman}, {Rebollido}, {Sohn}, {van der Marel}, \& {Watkins}}]{Grant+23}
{Grant}, D., {Lewis}, N.~K., {Wakeford}, H.~R., {et~al.} 2023, \apjl, 956, L32, \dodoi{10.3847/2041-8213/acfc3b10.3847/2041-8213/acfdab}

\bibitem[{Greaves {et~al.}(2018)Greaves, Scaife, Frayer, Green, Mason, \& Smith}]{Greaves+18}
Greaves, J.~S., Scaife, A. M.~M., Frayer, D.~T., {et~al.} 2018, Nature Astronomy, 2, 662

\bibitem[{{Guillois} {et~al.}(1999){Guillois}, {Ledoux}, \& {Reynaud}}]{Guillois+99}
{Guillois}, O., {Ledoux}, G., \& {Reynaud}, C. 1999, \apjl, 521, L133, \dodoi{10.1086/312199}

\bibitem[{Gulder {et~al.}(1996)Gulder, Snelling, \& Sawchuk}]{Gulder+96}
Gulder, O., Snelling, D., \& Sawchuk, R. 1996, Symposium (International) on Combustion, 26, 2351, \dodoi{https://doi.org/10.1016/S0082-0784(96)80064-6}

\bibitem[{{Habart} {et~al.}(2004){Habart}, {Testi}, {Natta}, \& {Carbillet}}]{Habart+04}
{Habart}, E., {Testi}, L., {Natta}, A., \& {Carbillet}, M. 2004, \apjl, 614, L129, \dodoi{10.1086/425867}

\bibitem[{Harris {et~al.}(2020)Harris, Millman, van~der Walt, Gommers, Virtanen, Cournapeau, Wieser, Taylor, Berg, Smith, Kern, Picus, Hoyer, van Kerkwijk, Brett, Haldane, del R{'{\i}}o, Wiebe, Peterson, G{'{e}}rard-Marchant, Sheppard, Reddy, Weckesser, Abbasi, Gohlke, \& Oliphant}]{harris2020NumPy}
Harris, C.~R., Millman, K.~J., van~der Walt, S.~J., {et~al.} 2020, Nature, 585, 357, \dodoi{10.1038/s41586-020-2649-2}

\bibitem[{Harris(1990)}]{Harris90}
Harris, S.~J. 1990, Applied Physics Letters, 56, 2298, \dodoi{10.1063/1.102946}

\bibitem[{Harris \& Goodwin(1993)}]{Harris&Goodwin93}
Harris, S.~J., \& Goodwin, D.~G. 1993, The Journal of Physical Chemistry, 97, 23

\bibitem[{Harris \& Weiner(1992)}]{Harris&Weiner92_C2H2}
Harris, S.~J., \& Weiner, A.~M. 1992, Thin Solid Films, 212, 201, \dodoi{https://doi.org/10.1016/0040-6090(92)90521-C}

\bibitem[{Harris {et~al.}(1991)Harris, Weiner, \& Perry}]{Harris&Weiner91_experiment}
Harris, S.~J., Weiner, A.~M., \& Perry, T.~A. 1991, Journal of Applied Physics, 70, 1385, \dodoi{10.1063/1.349546}

\bibitem[{{He} {et~al.}(2018{\natexlab{a}}){He}, {H{\"o}rst}, {Lewis}, {Yu}, {Moses}, {Kempton}, {McGuiggan}, {Morley}, {Valenti}, \& {Vuitton}}]{He+18}
{He}, C., {H{\"o}rst}, S.~M., {Lewis}, N.~K., {et~al.} 2018{\natexlab{a}}, \apjl, 856, L3, \dodoi{10.3847/2041-8213/aab42b}

\bibitem[{{He} {et~al.}(2018{\natexlab{b}}){He}, {H{\"o}rst}, {Lewis}, {Yu}, {Moses}, {Kempton}, {Marley}, {McGuiggan}, {Morley}, {Valenti}, \& {Vuitton}}]{He+18b}
---. 2018{\natexlab{b}}, \aj, 156, 38, \dodoi{10.3847/1538-3881/aac883}

\bibitem[{{He} {et~al.}(2019){He}, {H{\"o}rst}, {Lewis}, {Moses}, {Kempton}, {Marley}, {Morley}, {Valenti}, \& {Vuitton}}]{He+19_CO}
---. 2019, ACS Earth and Space Chemistry, 3, 39, \dodoi{10.1021/acsearthspacechem.8b00133}

\bibitem[{{He} {et~al.}(2020{\natexlab{a}}){He}, {H{\"o}rst}, {Lewis}, {Yu}, {Moses}, {McGuiggan}, {Marley}, {Kempton}, {Morley}, {Valenti}, \& {Vuitton}}]{He+20}
---. 2020{\natexlab{a}}, \psj, 1, 51, \dodoi{10.3847/PSJ/abb1a4}

\bibitem[{{He} {et~al.}(2020{\natexlab{b}}){He}, {H{\"o}rst}, {Lewis}, {Yu}, {Moses}, {McGuiggan}, {Marley}, {Kempton}, {Moran}, {Morley}, \& {Vuitton}}]{He+20_sulfur}
---. 2020{\natexlab{b}}, Nature Astronomy, 4, 986, \dodoi{10.1038/s41550-020-1072-9}

\bibitem[{{He} {et~al.}(2024){He}, {Radke}, {Moran}, {H{\"o}rst}, {Lewis}, {Moses}, {Marley}, {Batalha}, {Kempton}, {Morley}, {Valenti}, \& {Vuitton}}]{He+24}
{He}, C., {Radke}, M., {Moran}, S.~E., {et~al.} 2024, Nature Astronomy, 8, 182, \dodoi{10.1038/s41550-023-02140-4}

\bibitem[{{Helling} {et~al.}(2020){Helling}, {Kawashima}, {Graham}, {Samra}, {Chubb}, {Min}, {Waters}, \& {Parmentier}}]{Helling+20}
{Helling}, C., {Kawashima}, Y., {Graham}, V., {et~al.} 2020, \aap, 641, A178, \dodoi{10.1051/0004-6361/202037633}

\bibitem[{{Heng} {et~al.}(2012){Heng}, {Hayek}, {Pont}, \& {Sing}}]{Heng+12}
{Heng}, K., {Hayek}, W., {Pont}, F., \& {Sing}, D.~K. 2012, \mnras, 420, 20, \dodoi{10.1111/j.1365-2966.2011.19943.x}

\bibitem[{{Heng} {et~al.}(2016){Heng}, {Lyons}, \& {Tsai}}]{Heng+16}
{Heng}, K., {Lyons}, J.~R., \& {Tsai}, S.-M. 2016, \apj, 816, 96, \dodoi{10.3847/0004-637X/816/2/96}

\bibitem[{{H{\"o}rst} {et~al.}(2018){H{\"o}rst}, {He}, {Lewis}, {Kempton}, {Marley}, {Morley}, {Moses}, {Valenti}, \& {Vuitton}}]{Horst+18}
{H{\"o}rst}, S.~M., {He}, C., {Lewis}, N.~K., {et~al.} 2018, Nature Astronomy, 2, 303, \dodoi{10.1038/s41550-018-0397-0}

\bibitem[{Hosseini {et~al.}(2023)Hosseini, Tsolakis, Alagumalai, Mahian, Lam, Pan, Peng, Tabatabaei, \& Aghbashlo}]{Hosseini+23}
Hosseini, S.~H., Tsolakis, A., Alagumalai, A., {et~al.} 2023, Progress in Energy and Combustion Science, 98, 101100, \dodoi{https://doi.org/10.1016/j.pecs.2023.101100}

\bibitem[{Hsu(1988)}]{Hsu88}
Hsu, W.~L. 1988, Journal of Vacuum Science \& Technology A, 6, 1803, \dodoi{10.1116/1.575257}

\bibitem[{{Hu} {et~al.}(2013){Hu}, {Seager}, \& {Bains}}]{Hu+13}
{Hu}, R., {Seager}, S., \& {Bains}, W. 2013, \apj, 769, 6, \dodoi{10.1088/0004-637X/769/1/6}

\bibitem[{{Huang} {et~al.}(2024){Huang}, {Ormel}, \& {Min}}]{Huang+24}
{Huang}, H., {Ormel}, C.~W., \& {Min}, M. 2024, arXiv e-prints, arXiv:2409.18181, \dodoi{10.48550/arXiv.2409.18181}

\bibitem[{Hunter(2007)}]{Hunter2007matplotlib}
Hunter, J.~D. 2007, Computing In Science \& Engineering, 9, 90, \dodoi{10.1109/MCSE.2007.55}

\bibitem[{{Inglis} {et~al.}(2024){Inglis}, {Batalha}, {Lewis}, {Kataria}, {Knutson}, {Kilpatrick}, {Gagnebin}, {Mukherjee}, {Pettyjohn}, {Crossfield}, {Foote}, {Grant}, {Henry}, {Lally}, {McKemmish}, {Sing}, {Wakeford}, {Zapata Trujillo}, \& {Zellem}}]{Inglis+24_HD189}
{Inglis}, J., {Batalha}, N.~E., {Lewis}, N.~K., {et~al.} 2024, arXiv e-prints, arXiv:2409.11395, \dodoi{10.48550/arXiv.2409.11395}

\bibitem[{{Jacobs} {et~al.}(2023){Jacobs}, {D{\'e}sert}, {Gao}, {Morley}, {Arcangeli}, {Barat}, {Marley}, {Moses}, {Fortney}, {Bean}, {Stevenson}, \& {Panwar}}]{Jacobs+23}
{Jacobs}, B., {D{\'e}sert}, J.-M., {Gao}, P., {et~al.} 2023, \apjl, 956, L43, \dodoi{10.3847/2041-8213/acfee9}

\bibitem[{Jin \& Moustakas(1994)}]{Jin&Moustakas94}
Jin, S., \& Moustakas, T. 1994, Applied Physics Letters, 65, 403

\bibitem[{{Jones}(2012{\natexlab{a}})}]{Jones12a}
{Jones}, A.~P. 2012{\natexlab{a}}, \aap, 540, A1, \dodoi{10.1051/0004-6361/201117623}

\bibitem[{{Jones}(2012{\natexlab{b}})}]{Jones12b}
---. 2012{\natexlab{b}}, \aap, 540, A2, \dodoi{10.1051/0004-6361/201117624}

\bibitem[{{Jones}(2012{\natexlab{c}})}]{Jones12c}
---. 2012{\natexlab{c}}, \aap, 542, A98, \dodoi{10.1051/0004-6361/201118483}

\bibitem[{{Jones} {et~al.}(2004){Jones}, {d'Hendecourt}, {Sheu}, {Chang}, {Cheng}, \& {Hill}}]{Jones+04}
{Jones}, A.~P., {d'Hendecourt}, L.~B., {Sheu}, S.~Y., {et~al.} 2004, \aap, 416, 235, \dodoi{10.1051/0004-6361:20031708}

\bibitem[{{Jones} \& {Ysard}(2022)}]{Jones&Ysard22_diamond_optics}
{Jones}, A.~P., \& {Ysard}, N. 2022, \aap, 657, A128, \dodoi{10.1051/0004-6361/202141793}

\bibitem[{Kanai {et~al.}(2001)Kanai, Watanabe, \& Takakuwa}]{Kanai+01}
Kanai, C., Watanabe, K., \& Takakuwa, Y. 2001, Phys. Rev. B, 63, 235311, \dodoi{10.1103/PhysRevB.63.235311}

\bibitem[{{Kawashima} \& {Ikoma}(2018)}]{Kawashima&Ikoma18}
{Kawashima}, Y., \& {Ikoma}, M. 2018, \apj, 853, 7, \dodoi{10.3847/1538-4357/aaa0c5}

\bibitem[{{Kawashima} \& {Ikoma}(2019)}]{Kawashima&Ikoma19}
---. 2019, \apj, 877, 109, \dodoi{10.3847/1538-4357/ab1b1d}

\bibitem[{{Kawashima} \& {Min}(2021)}]{Kawashima&Min21}
{Kawashima}, Y., \& {Min}, M. 2021, \aap, 656, A90, \dodoi{10.1051/0004-6361/202141548}

\bibitem[{Kazakov {et~al.}(1995)Kazakov, Wang, \& Frenklach}]{Kazakov+95}
Kazakov, A., Wang, H., \& Frenklach, M. 1995, Combustion and Flame, 100, 111, \dodoi{https://doi.org/10.1016/0010-2180(94)00086-8}

\bibitem[{{Kempton} {et~al.}(2023){Kempton}, {Zhang}, {Bean}, {Steinrueck}, {Piette}, {Parmentier}, {Malsky}, {Roman}, {Rauscher}, {Gao}, {Bell}, {Xue}, {Taylor}, {Savel}, {Arnold}, {Nixon}, {Stevenson}, {Mansfield}, {Kendrew}, {Zieba}, {Ducrot}, {Dyrek}, {Lagage}, {Stassun}, {Henry}, {Barman}, {Lupu}, {Malik}, {Kataria}, {Ih}, {Fu}, {Welbanks}, \& {McGill}}]{Kempton+23}
{Kempton}, E. M.~R., {Zhang}, M., {Bean}, J.~L., {et~al.} 2023, arXiv e-prints, arXiv:2305.06240, \dodoi{10.48550/arXiv.2305.06240}

\bibitem[{{Khalafinejad} {et~al.}(2021){Khalafinejad}, {Molaverdikhani}, {Blecic}, {Mallonn}, {Nortmann}, {Caballero}, {Rahmati}, {Kaminski}, {Sadegi}, {Nagel}, {Carone}, {Amado}, {Azzaro}, {Bauer}, {Casasayas-Barris}, {Czesla}, {von Essen}, {Fossati}, {G{\"u}del}, {Henning}, {L{\'o}pez-Puertas}, {Lendl}, {L{\"u}ftinger}, {Montes}, {Oshagh}, {Pall{\'e}}, {Quirrenbach}, {Reffert}, {Reiners}, {Ribas}, {Stock}, {Yan}, {Zapatero Osorio}, \& {Zechmeister}}]{Khalafinejad+21}
{Khalafinejad}, S., {Molaverdikhani}, K., {Blecic}, J., {et~al.} 2021, \aap, 656, A142, \dodoi{10.1051/0004-6361/202141191}

\bibitem[{{Khare} {et~al.}(1984){Khare}, {Sagan}, {Arakawa}, {Suits}, {Callcott}, \& {Williams}}]{Khare+84}
{Khare}, B.~N., {Sagan}, C., {Arakawa}, E.~T., {et~al.} 1984, \icarus, 60, 127, \dodoi{10.1016/0019-1035(84)90142-8}

\bibitem[{{Kiefer} {et~al.}(2024){Kiefer}, {Samra}, {Lewis}, {Schneider}, {Min}, {Carone}, {Decin}, \& {Helling}}]{Kiefer+24}
{Kiefer}, S., {Samra}, D., {Lewis}, D.~A., {et~al.} 2024, \aap, 690, A244, \dodoi{10.1051/0004-6361/202450526}

\bibitem[{{Knutson} {et~al.}(2014){Knutson}, {Benneke}, {Deming}, \& {Homeier}}]{Knutson+14}
{Knutson}, H.~A., {Benneke}, B., {Deming}, D., \& {Homeier}, D. 2014, \nat, 505, 66, \dodoi{10.1038/nature12887}

\bibitem[{Kobashi {et~al.}(1988)Kobashi, Nishimura, Kawate, \& Horiuchi}]{Kobashi+88}
Kobashi, K., Nishimura, K., Kawate, Y., \& Horiuchi, T. 1988, Phys. Rev. B, 38, 4067, \dodoi{10.1103/PhysRevB.38.4067}

\bibitem[{{Komacek} {et~al.}(2019){Komacek}, {Showman}, \& {Parmentier}}]{Komacek+19}
{Komacek}, T.~D., {Showman}, A.~P., \& {Parmentier}, V. 2019, \apj, 881, 152, \dodoi{10.3847/1538-4357/ab338b}

\bibitem[{{Kouchi} {et~al.}(2005){Kouchi}, {Nakano}, {Kimura}, \& {Kaito}}]{Kouchi+05}
{Kouchi}, A., {Nakano}, H., {Kimura}, Y., \& {Kaito}, C. 2005, \apjl, 626, L129, \dodoi{10.1086/431758}

\bibitem[{{Kreidberg}(2018)}]{Kreidberg18}
{Kreidberg}, L. 2018, in Handbook of Exoplanets, ed. H.~J. {Deeg} \& J.~A. {Belmonte}, 100, \dodoi{10.1007/978-3-319-55333-7_100}

\bibitem[{{Kreidberg} {et~al.}(2014){Kreidberg}, {Bean}, {D{\'e}sert}, {Benneke}, {Deming}, {Stevenson}, {Seager}, {Berta-Thompson}, {Seifahrt}, \& {Homeier}}]{Kreidberg+14}
{Kreidberg}, L., {Bean}, J.~L., {D{\'e}sert}, J.-M., {et~al.} 2014, \nat, 505, 69, \dodoi{10.1038/nature12888}

\bibitem[{{Kuchner} \& {Seager}(2005)}]{Kuchner&Seager05}
{Kuchner}, M.~J., \& {Seager}, S. 2005, arXiv e-prints, astro, \dodoi{10.48550/arXiv.astro-ph/0504214}

\bibitem[{{Lavvas} \& {Arfaux}(2021)}]{Lavvas&Aufaux21}
{Lavvas}, P., \& {Arfaux}, A. 2021, \mnras, 502, 5643, \dodoi{10.1093/mnras/stab456}

\bibitem[{{Lavvas} \& {Koskinen}(2017)}]{Lavvas&Koskinen17}
{Lavvas}, P., \& {Koskinen}, T. 2017, \apj, 847, 32, \dodoi{10.3847/1538-4357/aa88ce}

\bibitem[{{Lavvas} {et~al.}(2019){Lavvas}, {Koskinen}, {Steinrueck}, {Garc{\'\i}a Mu{\~n}oz}, \& {Showman}}]{Lavvas+19}
{Lavvas}, P., {Koskinen}, T., {Steinrueck}, M.~E., {Garc{\'\i}a Mu{\~n}oz}, A., \& {Showman}, A.~P. 2019, \apj, 878, 118, \dodoi{10.3847/1538-4357/ab204e}

\bibitem[{{Lavvas} {et~al.}(2010){Lavvas}, {Yelle}, \& {Griffith}}]{Lavvas+11}
{Lavvas}, P., {Yelle}, R.~V., \& {Griffith}, C.~A. 2010, \icarus, 210, 832, \dodoi{10.1016/j.icarus.2010.07.025}

\bibitem[{Lee {et~al.}(1994)Lee, Minsek, Vestyck, \& Chen}]{Lee+94_Diamond}
Lee, S.~S., Minsek, D.~W., Vestyck, D.~J., \& Chen, P. 1994, Science, 263, 1596, \dodoi{10.1126/science.263.5153.1596}

\bibitem[{{Lewis} {et~al.}(1987){Lewis}, {Ming}, {Wacker}, {Anders}, \& {Steel}}]{Lewis+87}
{Lewis}, R.~S., {Ming}, T., {Wacker}, J.~F., {Anders}, E., \& {Steel}, E. 1987, \nat, 326, 160, \dodoi{10.1038/326160a0}

\bibitem[{{Libby-Roberts} {et~al.}(2020){Libby-Roberts}, {Berta-Thompson}, {D{\'e}sert}, {Masuda}, {Morley}, {Lopez}, {Deck}, {Fabrycky}, {Fortney}, {Line}, {Sanchis-Ojeda}, \& {Winn}}]{Libby-Roberts+20}
{Libby-Roberts}, J.~E., {Berta-Thompson}, Z.~K., {D{\'e}sert}, J.-M., {et~al.} 2020, \aj, 159, 57, \dodoi{10.3847/1538-3881/ab5d36}

\bibitem[{{Line} {et~al.}(2013){Line}, {Wolf}, {Zhang}, {Knutson}, {Kammer}, {Ellison}, {Deroo}, {Crisp}, \& {Yung}}]{Line+13}
{Line}, M.~R., {Wolf}, A.~S., {Zhang}, X., {et~al.} 2013, \apj, 775, 137, \dodoi{10.1088/0004-637X/775/2/137}

\bibitem[{{Lodge} {et~al.}(2024){Lodge}, {Wakeford}, \& {Leinhardt}}]{Lodge+24}
{Lodge}, M.~G., {Wakeford}, H.~R., \& {Leinhardt}, Z.~M. 2024, \mnras, 527, 11113, \dodoi{10.1093/mnras/stad3743}

\bibitem[{Lupu {et~al.}(2023)Lupu, Freedman, Gharib-Nezhad, Visscher, \& Molliere}]{Lupu+23}
Lupu, R., Freedman, R., Gharib-Nezhad, E., Visscher, C., \& Molliere, P. 2023, {Correlated k coefficients for H2-He atmospheres; 196 spectral windows and 1460 pressure-temperature points},  Zenodo, \dodoi{10.5281/zenodo.7542068}

\bibitem[{{Lyutovich} \& {Banhart}(1999)}]{Lyutovich&Banhart99}
{Lyutovich}, Y., \& {Banhart}, F. 1999, Applied Physics Letters, 74, 659, \dodoi{10.1063/1.122978}

\bibitem[{{Madhusudhan} {et~al.}(2012){Madhusudhan}, {Lee}, \& {Mousis}}]{Madhusudhan+12_diamond}
{Madhusudhan}, N., {Lee}, K. K.~M., \& {Mousis}, O. 2012, \apjl, 759, L40, \dodoi{10.1088/2041-8205/759/2/L40}

\bibitem[{{Mai} \& {Line}(2019)}]{Mai&Line19}
{Mai}, C., \& {Line}, M.~R. 2019, \apj, 883, 144, \dodoi{10.3847/1538-4357/ab3e6d}

\bibitem[{Marinelli {et~al.}(1994)Marinelli, Milani, Montuori, Paoletti, Tebano, Balestrino, \& Paroli}]{Marinelli+94}
Marinelli, M., Milani, E., Montuori, M., {et~al.} 1994, Journal of Applied Physics, 76, 5702, \dodoi{10.1063/1.357077}

\bibitem[{{Marley} \& {McKay}(1999)}]{Marley&McKay99}
{Marley}, M.~S., \& {McKay}, C.~P. 1999, \icarus, 138, 268, \dodoi{10.1006/icar.1998.6071}

\bibitem[{{Marley} \& {Robinson}(2015)}]{Marley&Robinson15}
{Marley}, M.~S., \& {Robinson}, T.~D. 2015, \araa, 53, 279, \dodoi{10.1146/annurev-astro-082214-122522}

\bibitem[{Martin \& Hill(1990)}]{Martin&Hill90}
Martin, L.~R., \& Hill, M.~W. 1990, Journal of materials science letters, 9, 621

\bibitem[{May(2000)}]{May00_review}
May, P.~W. 2000, Philosophical Transactions: Mathematical, Physical and Engineering Sciences, 358, 473.
\newblock \url{http://www.jstor.org/stable/2666865}

\bibitem[{May {et~al.}(2007)May, Ashfold, \& Mankelevich}]{May+07}
May, P.~W., Ashfold, M. N.~R., \& Mankelevich, Y.~A. 2007, Journal of Applied Physics, 101, 053115, \dodoi{10.1063/1.2696363}

\bibitem[{May \& Mankelevich(2008)}]{May&Mankelevich08}
May, P.~W., \& Mankelevich, Y.~A. 2008, The Journal of Physical Chemistry C, 112, 12432

\bibitem[{{Mbarek} \& {Kempton}(2016)}]{Mbarek&Kempton16}
{Mbarek}, R., \& {Kempton}, E. M.~R. 2016, \apj, 827, 121, \dodoi{10.3847/0004-637X/827/2/121}

\bibitem[{{McKay} {et~al.}(2001){McKay}, {Coustenis}, {Samuelson}, {Lemmon}, {Lorenz}, {Cabane}, {Rannou}, \& {Drossart}}]{McKay+01}
{McKay}, C.~P., {Coustenis}, A., {Samuelson}, R.~E., {et~al.} 2001, \planss, 49, 79, \dodoi{10.1016/S0032-0633(00)00051-9}

\bibitem[{{McKay} {et~al.}(1989){McKay}, {Pollack}, \& {Courtin}}]{McKay+89}
{McKay}, C.~P., {Pollack}, J.~B., \& {Courtin}, R. 1989, \icarus, 80, 23, \dodoi{10.1016/0019-1035(89)90160-7}

\bibitem[{{Molaverdikhani} {et~al.}(2020){Molaverdikhani}, {Henning}, \& {Molli{\`e}re}}]{Molaverdikhani+20}
{Molaverdikhani}, K., {Henning}, T., \& {Molli{\`e}re}, P. 2020, \apj, 899, 53, \dodoi{10.3847/1538-4357/aba52b}

\bibitem[{{Moran} {et~al.}(2020){Moran}, {H{\"o}rst}, {Vuitton}, {He}, {Lewis}, {Flandinet}, {Moses}, {North}, {Orthous-Daunay}, {Sebree}, {Wolters}, {Kempton}, {Marley}, {Morley}, \& {Valenti}}]{Moran+20}
{Moran}, S.~E., {H{\"o}rst}, S.~M., {Vuitton}, V., {et~al.} 2020, \psj, 1, 17, \dodoi{10.3847/PSJ/ab8eae}

\bibitem[{{Morley} {et~al.}(2013){Morley}, {Fortney}, {Kempton}, {Marley}, {Visscher}, \& {Zahnle}}]{Morley+13}
{Morley}, C.~V., {Fortney}, J.~J., {Kempton}, E.~M.-R., {et~al.} 2013, \apj, 775, 33, \dodoi{10.1088/0004-637X/775/1/33}

\bibitem[{{Morley} {et~al.}(2015){Morley}, {Fortney}, {Marley}, {Zahnle}, {Line}, {Kempton}, {Lewis}, \& {Cahoy}}]{Morley+15}
{Morley}, C.~V., {Fortney}, J.~J., {Marley}, M.~S., {et~al.} 2015, \apj, 815, 110, \dodoi{10.1088/0004-637X/815/2/110}

\bibitem[{{Moses} {et~al.}(2013){Moses}, {Madhusudhan}, {Visscher}, \& {Freedman}}]{Moses+13}
{Moses}, J.~I., {Madhusudhan}, N., {Visscher}, C., \& {Freedman}, R.~S. 2013, \apj, 763, 25, \dodoi{10.1088/0004-637X/763/1/25}

\bibitem[{{Moses} {et~al.}(2021){Moses}, {Tremblin}, {Venot}, \& {Miguel}}]{Moses+21}
{Moses}, J.~I., {Tremblin}, P., {Venot}, O., \& {Miguel}, Y. 2021, Experimental Astronomy, \dodoi{10.1007/s10686-021-09749-1}

\bibitem[{{Moses} {et~al.}(2011){Moses}, {Visscher}, {Fortney}, {Showman}, {Lewis}, {Griffith}, {Klippenstein}, {Shabram}, {Friedson}, {Marley}, \& {Freedman}}]{Moses+11}
{Moses}, J.~I., {Visscher}, C., {Fortney}, J.~J., {et~al.} 2011, \apj, 737, 15, \dodoi{10.1088/0004-637X/737/1/15}

\bibitem[{{Mukherjee} {et~al.}(2022){Mukherjee}, {Batalha}, {Fortney}, \& {Marley}}]{Mukherjee+22a}
{Mukherjee}, S., {Batalha}, N.~E., {Fortney}, J.~J., \& {Marley}, M.~S. 2022, arXiv e-prints, arXiv:2208.07836.
\newblock \doarXiv{2208.07836}

\bibitem[{{Murgas} {et~al.}(2020){Murgas}, {Chen}, {Nortmann}, {Palle}, \& {Nowak}}]{Murgas+20}
{Murgas}, F., {Chen}, G., {Nortmann}, L., {Palle}, E., \& {Nowak}, G. 2020, \aap, 641, A158, \dodoi{10.1051/0004-6361/202038161}

\bibitem[{{Murphy} {et~al.}(2024){Murphy}, {Beatty}, {Schlawin}, {Bell}, {Line}, {Greene}, {Parmentier}, {Rauscher}, {Welbanks}, {Fortney}, \& {Rieke}}]{Murphy+24}
{Murphy}, M.~M., {Beatty}, T.~G., {Schlawin}, E., {et~al.} 2024, arXiv e-prints, arXiv:2406.09863, \dodoi{10.48550/arXiv.2406.09863}

\bibitem[{{Mutschke} {et~al.}(2004){Mutschke}, {Andersen}, {J{\"a}ger}, {Henning}, \& {Braatz}}]{Mutschke+04}
{Mutschke}, H., {Andersen}, A.~C., {J{\"a}ger}, C., {Henning}, T., \& {Braatz}, A. 2004, \aap, 423, 983, \dodoi{10.1051/0004-6361:20034544}

\bibitem[{Nakano {et~al.}(2022)Nakano, Zhang, Kobayashi, Matsumoto, Inokuma, Yamasaki, Nebel, \& Tokuda}]{Nakano+22}
Nakano, Y., Zhang, X., Kobayashi, K., {et~al.} 2022, Diamond and Related Materials, 125, 108997, \dodoi{https://doi.org/10.1016/j.diamond.2022.108997}

\bibitem[{Neoh {et~al.}(1981)Neoh, Howard, \& Sarofim}]{Neoh+81_Soot_oxidization}
Neoh, K.~G., Howard, J.~B., \& Sarofim, A.~F. 1981, Soot Oxidation in Flames, ed. D.~C. Siegla \& G.~W. Smith (Boston, MA: Springer US), 261--282, \dodoi{10.1007/978-1-4757-6137-5_9}

\bibitem[{{Ohno} \& {Fortney}(2023)}]{Ohno&Fortney22a}
{Ohno}, K., \& {Fortney}, J.~J. 2023, \apj, 946, 18, \dodoi{10.3847/1538-4357/acafed}

\bibitem[{{Ohno} \& {Kawashima}(2020)}]{Ohno&Kawashima20}
{Ohno}, K., \& {Kawashima}, Y. 2020, \apjl, 895, L47, \dodoi{10.3847/2041-8213/ab93d7}

\bibitem[{{Ohno} \& {Okuzumi}(2017)}]{Ohno&Okuzumi17}
{Ohno}, K., \& {Okuzumi}, S. 2017, \apj, 835, 261, \dodoi{10.3847/1538-4357/835/2/261}

\bibitem[{{Ohno} \& {Okuzumi}(2018)}]{Ohno&Okuzumi18}
---. 2018, \apj, 859, 34, \dodoi{10.3847/1538-4357/aabee3}

\bibitem[{{Ohno} {et~al.}(2020){Ohno}, {Okuzumi}, \& {Tazaki}}]{Ohno+20}
{Ohno}, K., {Okuzumi}, S., \& {Tazaki}, R. 2020, \apj, 891, 131, \dodoi{10.3847/1538-4357/ab44bd}

\bibitem[{{Ohno} \& {Tanaka}(2021)}]{Ohno&Tanaka21}
{Ohno}, K., \& {Tanaka}, Y.~A. 2021, \apj, 920, 124, \dodoi{10.3847/1538-4357/ac1516}

\bibitem[{{Ohno} {et~al.}(2021){Ohno}, {Zhang}, {Tazaki}, \& {Okuzumi}}]{Ohno+21}
{Ohno}, K., {Zhang}, X., {Tazaki}, R., \& {Okuzumi}, S. 2021, \apj, 912, 37, \dodoi{10.3847/1538-4357/abee82}

\bibitem[{P.~G.~Partridge \& Ashfald(1994)}]{Partridge+94}
P.~G.~Partridge, P. W.~May, C. A.~R., \& Ashfald, M. N.~R. 1994, Materials Science and Technology, 10, 505, \dodoi{10.1179/mst.1994.10.6.505}

\bibitem[{{Palik}(1985)}]{Palik85}
{Palik}, E.~D. 1985, {Handbook of optical constants of solids}

\bibitem[{Pandey {et~al.}(2007)Pandey, Pundir, \& Panigrahi}]{Pandey+07}
Pandey, P., Pundir, B., \& Panigrahi, P. 2007, Combustion and Flame, 148, 249, \dodoi{https://doi.org/10.1016/j.combustflame.2006.09.004}

\bibitem[{{Parmentier} {et~al.}(2013){Parmentier}, {Showman}, \& {Lian}}]{Parmentier+13}
{Parmentier}, V., {Showman}, A.~P., \& {Lian}, Y. 2013, \aap, 558, A91, \dodoi{10.1051/0004-6361/201321132}

\bibitem[{Petherbridge {et~al.}(2001)Petherbridge, May, \& Ashfold}]{Petherbridge+01_Ternary}
Petherbridge, J.~R., May, P.~W., \& Ashfold, M. N.~R. 2001, Journal of Applied Physics, 89, 5219, \dodoi{10.1063/1.1360221}

\bibitem[{{Pont} {et~al.}(2013){Pont}, {Sing}, {Gibson}, {Aigrain}, {Henry}, \& {Husnoo}}]{Pont+13_hd189733_slope}
{Pont}, F., {Sing}, D.~K., {Gibson}, N.~P., {et~al.} 2013, \mnras, 432, 2917, \dodoi{10.1093/mnras/stt651}

\bibitem[{{Poser} {et~al.}(2019){Poser}, {Nettelmann}, \& {Redmer}}]{Poser+19}
{Poser}, A.~J., {Nettelmann}, N., \& {Redmer}, R. 2019, Atmosphere, 10, 664, \dodoi{10.3390/atmos10110664}

\bibitem[{{Poser} \& {Redmer}(2024)}]{Poser&Redmer24}
{Poser}, A.~J., \& {Redmer}, R. 2024, \mnras, 529, 2242, \dodoi{10.1093/mnras/stae645}

\bibitem[{{Powell} {et~al.}(2019){Powell}, {Louden}, {Kreidberg}, {Zhang}, {Gao}, \& {Parmentier}}]{Powell+19}
{Powell}, D., {Louden}, T., {Kreidberg}, L., {et~al.} 2019, \apj, 887, 170, \dodoi{10.3847/1538-4357/ab55d9}

\bibitem[{Prelas {et~al.}(1997)Prelas, Popovici, \& Bigelow}]{CVD_diamond_handbook}
Prelas, M., Popovici, G., \& Bigelow, L. 1997, Handbook of Industrial Diamonds and Diamond Films (Taylor \& Francis).
\newblock \url{https://books.google.co.jp/books?id=X3qe9jzYUAQC}

\bibitem[{{Price-Whelan} {et~al.}(2018){Price-Whelan}, {Sip{\H{o}}cz}, {G{\"u}nther}, {Lim}, {Crawford}, {Conseil}, {Shupe}, {Craig}, {Dencheva}, {Ginsburg}, {VanderPlas}, {Bradley}, {P{\'e}rez-Su{\'a}rez}, {de Val-Borro}, {Paper Contributors}, {Aldcroft}, {Cruz}, {Robitaille}, {Tollerud}, {Coordination Committee}, {Ardelean}, {Babej}, {Bach}, {Bachetti}, {Bakanov}, {Bamford}, {Barentsen}, {Barmby}, {Baumbach}, {Berry}, {Biscani}, {Boquien}, {Bostroem}, {Bouma}, {Brammer}, {Bray}, {Breytenbach}, {Buddelmeijer}, {Burke}, {Calderone}, {Cano Rodr{\'\i}guez}, {Cara}, {Cardoso}, {Cheedella}, {Copin}, {Corrales}, {Crichton}, {D{\textquoteright}Avella}, {Deil}, {Depagne}, {Dietrich}, {Donath}, {Droettboom}, {Earl}, {Erben}, {Fabbro}, {Ferreira}, {Finethy}, {Fox}, {Garrison}, {Gibbons}, {Goldstein}, {Gommers}, {Greco}, {Greenfield}, {Groener}, {Grollier}, {Hagen}, {Hirst}, {Homeier}, {Horton}, {Hosseinzadeh}, {Hu}, {Hunkeler}, {Ivezi{\'c}}, {Jain}, {Jenness}, {Kanarek}, {Kendrew}, {Kern}, {Kerzendorf}, {Khvalko},
  {King}, {Kirkby}, {Kulkarni}, {Kumar}, {Lee}, {Lenz}, {Littlefair}, {Ma}, {Macleod}, {Mastropietro}, {McCully}, {Montagnac}, {Morris}, {Mueller}, {Mumford}, {Muna}, {Murphy}, {Nelson}, {Nguyen}, {Ninan}, {N{\"o}the}, {Ogaz}, {Oh}, {Parejko}, {Parley}, {Pascual}, {Patil}, {Patil}, {Plunkett}, {Prochaska}, {Rastogi}, {Reddy Janga}, {Sabater}, {Sakurikar}, {Seifert}, {Sherbert}, {Sherwood-Taylor}, {Shih}, {Sick}, {Silbiger}, {Singanamalla}, {Singer}, {Sladen}, {Sooley}, {Sornarajah}, {Streicher}, {Teuben}, {Thomas}, {Tremblay}, {Turner}, {Terr{\'o}n}, {van Kerkwijk}, {de la Vega}, {Watkins}, {Weaver}, {Whitmore}, {Woillez}, {Zabalza}, \& {Contributors}}]{astropy:2018}
{Price-Whelan}, A.~M., {Sip{\H{o}}cz}, B.~M., {G{\"u}nther}, H.~M., {et~al.} 2018, \aj, 156, 123, \dodoi{10.3847/1538-3881/aabc4f}

\bibitem[{{Rossow}(1978)}]{Rossow78}
{Rossow}, W.~B. 1978, \icarus, 36, 1, \dodoi{10.1016/0019-1035(78)90072-6}

\bibitem[{{Rybicki} \& {Lightman}(1979)}]{Ribicki&Lightman79}
{Rybicki}, G.~B., \& {Lightman}, A.~P. 1979, {Radiative processes in astrophysics}

\bibitem[{{Sagan} \& {Khare}(1979)}]{Sagan&Khare79}
{Sagan}, C., \& {Khare}, B.~N. 1979, \nat, 277, 102, \dodoi{10.1038/277102a0}

\bibitem[{{Schlawin} {et~al.}(2024){Schlawin}, {Mukherjee}, {Ohno}, {Bell}, {Beatty}, {Greene}, {Line}, {Challener}, {Parmentier}, {Fortney}, {Rauscher}, {Wiser}, {Welbanks}, {Murphy}, {Edelman}, {Batalha}, {Moran}, {Mehta}, \& {Rieke}}]{Schlawin+24}
{Schlawin}, E., {Mukherjee}, S., {Ohno}, K., {et~al.} 2024, arXiv e-prints, arXiv:2406.15543, \dodoi{10.48550/arXiv.2406.15543}

\bibitem[{Schwander \& Partes(2011)}]{Schwander&Partes11_CVD_review}
Schwander, M., \& Partes, K. 2011, Diamond and Related Materials, 20, 1287, \dodoi{https://doi.org/10.1016/j.diamond.2011.08.005}

\bibitem[{Setaka(1989)}]{Setaka89}
Setaka, N. 1989, Journal of Materials Research, 4, 664

\bibitem[{{Sheu} {et~al.}(2002){Sheu}, {Lee}, {Lee}, \& {Chang}}]{Sheu+02}
{Sheu}, S.~Y., {Lee}, I.~P., {Lee}, Y.~T., \& {Chang}, H.~C. 2002, \apjl, 581, L55, \dodoi{10.1086/345519}

\bibitem[{{Sing} {et~al.}(2016){Sing}, {Fortney}, {Nikolov}, {Wakeford}, {Kataria}, {Evans}, {Aigrain}, {Ballester}, {Burrows}, {Deming}, {D{\'e}sert}, {Gibson}, {Henry}, {Huitson}, {Knutson}, {Lecavelier Des Etangs}, {Pont}, {Showman}, {Vidal-Madjar}, {Williamson}, \& {Wilson}}]{Sing+16}
{Sing}, D.~K., {Fortney}, J.~J., {Nikolov}, N., {et~al.} 2016, \nat, 529, 59, \dodoi{10.1038/nature16068}

\bibitem[{{Sing} {et~al.}(2024){Sing}, {Rustamkulov}, {Thorngren}, {Barstow}, {Tremblin}, {Alves de Oliveira}, {Beck}, {Birkmann}, {Challener}, {Crouzet}, {Espinoza}, {Ferruit}, {Giardino}, {Gressier}, {Lee}, {Lewis}, {Maiolino}, {Manjavacas}, {Rauscher}, {Sirianni}, \& {Valenti}}]{Sing+24}
{Sing}, D.~K., {Rustamkulov}, Z., {Thorngren}, D.~P., {et~al.} 2024, arXiv e-prints, arXiv:2405.11027, \dodoi{10.48550/arXiv.2405.11027}

\bibitem[{Spitsyn {et~al.}(1981)Spitsyn, Bouilov, \& Derjaguin}]{Spytsyn+81}
Spitsyn, B., Bouilov, L., \& Derjaguin, B. 1981, Journal of Crystal Growth, 52, 219, \dodoi{https://doi.org/10.1016/0022-0248(81)90197-4}

\bibitem[{{Steinrueck} {et~al.}(2023){Steinrueck}, {Koskinen}, {Lavvas}, {Parmentier}, {Zieba}, {Tan}, {Zhang}, \& {Kreidberg}}]{Steinrueck+23}
{Steinrueck}, M.~E., {Koskinen}, T., {Lavvas}, P., {et~al.} 2023, arXiv e-prints, arXiv:2305.09654, \dodoi{10.48550/arXiv.2305.09654}

\bibitem[{{Steinrueck} {et~al.}(2021){Steinrueck}, {Showman}, {Lavvas}, {Koskinen}, {Tan}, \& {Zhang}}]{Steinrueck+21}
{Steinrueck}, M.~E., {Showman}, A.~P., {Lavvas}, P., {et~al.} 2021, \mnras, 504, 2783, \dodoi{10.1093/mnras/stab1053}

\bibitem[{{STScI Development Team}(2013)}]{STScI}
{STScI Development Team}. 2013, {pysynphot: Synthetic photometry software package}, Astrophysics Source Code Library, record ascl:1303.023

\bibitem[{Sumlin {et~al.}(2018)Sumlin, Heinson, \& Chakrabarty}]{Sumlin+18}
Sumlin, B.~J., Heinson, W.~R., \& Chakrabarty, R.~K. 2018, Journal of Quantitative Spectroscopy and Radiative Transfer, 205, 127, \dodoi{https://doi.org/10.1016/j.jqsrt.2017.10.012}

\bibitem[{{Tan}(2022)}]{Tan22}
{Tan}, X. 2022, \mnras, 511, 4861, \dodoi{10.1093/mnras/stac344}

\bibitem[{{Taylor} {et~al.}(2021){Taylor}, {Parmentier}, {Line}, {Lee}, {Irwin}, \& {Aigrain}}]{Taylor+21}
{Taylor}, J., {Parmentier}, V., {Line}, M.~R., {et~al.} 2021, \mnras, 506, 1309, \dodoi{10.1093/mnras/stab1854}

\bibitem[{{Tazaki} \& {Tanaka}(2018)}]{Tazaki&Tanaka18}
{Tazaki}, R., \& {Tanaka}, H. 2018, \apj, 860, 79, \dodoi{10.3847/1538-4357/aac32d}

\bibitem[{{Teinturier} {et~al.}(2024){Teinturier}, {Charnay}, {Spiga}, {B{\'e}zard}, {Leconte}, {Mechineau}, {Ducrot}, {Millour}, \& {Cl{\'e}ment}}]{Teinturier+24}
{Teinturier}, L., {Charnay}, B., {Spiga}, A., {et~al.} 2024, \aap, 683, A231, \dodoi{10.1051/0004-6361/202347069}

\bibitem[{{Thorngren} {et~al.}(2019){Thorngren}, {Gao}, \& {Fortney}}]{Thorngren+19}
{Thorngren}, D., {Gao}, P., \& {Fortney}, J.~J. 2019, \apjl, 884, L6, \dodoi{10.3847/2041-8213/ab43d0}

\bibitem[{{Tielens} {et~al.}(1987){Tielens}, {Seab}, {Hollenbach}, \& {McKee}}]{Tielens+87}
{Tielens}, A.~G.~G.~M., {Seab}, C.~G., {Hollenbach}, D.~J., \& {McKee}, C.~F. 1987, \apjl, 319, L109, \dodoi{10.1086/184964}

\bibitem[{{Tsai} {et~al.}(2017){Tsai}, {Lyons}, {Grosheintz}, {Rimmer}, {Kitzmann}, \& {Heng}}]{Tsai+17}
{Tsai}, S.-M., {Lyons}, J.~R., {Grosheintz}, L., {et~al.} 2017, \apjs, 228, 20, \dodoi{10.3847/1538-4365/228/2/20}

\bibitem[{{Tsai} {et~al.}(2021){Tsai}, {Malik}, {Kitzmann}, {Lyons}, {Fateev}, {Lee}, \& {Heng}}]{Tsai+21}
{Tsai}, S.-M., {Malik}, M., {Kitzmann}, D., {et~al.} 2021, \apj, 923, 264, \dodoi{10.3847/1538-4357/ac29bc}

\bibitem[{{Vahidinia} {et~al.}(2024){Vahidinia}, {Moran}, {Marley}, \& {Cuzzi}}]{Vahidinia+24}
{Vahidinia}, S., {Moran}, S.~E., {Marley}, M.~S., \& {Cuzzi}, J.~N. 2024, arXiv e-prints, arXiv:2408.11215, \dodoi{10.48550/arXiv.2408.11215}

\bibitem[{{Valentine} {et~al.}(2024){Valentine}, {Wakeford}, {Challener}, {Batalha}, {Lewis}, {Grant}, {Mullens}, {Alderson}, {Goyal}, {MacDonald}, {May}, {Seager}, {Stevenson}, {Valenti}, {Allen}, {Espinoza}, {Glidden}, {Gressier}, {Huang}, {Lin}, {Long}, {Louie}, {Clampin}, {Perrin}, {van der Marel}, \& {Mountain}}]{Valentine+24}
{Valentine}, D., {Wakeford}, H.~R., {Challener}, R.~C., {et~al.} 2024, \aj, 168, 123, \dodoi{10.3847/1538-3881/ad5c61}

\bibitem[{{Van Kerckhoven} {et~al.}(2002{\natexlab{a}}){Van Kerckhoven}, {Tielens}, \& {Waelkens}}]{VanKerckhoven02}
{Van Kerckhoven}, C., {Tielens}, A.~G.~G.~M., \& {Waelkens}, C. 2002{\natexlab{a}}, \aap, 384, 568, \dodoi{10.1051/0004-6361:20011814}

\bibitem[{{Van Kerckhoven} {et~al.}(2002{\natexlab{b}}){Van Kerckhoven}, {Tielens}, \& {Waelkens}}]{VanKerckhoven+02}
---. 2002{\natexlab{b}}, \aap, 384, 568, \dodoi{10.1051/0004-6361:20011814}

\bibitem[{Virtanen {et~al.}(2020)Virtanen, Gommers, Oliphant, Haberland, Reddy, Cournapeau, Burovski, Peterson, Weckesser, Bright, {van der Walt}, Brett, Wilson, Millman, Mayorov, Nelson, Jones, Kern, Larson, Carey, Polat, Feng, Moore, {VanderPlas}, Laxalde, Perktold, Cimrman, Henriksen, Quintero, Harris, Archibald, Ribeiro, Pedregosa, {van Mulbregt}, \& {SciPy 1.0 Contributors}}]{2020SciPy-NMeth}
Virtanen, P., Gommers, R., Oliphant, T.~E., {et~al.} 2020, Nature Methods, 17, 261, \dodoi{10.1038/s41592-019-0686-2}

\bibitem[{{Visscher} \& {Moses}(2011)}]{Vischer&Moses11}
{Visscher}, C., \& {Moses}, J.~I. 2011, \apj, 738, 72, \dodoi{10.1088/0004-637X/738/1/72}

\bibitem[{{Wan} {et~al.}(1998){Wan}, {Zhang}, {Liu}, \& {Wang}}]{Wan+98}
{Wan}, Y.~Z., {Zhang}, D.~W., {Liu}, Z.~J., \& {Wang}, J.~T. 1998, Applied Physics A: Materials Science \& Processing, 67, 225, \dodoi{10.1007/s003390050762}

\bibitem[{Wang(2011)}]{Wang11_soot_review}
Wang, H. 2011, Proceedings of the Combustion Institute, 33, 41, \dodoi{https://doi.org/10.1016/j.proci.2010.09.009}

\bibitem[{Wang {et~al.}(1998)Wang, Wan, Liu, Wang, Zhang, \& Huang}]{Wang+98_Ternay}
Wang, J.-T., Wan, Y.-Z., Liu, Z.-J., {et~al.} 1998, Materials Letters, 33, 311, \dodoi{https://doi.org/10.1016/S0167-577X(97)00123-7}

\bibitem[{Wang \& Chung(2019)}]{Wang&Suk19_soot_review}
Wang, Y., \& Chung, S.~H. 2019, Progress in Energy and Combustion Science, 74, 152, \dodoi{https://doi.org/10.1016/j.pecs.2019.05.003}

\bibitem[{Wang {et~al.}(2015)Wang, Raj, \& Chung}]{Wang+15_soot}
Wang, Y., Raj, A., \& Chung, S.~H. 2015, Combustion and Flame, 162, 586, \dodoi{https://doi.org/10.1016/j.combustflame.2014.08.016}

\bibitem[{{Welbanks} {et~al.}(2024){Welbanks}, {Bell}, {Beatty}, {Line}, {Ohno}, {Fortney}, {Schlawin}, {Greene}, {Rauscher}, {McGill}, {Murphy}, {Parmentier}, {Tang}, {Edelman}, {Mukherjee}, {Wiser}, {Lagage}, {Dyrek}, \& {Arnold}}]{Welbanks+24}
{Welbanks}, L., {Bell}, T.~J., {Beatty}, T.~G., {et~al.} 2024, arXiv e-prints, arXiv:2405.11018, \dodoi{10.48550/arXiv.2405.11018}

\bibitem[{{Wong} {et~al.}(2022){Wong}, {Chachan}, {Knutson}, {Henry}, {Adams}, {Kataria}, {Benneke}, {Gao}, {Deming}, {L{\'o}pez-Morales}, {Sing}, {Alam}, {Ballester}, {Barstow}, {Buchhave}, {dos Santos}, {Fu}, {Garc{\'\i}a Mu{\~n}oz}, {MacDonald}, {Mikal-Evans}, {Sanz-Forcada}, \& {Wakeford}}]{Wong+22}
{Wong}, I., {Chachan}, Y., {Knutson}, H.~A., {et~al.} 2022, \aj, 164, 30, \dodoi{10.3847/1538-3881/ac7234}

\bibitem[{Xu {et~al.}(2020)Xu, Yan, Wang, \& Chung}]{Xu+20_soot_hydrogen}
Xu, L., Yan, F., Wang, Y., \& Chung, S.~H. 2020, Combustion and Flame, 213, 14, \dodoi{https://doi.org/10.1016/j.combustflame.2019.11.011}

\bibitem[{Yu \& Girshick(1994)}]{Yu&Girshick94}
Yu, B.~W., \& Girshick, S.~L. 1994, Journal of Applied Physics, 75, 3914, \dodoi{10.1063/1.356037}

\bibitem[{{Yu} {et~al.}(2021){Yu}, {Moses}, {Fortney}, \& {Zhang}}]{Yu+21}
{Yu}, X., {Moses}, J.~I., {Fortney}, J.~J., \& {Zhang}, X. 2021, \apj, 914, 38, \dodoi{10.3847/1538-4357/abfdc7}

\bibitem[{{Zahnle} {et~al.}(2009){Zahnle}, {Marley}, \& {Fortney}}]{Zahnle+09_soot}
{Zahnle}, K., {Marley}, M.~S., \& {Fortney}, J.~J. 2009, arXiv e-prints, arXiv:0911.0728, \dodoi{10.48550/arXiv.0911.0728}

\bibitem[{{Zaiser} \& {Banhart}(1997)}]{Zaiser&Banhart97}
{Zaiser}, M., \& {Banhart}, F. 1997, \prl, 79, 3680, \dodoi{10.1103/PhysRevLett.79.3680}

\bibitem[{{Zaiser} {et~al.}(2000){Zaiser}, {Lyutovich}, \& {Banhart}}]{Zaiser+00}
{Zaiser}, M., {Lyutovich}, Y., \& {Banhart}, F. 2000, \prb, 62, 3058, \dodoi{10.1103/PhysRevB.62.3058}

\bibitem[{{Zhang} {et~al.}(2020){Zhang}, {Chachan}, {Kempton}, {Knutson}, \& {Chang}}]{Zhang20_hd189733b}
{Zhang}, M., {Chachan}, Y., {Kempton}, E. M.~R., {Knutson}, H.~A., \& {Chang}, W.~H. 2020, \apj, 899, 27, \dodoi{10.3847/1538-4357/aba1e6}

\bibitem[{{Zhang}(2020)}]{Zhang20_review}
{Zhang}, X. 2020, Research in Astronomy and Astrophysics, 20, 099, \dodoi{10.1088/1674-4527/20/7/99}

\bibitem[{{Zhang} \& {Showman}(2018{\natexlab{a}})}]{Zhang&Showman18a}
{Zhang}, X., \& {Showman}, A.~P. 2018{\natexlab{a}}, \apj, 866, 1, \dodoi{10.3847/1538-4357/aada85}

\bibitem[{{Zhang} \& {Showman}(2018{\natexlab{b}})}]{Zhang&Showman18b}
---. 2018{\natexlab{b}}, \apj, 866, 2, \dodoi{10.3847/1538-4357/aada7c}

\bibitem[{Zhang {et~al.}(2023)Zhang, Zhang, Zhang, Yang, \& Gan}]{Zhang+23_diamond_etching}
Zhang, Y., Zhang, D., Zhang, L., Yang, B., \& Gan, Z. 2023, Advanced Theory and Simulations, 6, 2300371, \dodoi{https://doi.org/10.1002/adts.202300371}

\end{thebibliography}
\end{document}